\documentclass[namedreferences]{solarphysics}

\usepackage[urlcolor=blue,colorlinks=true,linkcolor=red,citecolor=magenta,breaklinks=true]{hyperref}
\usepackage[optionalrh,natbib]{spr-sola-addons} 
\usepackage{graphicx}        
\usepackage{color}           
\usepackage{breakurl}        
\usepackage{rotating}        
\usepackage{upgreek}         
\usepackage[flushleft]{threeparttable}

\graphicspath{{figs/}}

\renewcommand{\vec}[1]{{\mathbfit #1}}

\renewcommand{\deg}{^\circ}

\newcommand{\Bsun}{\overline{B_\odot}} 

\newcommand{\fov}{field of view{\ }}
\newcommand{\fovnospace}{field of view}
\newcommand{\ie}{{i.e.}{\ }}
\newcommand{\eg}{{e.g.}{\ }}

\newcommand{\adsurl}[1]{\href{http://adsabs.harvard.edu/abs/#1}{ADS}}
\newcommand{\doiurl}[1]{\href{http://dx.doi.org/#1}{DOI}}

\begin{document}

\begin{article}

\begin{opening}

\title{Coronal Photopolarimetry with the LASCO-C3 Coronagraph over 24 Years [1996--2019]}
\subtitle{Application to the K/F Separation and to the Determination of the Electron Density}

\author[addressref={aff1},corref,email={philippe.lamy@latmos.ipsl.fr}]{\inits{P.}\fnm{Philippe}~\lnm{Lamy}\orcid{0000-0002-2104-2782}}
\author[addressref={aff1},email={hugo.gilardy@latmos.ipsl.fr}]{\inits{H. }\fnm{Hugo }\lnm{Gilardy}} 
\author[addressref={aff2},email={antoine.llebaria@lam.fr}]{\inits{A. }\fnm{Antoine }\lnm{Llebaria}} 	
\author[addressref={aff1},email={eric.quemerais@latmos.ipsl.fr}]{\inits{E. }\fnm{Eric }\lnm{Quemerais}} 
\author[addressref={aff2},email={fabrice.ernandes@hotmail.com}]{\inits{F. }\fnm{Fabrice }\lnm{Ernandez}} 	

\address[id=aff1]{Laboratoire Atmosph\`eres, Milieux et Observations Spatiales, CNRS \& UVSQ, 11 Bd d'Alembert, 78280 Guyancourt, France}
\address[id=aff2]{Laboratoire d'Astrophysique de Marseille, CNRS \& Aix-Marseille Universit\'e, 38 rue Fr\'ed\'eric Joliot-Curie, 13388 Marseille cedex 13, France}
       
\runningauthor{Lamy {\it et al.}}
\runningtitle{Coronal Photopolarimetry}
 
\begin{abstract}

We present an in-depth characterization of the polarimetric channel of the Large-Angle Spectrometric COronagraph/LASCO-C3 onboard the Solar and Heliospheric Observatory (SOHO).
The polarimetric analysis of the white-light images makes use of polarized sequences composed of three images obtained through three polarizers oriented at $+60\deg$, $0\deg$, and $-60\deg$, complemented by a neighboring unpolarized image.
However, the degradation of the $0\deg$ polarizer noticed in 1999 compelled us to reconstruct the corresponding images from those obtained with the two other polarizers and the unpolarized ones thereafter.
The analysis closely follows the method developed for LASCO-C2 (Lamy, P., et al. {\it Sol Phys} 295, 89 (2020)) and implements the formalism of Mueller, albeit with additional difficulties notably the presence of a non-axially symmetric component of stray light. 
Critical corrections were derived from a SOHO roll sequence and from consistency criteria (\eg the ``tangential'' direction of polarization). 
The quasi-uninterrupted photopolarimetric analysis of the outer corona over two complete Solar Cycles 23 and 24 was successfully achieved and our final results encompass the characterization of its polarization, of its polarized radiance, of the two-dimensional electron density, and of the K-corona.
Comparison between the C3 and C2 results in the region where their \fov overlaps shows an overall agreement. 
The C3 results are further in agreement with those of eclipses and radio ranging measurements to an elongation of $\approx$10\,R${}_\odot$ but tend to diverge further out. 
Although the coronal polarization out to 20\,R${}_\odot$ is still highly correlated with the temporal variation of the total magnetic field, this divergence probably results from the increasing polarization of the F-corona with increasing solar elongation.

\end{abstract}
\end{opening}

\section{Introduction}

Whereas the observation of the inner solar corona (defined here as extending to $\approx$3 R${}_\odot$ from the center of the solar disk) in polarized white-light has been actively pursued for decades with different purposes as summarized by \cite{Lamy2020}, the outer corona has received far less attention.
This is readily explained by the faintness of the brightness and of the polarization compared with the inner region so that stray components (stray light from the instrument and from the Earth atmosphere in the case of ground-based eclipse observations) become a major concern.
Ground-based coronagraphs even located at high altitude such as the Mark III, Mark IV, and K-Cor at the Mauna Loa Solar Observatory (Hawaii) have not helped as their polarized brightness $pB$ measurements are limited to a typical elongation of $\approx$1.5\,R${}_\odot$ from the center of the Sun.
The space coronagraphs of the first generation such as that flown aboard OSO-7 \citep{Koomen1975} and the Solar Maximum Mission (SMM) Coronagraph/Polarimeter  \citep{MacQueen1980} were hampered by the poor radiometric performances of their vidicon detector.
More recent space instruments obtained very valuable white-light images of the outer corona, notably the star tracker camera of the Clementine spacecraft over elongations of $3\deg \lesssim \epsilon \lesssim 30\deg$ \citep{Hahn2002} and the STEREO-A SECCHI/HI-1 heliospheric imager over elongations of $5\deg \lesssim \epsilon \lesssim 24\deg$ \citep{Stenborg2018} but they had no polarimetric capability.

We are therefore left with the eclipse measurements performed in the years 1952--1973 as presented in the review article of \cite{KoutchmyLamy1985}.
Table~\ref{Tab:Past_results} summarizes the information relevant to the six campaigns that took place during this time interval.
In comparison with Figure~4 of \cite{KoutchmyLamy1985}, we include the results of \cite{Michard1954a} but disregard those of \cite{Pepin1970} since this author displays a map of only a few contours of constant polarization (his Figure~3) which is so puzzling that it is retrospectively difficult to construct reliable equatorial and polar profiles.
With the exception of the results of \cite{Michard1965}, all other profiles along the equatorial direction agree on a continuous decrease from $\approx$0.18 at 3\,R${}_\odot$ to $\approx$0.08 at 6\,R${}_\odot$.
They diverge beyond, an effect that \cite{KoutchmyLamy1985} possibly attributed to the presence or absence of a window in aircraft observations.
The polarization along the polar direction is much fainter and the bulk of the measurements does not go beyond 5\,R${}_\odot$.

\begin{figure}[htpb!]
	\centering
	\includegraphics[width=\textwidth]{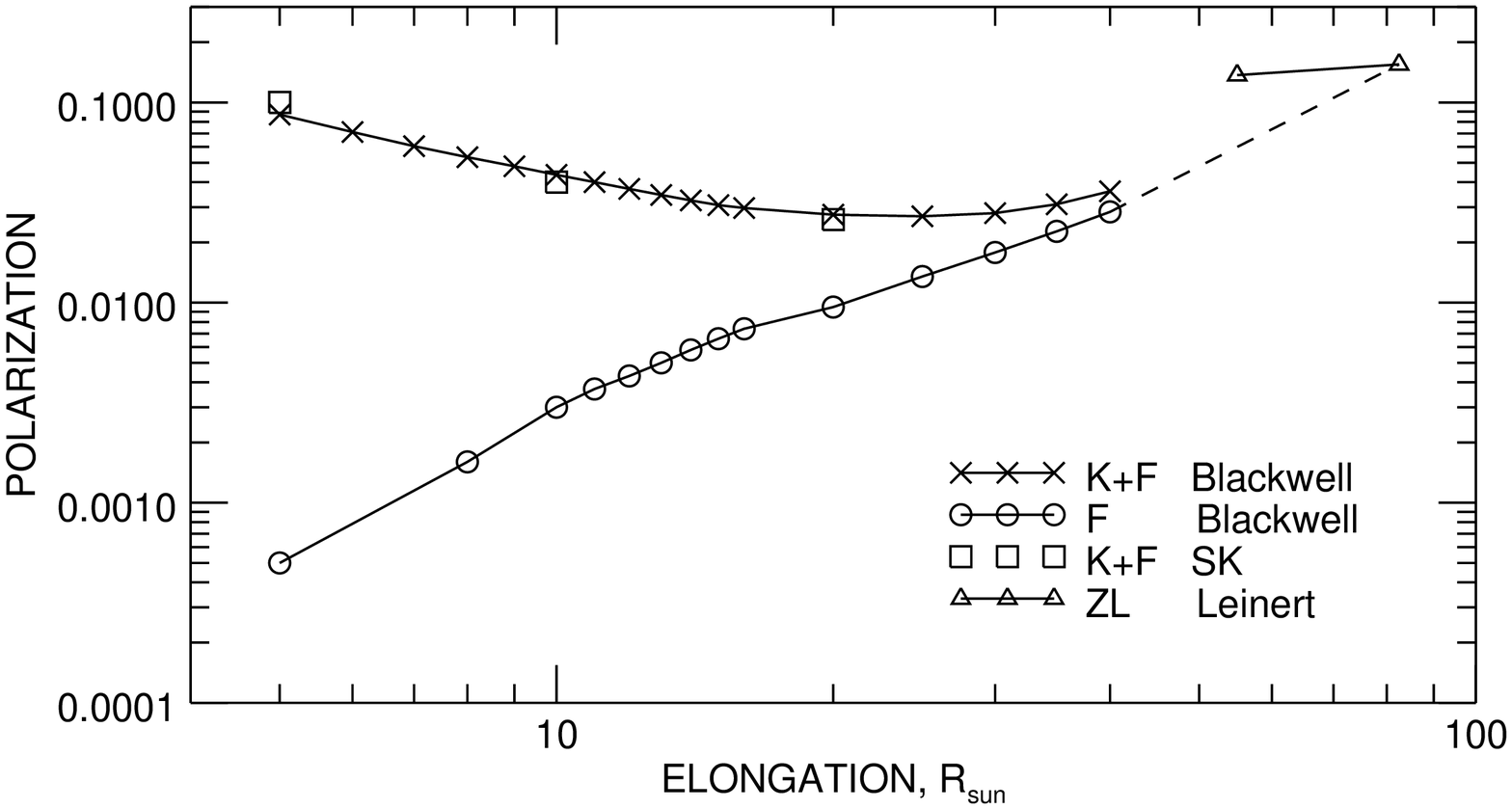}
	\caption{Polarization of the K+F and F coronae as a function of solar elongation in the equatorial direction from different sources.
	The ``Blackwell'' K+F and F data are from \cite{Blackwell1966}.
	The ``SK'' data are the compilation of two sources, \cite{Saito1972}, and \cite{Koutchmy1977} presented by S. Koutchmy in Section 14.8 ``Corona'' of the Fourth Edition of Allen's Astrophysical Quantities \citep{Cox2015}.
	Additional values for the Zodiacal Light are from \cite{Leinert1976}.
	The broken line is intended to show the satisfactory continuity between the data of \cite{Blackwell1966} and the second value of \cite{Leinert1976} at 82.5\,R${}_\odot$.}
	\label{Fig:Black1966}
\end{figure}

Whereas the polarization in the inner corona is determined  by the K component, in the outer corona it depends upon the interplay of the K and F components according to their respective weight as the elongation increases.
This implies that, at some elongation, there must be a turnover from an inner region where the polarization is dominated by the K component to an outer region where it is dominated by the F component according to the following equation:
\begin{equation}
p=\frac{p_K\;B_K+p_F\;B_F}{B_K+B_F}
\end{equation} 
and connecting to the Zodiacal Light further out.
In other words, we expect the radial variation of the polarization of the corona to slowly decrease with increasing elongation and then rise to progressively match the classical values of the Zodiacal Light at elongations $\epsilon\geq30\deg$, see for instance Figure~8\,c of \cite{Lamy1986}.
In between, there are presently only two measurements from the rocket flight of \cite{Leinert1974} corrected by \cite{Leinert1976}: $p=0.137$ at $\epsilon=14.3\deg$ (55\,R${}_\odot$) and $p=0.155$ at $\epsilon=21.0\deg$ (82.5\,R${}_\odot$).
Figure~\ref{Fig:Black1966} is intended to offer a view of this situation.
It regroups the above result of \cite{Leinert1976}, the polarization of an homogeneous, minimum K+F corona between 5 and 20\,R${}_\odot$ compiled in Allen's Astrophysical Quantities \citep{Cox2015} and the model of the outer corona constructed by \cite{Blackwell1966} which we briefly summarize.
The K/F separation was achieved based on measurements of the depth of the H$\alpha$ absorption line in the spectrum of the corona to an elongation of 16\,R${}_\odot$ and the resulting radial profile of the K-corona along the equatorial direction was inverted to yield a profile of the electron density which was subsequently smoothly extrapolated to 1 AU.
This latter profile was then used to calculate the brightness and polarization of the K-corona.
Combining with values of the K+F brightness from eclipse observations complemented by Zodiacal Light data, \cite{Blackwell1966} were able to obtain the radial variations of the brightness and the polarization of the K, F, and K+F coronae between 5 and 40\,R${}_\odot$ (their Table III).
An extrapolation of the K+F polarization curve of \cite{Blackwell1966} suggests a reasonable connection with the value of \cite{Leinert1976} at 82.5\,R${}_\odot$ whereas that at 55\,R${}_\odot$ is probably incorrect (quasi identical polarization values at so different elongations are extremely unlikely).
This model predicts i) a polarization of the F-corona remaining extremely low (less than 0.01) up to $\approx$20\,R${}_\odot$, and ii) a leveling-off of the total polarization $p$ followed by a turnover starting at about 30\,R${}_\odot$.
In reality, the location of this turnover depends upon several factors, prominently the activity of the Sun which directly influences the radiance and the geometry of the K-corona. 
As a matter of fact, the photometers at 16$\deg$ to the ecliptic plane aboard the Helios spacecraft detected variations of the polarization of the Zodiacal Light correlated with solar activity, prominently in the polar regions \citep{Leinert1989}.
The eclipse results listed in Table~\ref{Tab:Past_results} will be further illustrated and discussed when we will compare them with those coming from the LASCO-C3 observations.

\begin{table*}
\begin{threeparttable}
\caption{Summary of past results on the polarization of the outer corona along the equatorial and polar directions.}
\vspace{0.2cm}
\label{Tab:Past_results}
\begin{tabular}{ccccccc}
\hline 
Eclipse 		& Solar 			& Setup 			& Equ 				& Pol 				&	Wave-			& Ref	\\
						& activity 		&  						& range 			& range				& length		&			\\
						& ($\ast$)		& 		 				& R${}_\odot$	& R${}_\odot$	& nm 				&			\\
\hline 
1952-02-25	& -17					& Ground			& 3.0 -- 10.	& 3.0 -- 4.75	& 640				& a 	\\
1954-06-30	& 0					  & Aircraft		& 3.0 -- 20.	& 3.3 -- 6.6	& 630				& b 	\\
						&							& Windowless	&							&													&			\\
1955-06-20	& +12					& Aircraft		& 3.0 -- 16.	& 3.0 -- 6.0	& 556				& c 	\\
						&							& Windowless	&							&													&			\\
1961-02-15	& -41					& Ground			& 3.0 -- 10.	& 3.0 -- 4.0	& 640				& d 	\\
1963-07-20	& -13					& Aircraft		& 5.0 -- 40.	& -- 					& 664				& e 	\\
						&							& + Window		&							&													&			\\
1973-06-30	& -20					& Aircraft		& 3.0 -- 10.	& 3.0 -- 8.0	& ? 				& f 	\\
						&							& + Window		&							&													&			\\					
\hline
\end{tabular}
\begin{tablenotes}
\item ($\ast$) Time offset in months with respect to a minimum of solar activity
\item (a) \cite{Michard1954a}
\item (b) \cite{Blackwell1955}
\item (c) \cite{Michard1956}
\item (d) \cite{Michard1965}
\item (e) \cite{Blackwell1966}
\item (f) \cite{Mutschlecner1976}
\end{tablenotes}
\end{threeparttable}
\end{table*}

The Large-Angle Spectrometric COronagraph (LASCO), a set of three coronagraphs \citep{Brueckner1995} onboard the Solar and Heliospheric Observatory (SOHO) opened an entirely new era in the domain of photometric and polarimetric investigations of the white-light corona.
LASCO-C3, one of the two externally occulted coronagraphs of direct interest to this present article, has been in nearly continuous operation since January 1996, recording the solar corona from 3.7 to 30\,R${}_\odot$. 
Equipped with a CCD camera, it offers a good potential for polarization measurements although the selected technique, a set of three polarizers oriented at $+60\deg$, $0\deg$, $-60\deg$ and mounted on a wheel, is far from being ideal for the corona.
In spite of this potential and of the limited knowledge of the corona in this region, these C3 data have not received particular attention. 
Encouraged by our achievements with the C2 data as described in \cite{Lamy2020} (hereafter Paper I), we decided to analyze the C3 polarization data.
Since we implement the same technique, we presently limit its presentation to a summary and refer the interested reader to the aforementioned article for background information on the coronal polarization and for detail of the analysis.

Our article is organized as follows. 
Section~\ref{Sec:C3Op} describes the observations and the processing of the data.
Section~\ref{Sec:Analysis} presents the method of polarization analysis based on the Mueller formalism, the determination of the Mueller matrices and the calibration of the instrument.
The practical implementation of the analysis, the initial results and critical tests then reveals the necessity of additional corrections which are dealt with in Section~\ref{Sec:Improve}.
Final results for the polarization, the polarized radiance $pB$, the electron density, and the radiance of the K-corona over 24 years are presented in Section~\ref{Sec:Results} and
uncertainties are discussed in Section~\ref{Sec:Uncertainty}.
These results are subsequently compared with those coming from 
LASCO-C2 (Section~\ref{Sec:C2C3}), 
solar eclipses (Sections~\ref{Sec:Pol-Ecl} and \ref{Sec:K-Ecl}), 
and radio ranging measurements in Section~\ref{Sub:Ne-Ecl}.
We finally conclude in Section~\ref{Sec:Conclusion}.

\section{LASCO-C3 Observations and Data Processing}
\label{Sec:C3Op} 

\subsection{Observations}	
\label{Sub:Obs}

A LASCO polarization sequence is composed of three polarized images of the corona successively obtained with three polarizers oriented at +60${}^o$, 0${}^o$ and -60${}^o$ with respect to the direction of the rows ($\vec{X}$-axis) of the CCD detector and an unpolarized image altogether forming a quadruplet.
The polarizers themselves are a sandwich of a Polaroid foil (Kodak HN22) cemented in between two polished glass plates.
The polarization sequences are prominently taken with the ``orange'' filter (bandpass of 540-640\,nm) in the binned format of 512$\times$512 pixels in order to improve the signal-over-noise ratio of the polarized images.
A few sequences were taken in the full format of 1024$\times$1024 pixels and were subsequently rebinned to 512$\times$512 pixels.

\begin{figure}[htpb!]
	\centering
	\includegraphics[width=\textwidth]{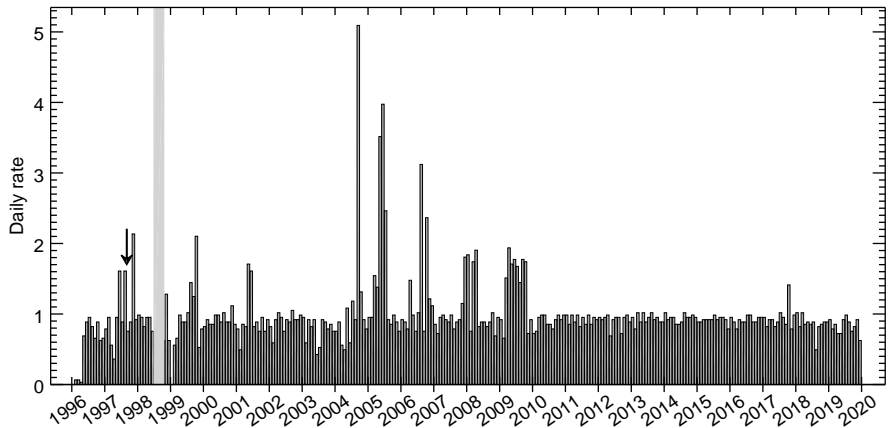}
	\caption{Monthly averaged daily rate of the polarization sequences obtained with the LASCO-C3 coronagraph with the orange filter in the binned format of 512$\times$512 pixels.
	  The arrow indicates the roll sequence of the spacecraft used for calibration.} 
	\label{Fig:ImRate}
\end{figure}

Figure~\ref{Fig:ImRate} displays the chronogram of the polarization sequences, typically one per day.
This routine program was affected by two major interruptions due to: i) the accidental loss of SOHO during a roll maneuver on 25 June 1998 which resulted in a long data gap until recovery on 22 October 1998, and ii) the failure of the gyroscopes which caused another gap from 21 December 1998 to 6 February 1999 when nominal operation resumed. 
Special polarization sequences took place a few times as shown by the peaks in the chronogram.
During its first years of operation, the attitude of SOHO was set such that its reference axis was aligned along the sky-projected direction of the solar rotational axis resulting in this direction being ``vertical'' (\ie along the column or $\vec{Y}$-axis of the CCD detector) with solar north up on the LASCO images.
Starting on 10 July 2013 and following the failure of the motor steering its antenna, SOHO was periodically (every three months) rolled by 180$\deg$ to maximize telemetry transmission to Earth.
On 29 October 2010 and still on-going, the attitude of SOHO was changed to simplify operation, the reference orientation being fixed to the perpendicular to the ecliptic plane causing the projected direction of the solar rotational axis to oscillate between $\pm$7$\deg$ 15' around the ``vertical'' direction on the LASCO images.

\subsection{Preprocessing of the C3 Images}	
\label{Sub:DataProc}

The original data stream coming from the spacecraft represents the lowest level data, known as Level-0.
As explained in Paper I, they were processed by the LASCO team at the Naval Research Laboratory to produce FITS files with documented headers forming the Level-0.5 data set of individual images; no correction was applied at this stage.
This process was slowed down in 2015 and eventually terminated.
As a consequence, we decided in October 2015 to switch to the ``quick look data'' produced at the Goddard Space Flight.
Strictly speaking, these two data sets Level-0.5 and ``quick look'' are identical except for slightly less missing telemetry blocks in the former data set.

The LASCO team at the Laboratoire d'Astrophysique de Marseille (formerly Laboratoire d'Astronomie Spatiale) developed a two-stage procedure which corrects for all instrumental effects and process the raw data to scientific images of the corona.
The first stage consists in a preprocessing applied to all images and it performs the following tasks.
\begin{itemize}
\item Bias correction.
The bias level of the CCD detector evolves with time; it is continuously monitored using specific blind zones and systematically subtracted from the images.
\item Exposure time equalization.
Small random errors in the exposure times are corrected using a method developed by \cite{Llebaria2001} in which relative and absolute correction factors are determined.
This method works extremely well for the routine (unpolarized) images because of their high cadence but less so for the less frequent polarized images (one per day). 
The Naval Research Laboratory later developed an alternative method with similar performances \citep{Morrill2006}.
\item Missing block correction.
Telemetry losses result in blocks of $32 \times 32$ pixels sometime missing in the images.
Different solutions are implemented to restore the missing signal depending upon the location of these blocks \citep{Pagot2014}.
\item Cosmic rays correction.
The impacts of cosmic rays (and stars as well) are eliminated from the images using the procedure of opening by morphological reconstruction developed by \cite{Pagot2014}.
\end{itemize}
The second stage is applied to the polarization sequences and is described in the next section.

\subsection{Processing of the C3 Polarized Images}	
\label{Sub:DataProcPol}

The polarized images further undergo the following processes.
\begin{itemize}	
	\item   
	Rebinning to 512$\times$512 pixels. This practically applies to the few 1024$\times$1024 pixels images to bring them to a common format for polarization analysis. 
	\item
	Correction for the stray light ramp.
	Unlike C2, C3 is affected by a non-axially symmetric stray light pattern best defined as a ramp roughly aligned with the NW-SE diagonal of the square \fovnospace .
	This effect was thoughtfully investigated by \cite{Morrill2006} who conjectured that it most likely results from a reflection from the CCD surface.
	We did not make use of the map of the ramp that they determined (their Figure~20) as we later realized that their determination suffers from a major flaw that led us to re-consider the problem and look for a correct solution.
	This is described in detail in Appendix I.
	\item
	Polarimetric analysis based on the Mueller procedure.
	It is applied to each triplet of polarized images and returns images of the total radiance, the polarized radiance, the polarization and the angle of polarization.	
	\item Vignetting correction.
	This vignetting introduced by the occulters is removed from the radiance images using a geometric model of the two-dimensional vignetting function of C3 \citep{Llebaria2004}.
	It is further combined with a cos$^4(\phi)$ correction for the optical vignetting where $\phi$ is the incidence angle of any light ray corresponding to a pixel. 
	\item
	Absolute calibration of the total radiance derived from the photometric measurements of stars present in the C3 \fov \citep{Thernisien2006}.	 
	\item
	K/F separation.
	\item
	Two-dimensional inversion of the polarized brightness to retrieve the electron density.
\end{itemize}

\section{Analysis of the LASCO-C3 Polarized Images}
\label{Sec:Analysis} 

We strictly follow the method implemented in Paper I for the analysis of the LASCO-C2 polarized images as briefly summarized below.
It relies on the formalism of \cite{Mueller1943} which relates the vector of the four Stokes parameters I, Q, U, V of the incident beam $\vec{S_{in}}$ to that of the
output beam $\vec{S_{out}}$ via:
\begin{equation}
\vec{S_{out}}= \vec{M}\;\vec{S_{in}}
\end{equation}
where the $4\times4$ Mueller matrix $\vec{M}$ characterizes the optical system.
In the case of the corona and since its polarization is linear and tangential, the analysis requires only three coefficients $m_{i,j}$ (instead of sixteen) for each of the three configurations corresponding to the three polarizers, that is a total of nine coefficients.

Let us introduce the fixed coordinate system defined by the center of the Sun and the direction of the rows ($\vec{X}$-axis) and that of the columns ($\vec{Y}$-axis) of the CCD detector.
By construction, the principal axis of maximum transmittance of the polarizer oriented at $0\deg$ is closely aligned with the $\vec{X}$-axis which is itself closely aligned with the equatorial direction when the SOHO roll angle is either $0\deg$ or $180\deg$.
Conversely, the $\vec{Y}$-axis is closely aligned with the polar direction under the same circumstances, hence the notation (O, $\vec{X_{equ}}$, $\vec{Y_{pol}}$) adopted for this coordinate system in Paper I.
Let $\vec{S_{cp}}=[I_{cp},Q_{cp},U_{cp},0]^{-T}$ and $\vec{S_{p}}$ be the Stokes vectors of the incoming and outgoing coronal light, respectively  expressed in this coordinate system.
We can then write:  
\begin{equation}
\vec{S_{p}}=\vec{M}\;\vec{S_{cp}}
\end{equation}
and in particular for the total intensity:
\begin{equation}
I_{p}=m_{11}\;I_{cp}+m_{12}\;Q_{cp}+m_{13}\;U_{cp}.
\end{equation}
When using the three polarizers successively, we secure three images of the polarized radiance of the corona such that we have at each pixel:
\begin{eqnarray}
\nonumber I_{1} & = & m_{11}(0)\;I_{cp} +m_{12}(0)\;Q_{cp}+m_{13}(0)\;U_{cp} \\ 
I_{2} & = & m_{11}(-60)\;I_{cp} +m_{12}(-60)\;Q_{cp}+m_{13}(-60)\;U_{cp} \\
 \nonumber I_{3} & = & m_{11}(+60)\;I_{cp} +m_{12}(+60)\;Q_{cp}+m_{13}(+60)\;U_{cp}
\end{eqnarray}
where $0$, $-60$, and $+60$ correspond to the orientation of the three polarizers.
$I_{cp}$, $Q_{cp}$, and $U_{cp}$ are all in units of intensity, as are $I_{1}$  $I_{2}$, and $I_{3}$. 
These latter intensities are known once instrument calibrations are known, if needed.
They can be counts such as DN/sec if absolute calibrations are not needed as it is the case for polarimetry.

Let us introduce the so-called IPMV (Intensity Polarization Modification Vector) matrix:
\begin{equation}
\vec{\chi}=\left( \begin{array}{ccc}
m_{11}(0) & m_{12}(0) &  m_{13}(0)\\
m_{11}(-60) & m_{12}(-60) & m_{13}(-60) \\
m_{11}(+60) & m_{12}(+60) & m_{13}(+60)
\end{array} \right)
\end{equation}
The problem simplifies to inverting $\vec{\chi}$ so as to determine the Stokes vector $\vec{S_{cp}}$ via:
\begin{equation}
\left( \begin{array}{c}
I_{cp} \\ Q_{cp} \\ U_{cp}
\end{array} \right) =
\chi^{-1}\left( \begin{array}{c}
I_{1} \\ I_{2} \\ I_{3}
\end{array} \right)
\end{equation}
from which we obtain the polarization $p_{cp}$ and its local angle $\alpha_{c}$ via:
\begin{equation}
p_{cp} = \sqrt{\frac{Q_{cp}^{2}+U_{cp}^{2}}{I_{cp}^{2}}} \nonumber\\
\end{equation}
\begin{equation}
\tan 2\;(\alpha_{c}+\varphi) = \frac{U_{cp}}{Q_{cp}}
\end{equation}
where $\varphi$ is the position angle of the considered pixel (see Figure~1 of Paper I).
We thus determine $\alpha_{c}$ and compare it to its theoretical values $\alpha_{c}=90\deg$ as a test of the quality of the measurements.

Up to now, it is implicitly assumed that all quantities are expressed in absolute unit of radiance and therefore the Mueller and the IPMV matrices must have absolute, dimensionless coefficients which correctly relate radiances. 
Their determination requires that both the input and output Stokes vectors be expressed in absolute unit of radiance such as $W\,m^{-2}\,st^{-1}\,\mu{}m^{-1}$, a very challenging calibration task further complicated by the vignetting inherent to externally occulted coronagraphs. 
We explain in the next section that it is in practice possible to work with relative coefficients and perform at the end, a global calibration of the total radiance.

\subsection{Determination of the Mueller Matrix}
\label{Sub:MuellerMatrix}

Paper I explained the difficulties of a global calibration of the instrument to determine the required nine coefficients of the Mueller matrix and that led to favoring the alternative approach of component calibration.
Contrary to C2, C3 has no folding mirrors so that its Mueller matrix is simply that of a linear polarizer: 
\begin{equation}
{\scriptsize
\vec{M_{P}(\theta)}=\frac{k_{1}}{2}\;
\left[ \begin{array}{cccc}
1+\varepsilon & (1-\varepsilon)\cos 2\theta &  (1-\varepsilon)\sin 2\theta & 0 \\
(1-\varepsilon)\;\cos 2\theta & (1-\sqrt{\varepsilon})^{2}\;\cos^{2} 2\theta+2\sqrt{\varepsilon} & (1-\sqrt{\varepsilon})^{2}\;\cos 2\theta\;\sin 2\theta & 0 \\
(1-\varepsilon)\;\sin 2\theta & (1-\sqrt{\varepsilon})^{2}\;\cos 2\theta\;\sin 2\theta & (1-\sqrt{\varepsilon})^{2}\;\sin^{2} 2\theta+2\sqrt{\varepsilon} & 0 \\
0 & 0 & 0 & 2\sqrt{\varepsilon}
\end{array} \right]}
\end{equation}
\noindent where $\varepsilon$ is the ratio of the principal transmittances of the polarizer k$_{1}$ and k$_{2}$: $\varepsilon=k_{1}/k_{2}$. 
The three Mueller matrices characterizing the three linear polarizers in a given spectral band are therefore given by setting $\theta$ equal to the three angles $0\deg$, $-60\deg$, and $+60\deg$.

As shown in Figure~2 of Paper I, the ``orange'' filter is broad enough that the spectral variation of the principal transmittances $k_{1}$ and $k_{2}$ of the polarizers must be taken into account.
Therefore the Mueller coefficients were averaged by considering the transmissions of the optics $T_{0}(\lambda)$, of the filters $T_{f}(\lambda)$, the quantum efficiency of the CCD $\eta(\lambda)$, and the spectrum of the coronal light which is nearly similar to that of the Sun $B_{\odot}(\lambda)$:
\begin{equation}
\overline{m_{ij}}=\frac{\int_{\lambda_{1}}^{\lambda_{2}}\;m_{ij}\;T_{0}(\lambda)\;T_{f}(\lambda)\;\eta(\lambda)\;B_{\odot}(\lambda)\;d\lambda}{\int_{\lambda_{1}}^{\lambda_{2}}\;T_{0}(\lambda)\;T_{f}(\lambda)\;\eta(\lambda)\;B_{\odot}(\lambda)\;d\lambda}
\end{equation}
where the units of $B_{\odot}(\lambda)$ must involve photons (\eg photon\,sec$^{-1}$\,cm$^{-2}$\,st$^{-1}$).
Note that we neglect the slight reddening of the F-corona.
Table~\ref{table:mcoeff} displays the first three coefficients of the Mueller matrix for the ``orange'' filter and the three polarizers of LASCO-C3.

As shown in Paper I, the principal transmittances of the polarizers may not be uniform over their surfaces and those of C2 exhibited variations up to 8\%.
Unfortunately maps of the $k_{1}$ transmittance such as those displayed in Figure~6 of Paper I are not available for the C3 polarizers: either they were not calibrated in the laboratory or the data were lost.

\begin{table}
\caption{The first three coefficients of the Mueller matrix for the ``orange'' filter and the three polarizers of LASCO-C3.}
\begin{tabular}{ccccc} 
\hline
Filter & Polarizer & $\mbox{m}_{11}$ & $\mbox{m}_{12}$ & $\mbox{m}_{13}$\\[3pt]
\hline
Orange & $0\deg$ & 0.246 & 0.246 & 0. \\
Orange & $-60\deg$ &  0.246 & -0.123 & -0.213 \\
Orange & $+60\deg$ & 0.246 & -0.123 & 0.213\\
\hline
\end{tabular}
\label{table:mcoeff}
\end{table}

\subsection{Calibration of the Total Radiance}
\label{Sub:CalibRadiance}

The calibration of $I_{cp}$ follows the procedure developed in Paper I and relies on the routine unpolarized images $I_{0}$ systematically taken before or after the three polarized images. 
The calibration of these images $I_{0}$ is derived from that of the routine images taken in the full format of 1024$\times$1024 pixels with the same filter.
The calibration factors relating counts (in DN/sec) to the units of mean solar radiance $\Bsun$ were determined using stars present in the \fov of these images by \cite{Thernisien2006} who reported their values averaged over the first 7.5 years of LASCO operation (their Table~II).
Unfortunately, this work has not been extended to the following years and the question arises as to whether there is any temporal evolution.
\cite{Thernisien2006} did consider this question for the ``clear'' filter and found a constant degradation of $0.44\pm0.1$\% per year. 
It is of course unclear whether the orange filter has experienced the same trend (which would point to a progressive reduction of the sensitivity of the CCD detector) or another trend. 
As a conservative approach, we adopted the averaged calibration factors reported by \cite{Thernisien2006} and will later check their validity in Section~\ref{Sec:C2C3} by comparing the C3 images with those of LASCO-C2 whose calibration is continuously monitored (see Figure~9 of Paper I).

\begin{figure}[htbp!]
	\centering
	\includegraphics[width=\textwidth]{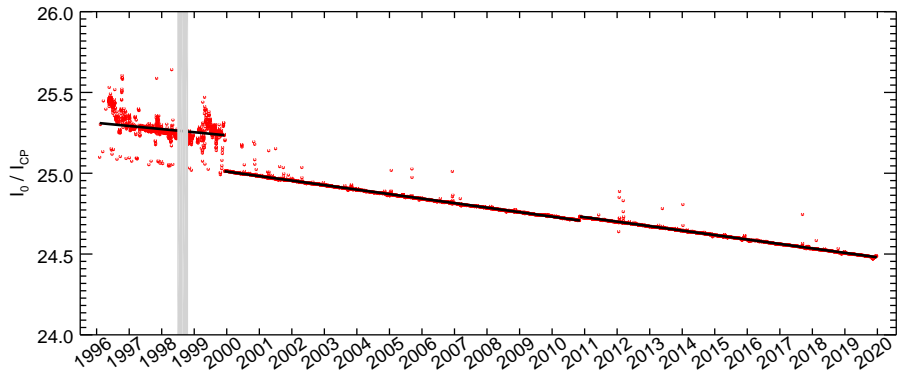}
	\caption{Temporal variations of the ratio $I_{0}/I_{cp}$ of the intensity of the unpolarized image to that resulting from the polarization analysis.}
	\label{Fig:C3CalCoef}
\end{figure}

Then for each quadruplet $I_{0}$, $I_{1}$, $I_{2}$, and $I_{3}$, we calculated the mean value of the ratio $I_{0}/I_{cp}$ where $I_{cp}$ results from the polarization analysis of $I_{1}$, $I_{2}$, and $I_{3}$. 
Figure~\ref{Fig:C3CalCoef} displays the temporal variations of this ratio and reveals a continuous decrease which reflects the global degradation of the transmittance of the three polarizers which amounts to a modest 1.8\% over 20 years.
Combining the calibrations of $I_{0}$ and of $I_{0}/I_{cp}$ allows calibrating the polarized radiance $pB$ of the corona and the radiance of the K-corona as observed by LASCO-C3 in units of $\Bsun$.

\subsection{Separation of the K and F Coronae}
\label{Sub:Separation}

We performed the K/F separation on the basis of the classical assumption that the F-corona is unpolarized ($p_{F}=0$) as well as the stray light ($p_{S}=0$).
The second assumption is justified on the basis that the stay light mostly results from light diffracted by the various occulters, apertures and stops, and is therefore axially symmetric (except for a narrow sector corresponding to the pylon holding the occulters) and unpolarized to first order.
It is generally considered that the first assumption holds up to an elongation of $\approx$ 6\,R${}_\odot$ as confirmed by the results of Paper I.
The windowless aircraft measurements during eclipses and the model derived by \cite{Blackwell1966} as presented in the Introduction suggest that the polarization of the F-corona remains very low beyond, possible up to 20\,R${}_\odot$.
In fact, it is quite interesting to test to what extend the assumption $p_{F}\approx0$ holds and may be used for the K/F separation allowing us writing:
\begin{equation}
pB=p_K\;B_K.
\end{equation}
We then followed the first route proposed in Paper I where $B_K$ is calculated by taking advantage of the robust ``asymptotic'' behaviour of $p_{K}(r)$ beyond $\approx$ 2.2\,R${}_\odot$, that is $p_{K}(r)\approx0.64$.

\section{Improvements of the Polarization Analysis}
\label{Sec:Improve}

\begin{figure}[htbp!]
	\centering
	\includegraphics{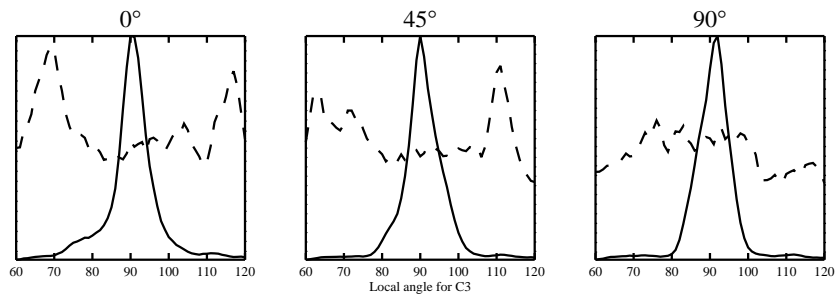}
	\caption{Histograms of the local angle of polarization calculated from images obtained during the roll sequence of September 1997.
	The three panels correspond to the three roll angles $0\deg$, $45\deg$ and $90\deg$ as indicated.
	Dashed lines: initial results. Solid lines: results after correcting for the global transmittance of the polarizers. 
	The histograms have been arbitrarily scaled for better legibility.}
	\label{Fig:HistC3}
\end{figure}

\begin{figure}[htbp!]
	\centering
	\includegraphics[width=\textwidth]{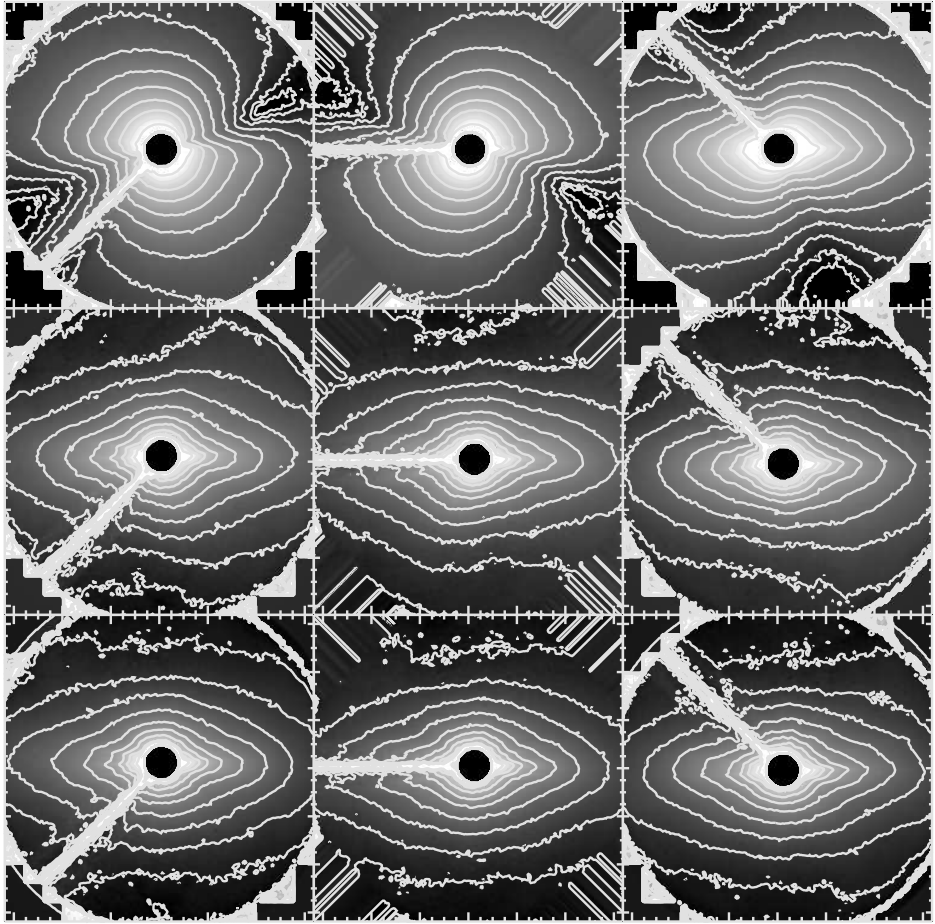}
	\caption{LASCO-C3 images of the radiance $B_K$ of the K-corona calculated from polarization sequences secured during the SoHO roll sequence of September 1997. 
	The three columns correspond to roll angles of $0\deg$ (left), $45\deg$ (middle), and $90\deg$ (right). 
	Upper row : basic polarization analysis with no corrections. 
	Second row : the correction for the global transmission of the polarizers has been introduced.
	Third row : the correction $S(x,y)$ has been further introduced.}
	\label{Fig:BkC3}
\end{figure}

The first run of the polarization analysis of the C3 images revealed problems similar to, and in fact worse, than those encountered with the C2 images.
First, the distributions of the local angle of polarization was extremely broad and not even centered at $90\deg$, the canonical value for tangential polarization of both the K and F coronae (Figure~\ref{Fig:HistC3}).
Second, the crucial test of the shape of the K-corona derived from the observations secured during the roll sequence of September 1997 when SOHO was allowed to dwell at specified roll angles (Table~\ref{Table:rollsequence}) revealed inconsistent images of the K-corona totally at odd with a corona of the minimum type (Figure~\ref{Fig:BkC3}, upper row). 
Although this roll sequence extended over $\approx20$ hours, the large scale corona was not expected to change much at a time close to the minimum of activity so this cannot explain the pronounced discrepancies.

\begin{table}
\caption{Journal of the C3 images taken during the roll sequence of September 1997.}
\label{Table:rollsequence}
\begin{tabular}{c c c c c}
\hline
Date & Time & Roll angle  & Polarizer & Exp. Time (sec) \\[3pt]
\hline
1997 Sep. 02 & 22:38:09 UT & $0\deg$    &  None    & 90.09 \\
1997 Sep. 02 & 22:41:43 UT & $0\deg$    &  -60 Deg  & 300.09 \\
1997 Sep. 02 & 22:48:46 UT & $0\deg$    &  0 Deg    & 300.29  \\
1997 Sep. 02 & 22:55:50 UT & $0\deg$    &  +60 Deg  & 300.09 \\
1997 Sep. 03 & 09:21:40 UT & $45\deg$   &  None    & 90.09 \\
1997 Sep. 03 & 09:25:14 UT & $45\deg$   &  -60 Deg  & 300.29 \\
1997 Sep. 03 & 09:32:18 UT & $45\deg$   &  0 Deg    & 300.09 \\
1997 Sep. 03 & 09:39:23 UT & $45\deg$   &  +60 Deg  & 300.09 \\
1997 Sep. 03 & 18:12:36 UT & $90\deg$   &  None    & 90.09 \\
1997 Sep. 03 & 18:16:09 UT & $90\deg$   & -60 Deg  & 300.095 \\
1997 Sep. 03 & 18:23:13 UT & $90\deg$   & 0 Deg & 300.093 \\
1997 Sep. 03 & 18:30:17 UT & $90\deg$   & +60 Deg & 300.091 \\
\hline
\end{tabular}
\end{table}

\subsection{Adjustment of the Transmission of the Polarizers}
Likewise C2, slightly different transmissions of the polarizers were found to be the cause of the above problems as also proposed by \cite{Moran2006} who derived correction factors for the transmission of the C3 polarizers.
Following Paper I, we determined these factors by minimizing the width of the histograms of the local angle of polarization and considered not a single image as done by \cite{Moran2006}, but the whole set of the three polarization sequences obtained during the roll maneuver of September 1997.
In addition to that obtained at a roll angle of $0\deg$, the other two offer a critical test since they were secured at $45\deg$ and $90\deg$, two angles markedly different from the  orientations of the polarizers (Table \ref{Table:rollsequence}). 
The corrections turned out to be more severe than those found in the case of C2 (for which only one transmission was decreased by 2\%) and the optimal correction factors are given in Table~\ref{table:coeffadj}.
They are consistent with those obtained by \cite{Moran2006}, but one should realize the extreme sensitivity of the factors up to the third decimal places which are therefore significant.
Our correction turned out to be extremely efficient in reducing the full width at half maximum of the distributions to $\approx 8\deg$ (Figure~\ref{Fig:HistC3}).
Figure~\ref{Fig:BkC3} dramatically illustrates the improvement of the shape of the K-corona resulting from our corrections when comparing the first and second rows. 
The north-south  distortions have almost disappeared and the three $B_K$ images obtained at the three different roll angles in September 1997 are close to identical.
There does however remain some discrepancies, for instance in the north-east quadrant, and we explain in the next section how they were corrected.

\begin{table}
\caption{Correction factors for the global transmissions of the C3 polarizers.
The fourth column displays the results of \cite{Moran2006} and the fifth column the ratio of the two results.}
\begin{tabular}{ccccc} 
\hline
Filter 	& Polarizer & Factor& Factor& Ratio \\[3pt]
\noalign{\vskip -2mm}
 & & This work & Moran & \\[3pt]
\hline
Orange 	& $+60\deg$		& 1.083	 	& 1.080		& 1.0028\\
Orange	& $0\deg$			& 0.967		& 0.965		& 1.0021\\
Orange 	& $-60\deg$		& 0.994		& 1.000		& 0.9940\\
\hline
\label{table:coeffadj}
\end{tabular}
\end{table}

\subsection{Global Correction}
\begin{figure}[htpb!]
	\centering
	\includegraphics[width=0.7\textwidth]{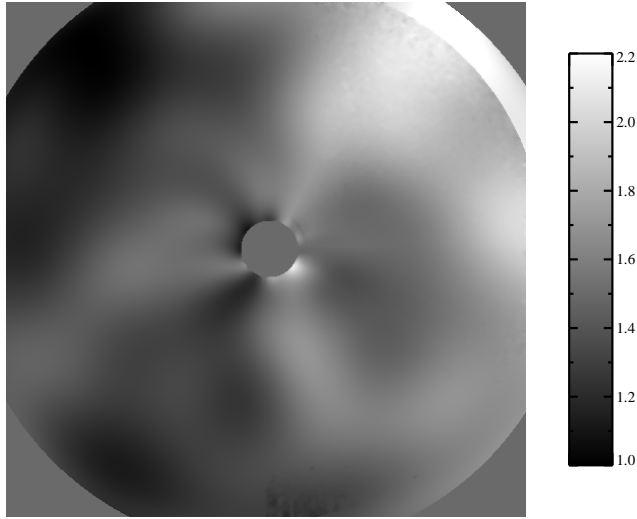}
	\caption{The LASCO-C3 correction pattern $S(x,y)$ built from the images taken during the roll maneuver of September 1997.}
	\label{Fig:CrossC3}
\end{figure}

Here again, we follow Paper I in introducing a global function $S(x,y)$ correcting for the remaining discrepancies that are difficult to track to a specific problem or problems with the optical components.
This function $S(x,y)$ is intended to be directly applied to the images produced by the polarization analysis in the ($\vec{X}$, $\vec{Y}$) reference frame introduced  in Section~\ref{Sec:Analysis}.
It was derived from the roll sequence of September 1997 according to a procedure described in Appendix I of Paper I which is therefore not repeated here and its image is displayed in Figure~\ref{Fig:CrossC3}.
The three radial structures are roughly reminiscent of the orientations of the three polarizers but the most important correction affects the north-west quadrant. 
The lower row in Figure~\ref{Fig:BkC3} clearly shows that the introduction of the $S(x,y)$ correction improves all image, but particularly those at roll angles of $0\deg$ and $90\deg$.

\subsection{Degradation of the $0\deg$ Polarizer}
\begin{figure}
	\centering
	\includegraphics{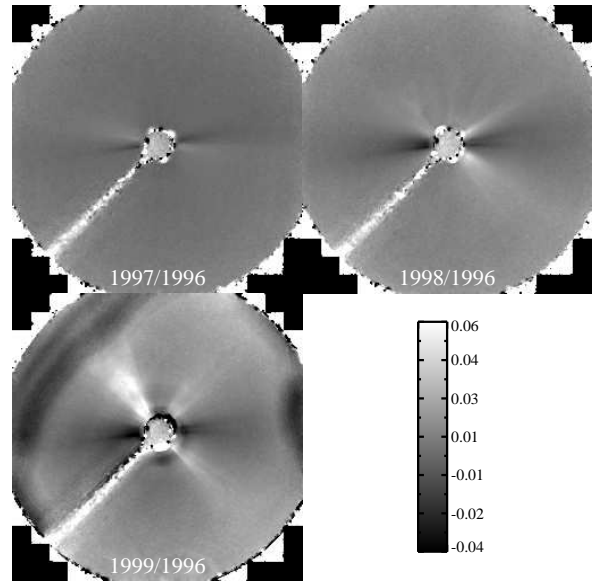}
	\caption{Temporal evolution of the LASCO-C3 polarizer oriented at $0\deg$. 
	Successive ratios of images taken with this polarizer using a reference image taken in 1996 are displayed with a logarithmic scale (see text for detail). 
	The last ratio ``1999/1996'' clearly shows the degradation of the polarizer.}
	\label{Fig:C30degDefault}
\end{figure}

In the course of testing the polarization results, we discovered a progressive degradation of the C3 polarizer oriented at $0\deg$ to the point of making it useless and terminating the acquisition of the corresponding images.
This is illustrated in Figure~\ref{Fig:C30degDefault} where we display successive ratios of images taken with this polarizer using a reference image taken in 1996.
All these images were obtained at the same period of the year, namely the June nodes (\ie when SOHO crosses the plane of symmetry of the Zodiacal Cloud) and typically ten images were averaged at each node.
The origin of the degradation was possibly traced to the loss of SOHO in 1998 which caused the temperature to severely drop well below specifications.
A slight defect of this polarizer, which otherwise would have remained unnoticed, could have triggered a separation inside the polarizer sandwich.

We outline below a method implemented after 1 August 1998 to generate the missing $0\deg$ polarized image $I_{2}$ (using our notations) from the other two polarized images $I_{1}$ and $I_{3}$ and the associated unpolarized image $I_{0}$ systematically taken before or after the polarized images.
First, using all quadruplets obtained until June 1998, we applied the appropriate corrections to these images (in particular for the ramp and for the global transmission of the polarizers), and calculate the mean value of the ratio $I_0/(I_1+I_2+I_3)$.
Note that this ratio is different from that introduced in Section~\ref{Sub:CalibRadiance}.
An analysis of the temporal variation of this ratio indicated that it remained nearly constant at a value of 1.32 with pseudo-periodic oscillations whose amplitude did not exceed $\pm1.5\% $ that we consequently neglected.
Assuming that this result holds true during the following years, it is straightforward to calculate $I_{2}$ from the above equation.
This image obviously does not need any correction and can directly be used in the polarization analysis along with the corrected $I_{1}$ and $I_{3}$ images.
We had of course no mean to re-derive the $S(x,y)$ correction function and therefore used that obtained from the September 1997 roll sequence.
As a test, we analyzed the polarized images obtained during the node of June 1996 using the original $I_{2}$ images on the one hand and the reconstructed $I_{2}$ images on the other hand.
The agreement is satisfactory, but we however observe that the latter approach leads to slightly larger polarization values, typical differences amounting to 0.005, and reaching 0.01 in the worst case at the edge of the C3 \fovnospace.
The ratio $I_{0}/I_{cp}$ needed to calibrate the total radiance (Section~\ref{Sub:CalibRadiance}) whose temporal variation is displayed in Figure~\ref{Fig:C3CalCoef} was naturally calculated using the reconstructed $I_{2}$ image when the above procedure was implemented.

\subsection{Possible Degradation of the $+60\deg$ and $-60\deg$ Polarizers}
\begin{figure}
	\centering
	\includegraphics[width=0.75\textwidth]{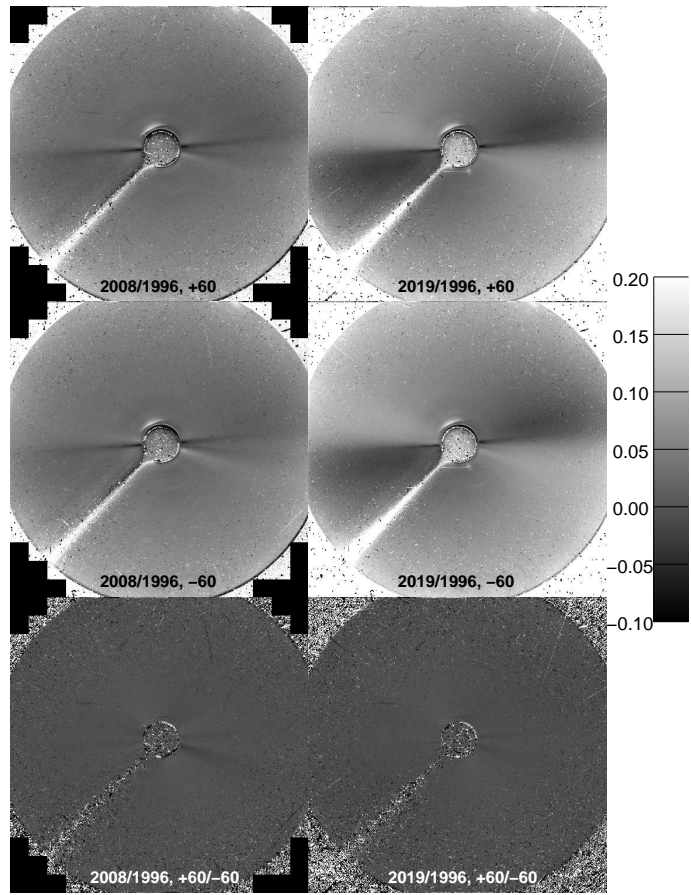}
	\caption{Temporal evolution of the LASCO-C3 polarizers oriented at $+60\deg$ and $-60\deg$. 
	Successive ratios of images taken during solar minima using a reference image taken in 1996 are displayed in the upper row ($+60\deg$) and middle row ($-60\deg$).
	The lower row displays super-ratio images ``$+60/-60$'' (see text for detail).
	All images are displayed using a logarithmic scale.}  
	\label{Fig:ratio_p60m60}
\end{figure}

In view of the above problem with the $0\deg$ polarizer, we wondered about the behaviour of the other two polarizers and investigated this question using the same procedure.
We considered the successive solar minima to minimize the influence of the corona and constructed ratio images ``2008/1996'' and ``2019/1996'' for the $+60\deg$ and $-60\deg$ polarizers as diplayed in Figure~\ref{Fig:ratio_p60m60}.
Whereas the ``2008/1996'' image ratios are fairly uniform, this is less so for the ``2019/1996'' case although the patterns of sectors of different brightness look similar.
To further investigate this aspect, we constructed super-ratio images ``$+60/-60$'' which unambiguously demonstrate the absence of differential effects between the two polarizers until 2019. 
This leaves open the possibility that the two polarizers underwent a strictly similar temporal evolution and indeed, similar aging cannot be excluded.
But this would only affect the absolute calibration of the radiance and not the polarimetric analysis, an important point coming out of this exercise.

\subsection{Results for the Improved Polarization Analysis}
\begin{figure}[htbp!]
	\centering
	\includegraphics[width=\textwidth]{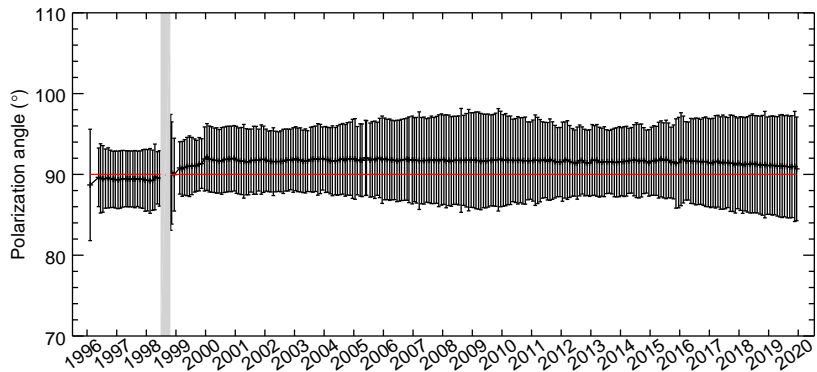}
	\caption{Temporal evolution of the local angle of polarization over 24 years. 
	The error bars represent the standard deviations of the monthly values.}
	\label{Fig:EvolHistC3}
\end{figure}

\begin{figure}[htbp!]
	\centering
	\includegraphics[width=\textwidth]{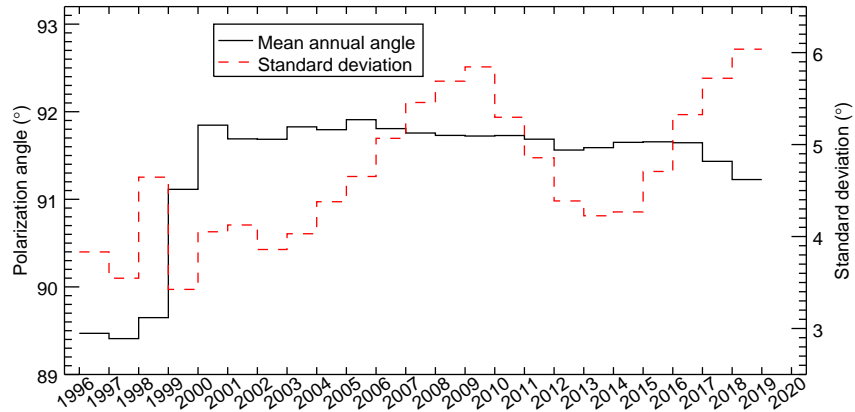}
	\caption{Temporal evolution of the annually averaged value of the local angle of polarization (left scale) and of its standard deviation (right scale) over 24 years.}
	\label{Fig:StatParamPolC3}
\end{figure}

Figures~\ref{Fig:EvolHistC3} and \ref{Fig:StatParamPolC3} respectively display the monthly and annually averaged values of the local angle of polarization $\overline{\alpha_c}$ and their standard deviations throughout the 24 years of LASCO-C3 observation.
The corrections are extremely effective during the first three years [1996--1998] during which the local polarization angle is just half a degree off the theoretical value of 90$\deg$.
Thereafter, this deviation slightly increases to a constant value of nearly 2$\deg$ until 2016 when a progressive decrease is observed, an effect probably related to the evolution of the remaining two polarizers.
We felt that this is a quite satisfactory achievement in view of the loss of one polarizer and the numerous corrections (and underlying assumptions) that had to be applied.
Consequently, we decided that the complex and lengthy optimization procedure introduced to bring the local angle of polarization to strictly 90$\deg$ as implemented in the case of C2 was not warranted in the case of C3.
It is interesting to note that, except for the first years of operation, the standard deviation varies in opposition with the solar cycle as found in the case of C2.
The same interpretation naturally holds: in phases of high activity, the increased number of highly polarized streamers improves the signal-over-noise ratio.
Consequently, the standard deviation of values is anti-correlated with the solar cycle as conspicuously seen in Figure~\ref{Fig:StatParamPolC3}.

\section{Results for the Photopolarimetric Properties of the Corona}
\label{Sec:Results}

Slightly over 8500 sequences of polarization have been accumulated by LASCO-C3 at the end of 2019 and it is quite challenging to present such a large amount of data.
We present below a synthesis aimed at giving an overview of the two-dimensional photopolarimetric properties of the corona and the derived science products over two solar cycles.

\begin{itemize}
	\item Maps at different phases of solar activity during each of the two solar cycles SC 23 and SC 24 as well as the corresponding profiles along the equatorial and polar directions.
	\item Monthly averaged temporal variations at 10\,R${}_\odot$, and additionally 20\,R${}_\odot$ in the case of the polarization and the polarized radiance. 
	These curves were constructed by extracting circular profiles from the images at these elongations, averaging the data over the whole range of position angle [0--360$\deg$] and further averaging over Carrington rotations (hence ``monthly'' refers to Carrington ``months''). 
They are compared with the temporal variation of the total photospheric magnetic flux (TMF) as this proxy of solar activity was found by \cite{Barlyaeva2015} to best match the integrated radiance of the K-corona.
The TMF was calculated from the Wilcox Solar Observatory photospheric field maps by Y.-M.~Wang according to a method described by \cite{Wang2003}.
\end{itemize}

\subsection{Polarization of the corona}
\label{Sub:Polar}

\begin{figure}[htpb!]
	\centering
	\includegraphics{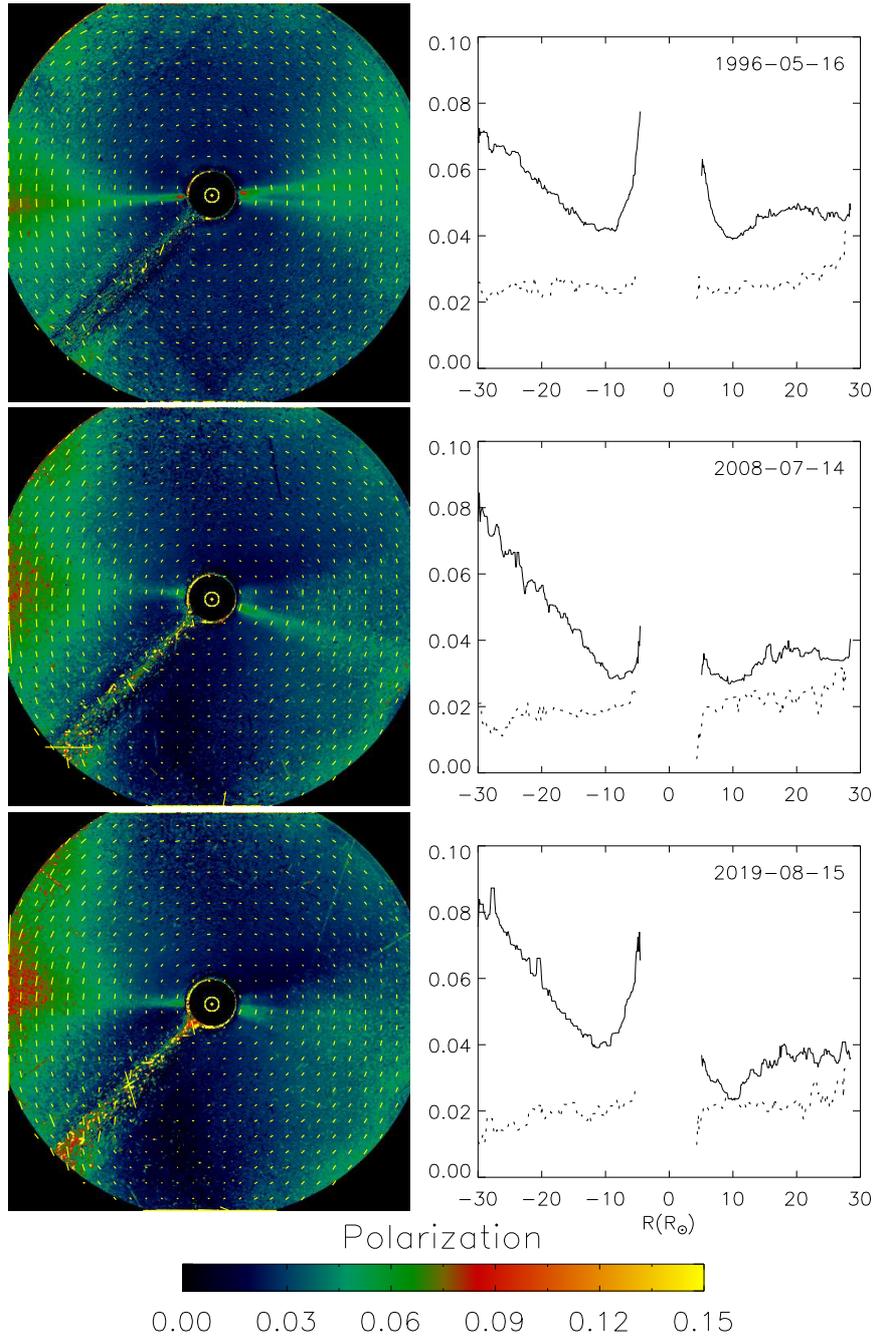}
	\caption{Maps and radial profiles of the polarization in the LASCO-C3 \fov at three minima of activity: SC22/SC23 (16 May 1996, upper panels), SC23/SC24 (14 Jul. 2008, middle panels), and SC24/SC25 (15 Aug. 2019, lower panels).
	The direction of polarization is indicated by yellow bars whose length is scaled to the polarization.
	The yellow circles represent the solar disk and solar north is down except for the upper image where it is up.
	The profiles are extracted along the equatorial (solid lines) and polar (dotted lines) directions.}
	\label{Fig:polar_vec_C3_MinActiv}
\end{figure}
\begin{figure}[htpb!]
	\centering
	\includegraphics{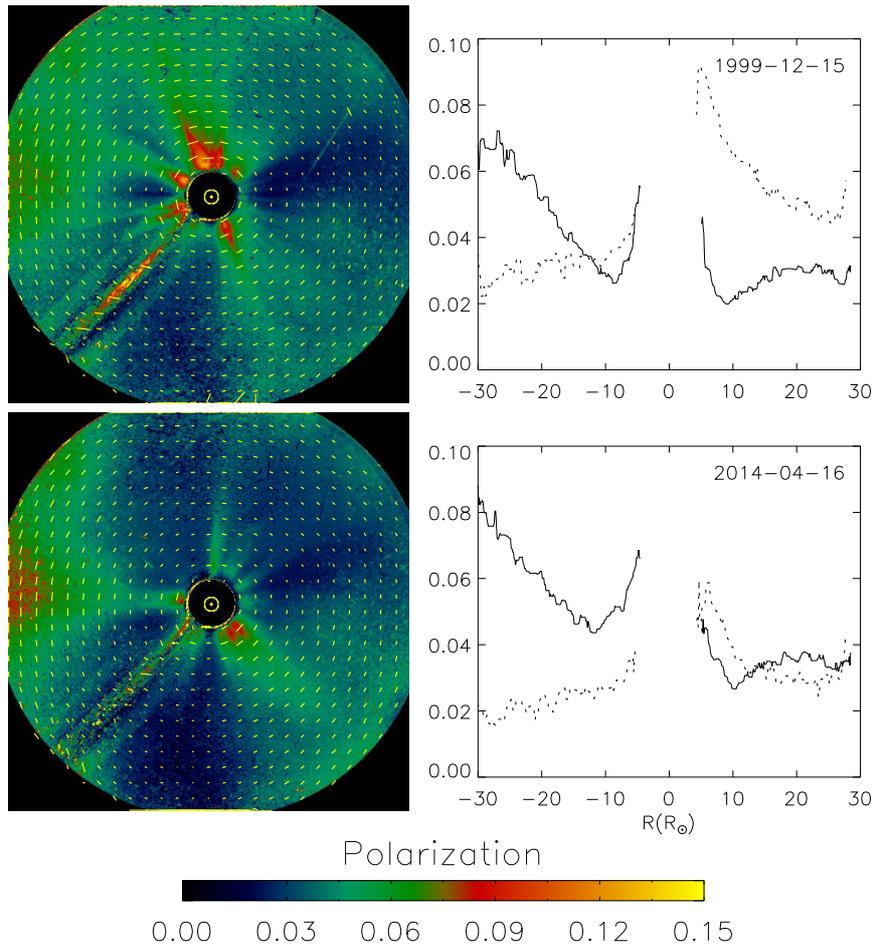}
	\caption{Maps and radial profiles of the polarization in the LASCO-C3 \fov at the two maxima of SC 23 (15 Dec. 1999) and SC 24 (16 Apr. 2014).
	Solar north is up in the upper image and down in the lower one.
	See Figure~\ref{Fig:polar_vec_C3_MinActiv} for further explanations.}
	\label{Fig:polar_vec_C3_MaxActiv}
\end{figure}
\begin{figure}[htpb!]
	\centering
	\includegraphics{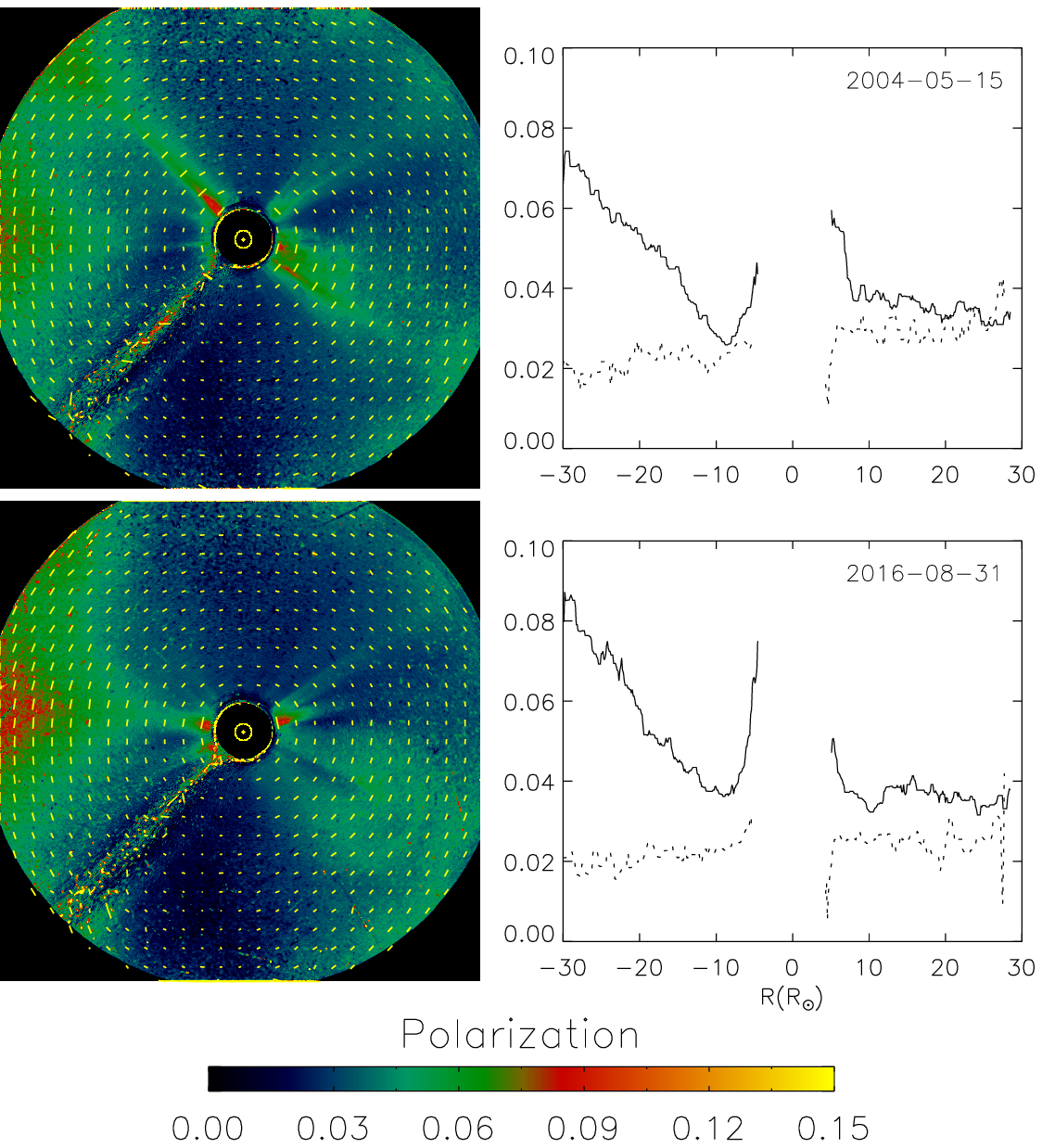}
	\caption{Maps and radial profiles of the polarization in the LASCO-C3 \fov at two dates of the declining phase of SC 23 (15 May 2004) and SC 24 (31 Aug. 2016).
	Solar north is up in the two images.
	See Figure~\ref{Fig:polar_vec_C3_MinActiv} for further explanations.}
	\label{Fig:polar_vec_C3_DecliningActiv}
\end{figure}

The results for the polarization are presented in three figures where all maps are shown with the same color scale and all profiles with the same scale to facilitate the intercomparison.
Figure~\ref{Fig:polar_vec_C3_MinActiv} displays the results at three dates of the three minima of solar activity observed by LASCO: SC22/SC23, SC23/SC24, and SC24/SC25, Figure~\ref{Fig:polar_vec_C3_MaxActiv} at two dates of the maximum of SC 23 and SC 24, and Figure~\ref{Fig:polar_vec_C3_DecliningActiv} at two dates of the declining phase of these two cycles.

A striking feature common to all images is the reinforcement at the left edge of their \fov -- also conspicuous on the equatorial profiles -- resulting in a strong east-west asymmetry which further increases with time.
A natural explanation would be a defect affecting at least one polarizer which keeps developing  with time but this is in contradiction with our investigation of this question in Section~\ref{Sec:Improve}. 
In addition, the local angle of polarization remains close to the theoretical value of 90$\deg$ in this region. 
We have no explanation for this problem, but consider that the polarization results in this region defined by a $\approx$90$\deg$ sector centered on the equatorial direction and originating at $\approx$10\,R${}_\odot$ on the left side of the \fov are questionable.
Much less severe is a hint of a similar problem at the edge of the \fov in the upper left quadrant, but here again, the local angle of polarization remains close to 90$\deg$.
A very slight enhancement appears to be present at the upper edge of the \fov so that the asymmetric increases of polarization seen along the corresponding profiles beyond $\approx$20\,R${}_\odot$ must be viewed with cautions.
Apart from these regions, the polarization patterns in the large coronal holes characteristic of solar minima are very consistent with a polarization of $\approx$0.02 (Figure~\ref{Fig:polar_vec_C3_MinActiv}).
The equatorial regions are dominated by very flat streamer belts of higher polarization, typically of the order of 0.04.
This contrasts with the configuration of the two maxima of activity (Figure~\ref{Fig:polar_vec_C3_MaxActiv}), particularly the stronger maximum of SC 23 compared with SC 24, where highly polarized streamers pervade both hemispheres. 
As expected, intermediate configurations prevail during the declining phases (Figure~\ref{Fig:polar_vec_C3_DecliningActiv}) with streamers still present at high latitudes but generally weaker and therefore less polarized than during the maxima.

\begin{figure}
\centering
\includegraphics[width=\textwidth]{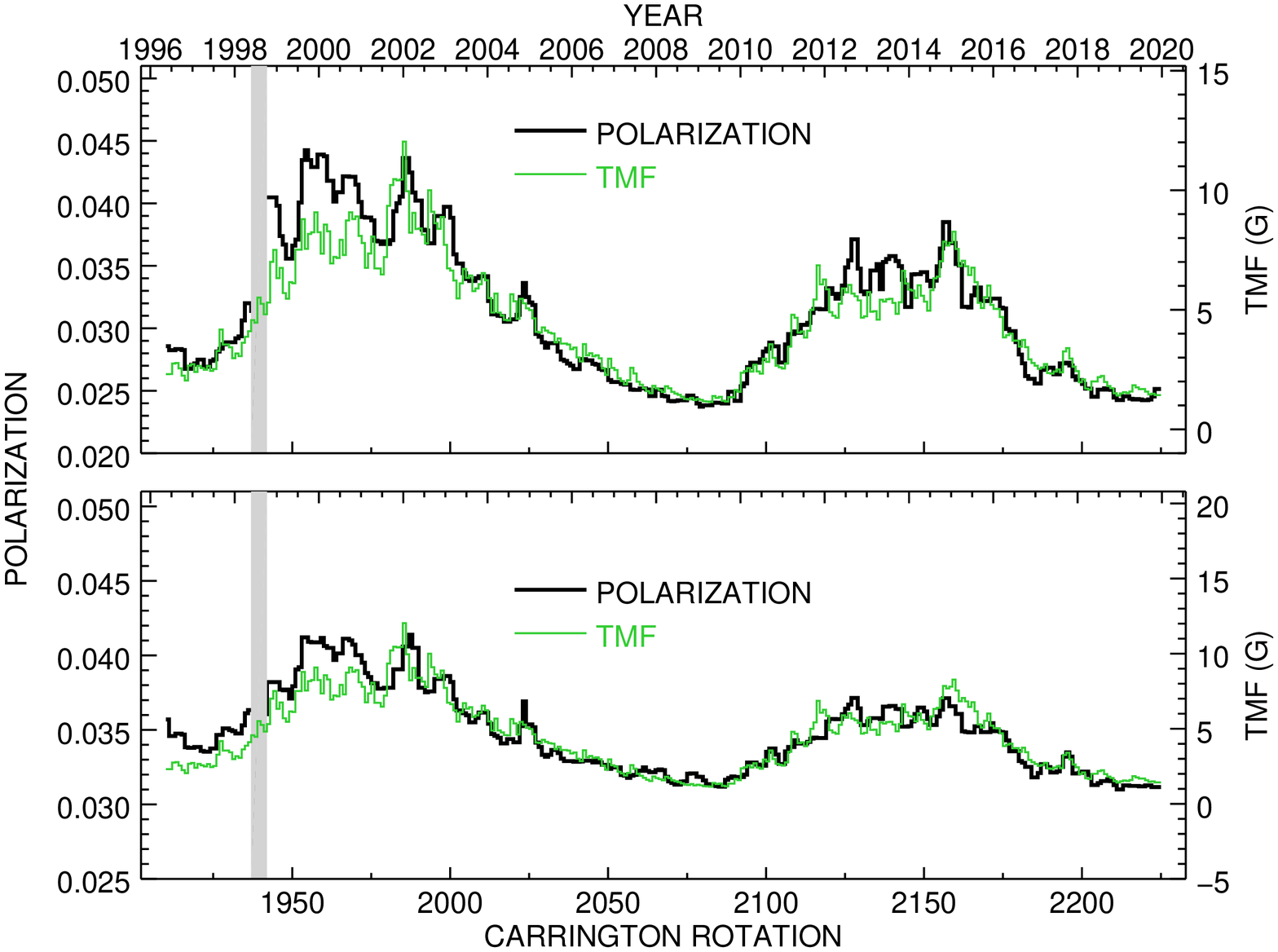}
\caption{Temporal variation of the polarization at 10\,R${}_\odot$ (upper panel) and at 20\,R${}_\odot$ (lower panel) averaged over Carrington rotations compared with that of the total magnetic field (TMF, right scale).
The gray bands correspond to missing data when SOHO lost its pointing.}			
\label{Fig:PActiv10_20}
\end{figure}

A more detailed quantitative view of the coronal polarization is offered by its temporal variation at 10 and 20\,R${}_\odot$ displayed in Figure~\ref{Fig:PActiv10_20}. 
It is quite amazing that, even at 20\,R${}_\odot$, the large scale variations are strongly correlated with those of the TMF, notably the relative strength of the two cycles.
However, smaller scale variations are not correlated as faithfully as it was the case of the LASCO-C2 polarization (Paper I).
Several peaks do coincide in 2002, 2005, 2014/2015, and 2018, and the third case is worth emphasizing: it extends from late 2014 to beginning of 2015 and is connected to the anomalous surge of the radiance of the corona discovered on LASCO-C2 images \citep{Lamy2017} and can be traced as far as 20\,R${}_\odot$.
However, the C3 polarization exceeds the values predicted by the TMF during the first broad peak of SC 23, from approximately 1999 to 2001.5 (and even from 1997 in the case of the 20\,R${}_\odot$ profile).
A similar, but more modest trend is observed during the first part of the maximum of SC 24 (2012--2014), best perceived at 10\,R${}_\odot$. 
There is no obvious explanation(s) for these slight anomalies.
It is interesting to note the difference of the low polarizations during the deep SC23/SC24 minimum, 0.024 at 10\,R${}_\odot$ and 0.032 at 20\,R${}_\odot$, a likely consequence of the rising level of the polarization of the F-corona with increasing elongations.


\subsection{Polarized Radiance of the Corona}   

\begin{figure}[htpb!]
\centering
\includegraphics[width=\textwidth]{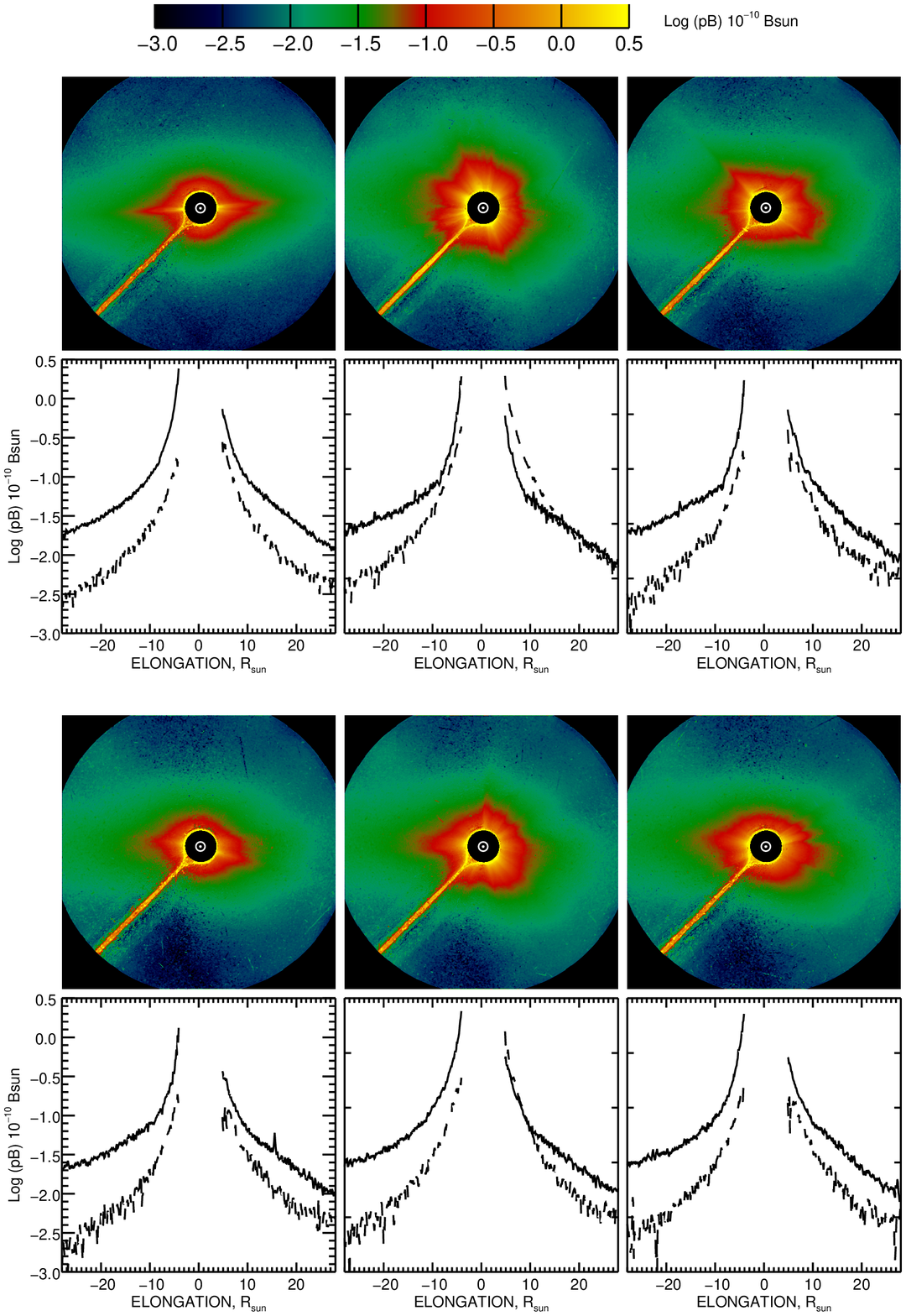}
\caption{Maps and radial profiles of the polarized radiance of the corona in the LASCO-C3 \fov at three phases of activity of SC 23 (upper two rows) and SC 24 (lower two rows).
The dates are identical to those of Figures\,\ref{Fig:polar_vec_C3_MinActiv}, \ref{Fig:polar_vec_C3_MaxActiv}, and \ref{Fig:polar_vec_C3_DecliningActiv}: 16 May 1996, 15 Dec. 1999, and 15 May 2004 for SC 23 and 14 Jul. 2008, 16 Apr. 2014, and 31 Aug. 2016 for SC 24. 
The white circles represent the solar disk and solar north is up except for the 14 Jul. 2008, and 16 Apr. 2014 images where it is down.
The profiles are extracted along the equatorial (solid lines) and polar (dotted lines) directions.}
	\label{Fig:C3_pBs}
\end{figure}

The presentation of the results for the polarized radiance $pB$ of the corona follows that of the polarization, except for a more compact format in the case of the images and profiles as illustrated in Figure~\ref{Fig:C3_pBs}.  
The dates are identical except for the deletion of the last minimum (15 Aug. 2019) to simplify the make-up of the figure. 
The reinforcement at their eastern edge of the \fov affecting the polarization is naturally perceptible on the images and more clearly on the equatorial profiles of the polarized radiance with the east-west asymmetry increasing with time. 
The temporal variations of the polarized radiance at 10 and 20\,R${}_\odot$ are displayed in Figure~\ref{Fig:PActiv10_20} and likewise the polarization, they closely track the large scale variations of the TMF.
At 10\,R${}_\odot$, the tracking is impressive to the level of small scale variations which was not the case of the polarization.
In particular, the polarized radiance tracks the first broad peak of the TMF during SC 23 unlike the polarization.
At 20\,R${}_\odot$, discrepancies appear as the two cycles cannot be fitted simultaneously and Figure~\ref{Fig:PActiv10_20} favors the fit during SC 23.
This problem may be understood as resulting from the very low values of $pB$ at this elongation, typically $10^{-14}\Bsun$, so that weak stray effects become important.
The progressive temporal increase of the reinforcement on the left side of the \fov probably explains why, at 20\,R${}_\odot$, the amplitude of the $pB$ variations over the two solar cycles are comparable whereas it is well-known that SC 24 was weaker than SC 23.

\begin{figure}[htpb!]
      \centering
      \includegraphics[width=\textwidth]{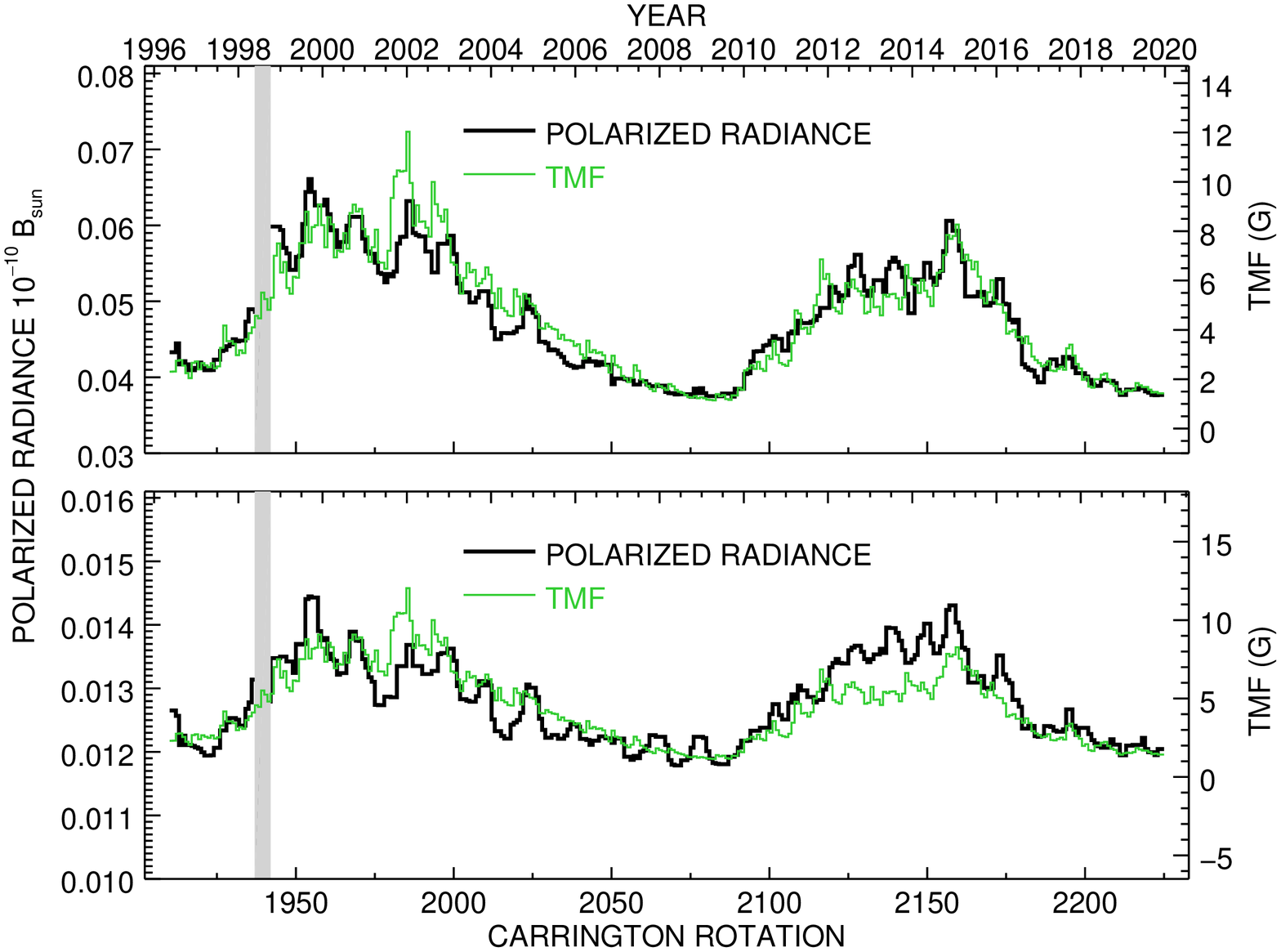}
      \caption{Temporal variation of the polarized radiance at 10\,R${}_\odot$ (upper panel) and at 20\,R${}_\odot$ (lower panel) averaged over Carrington rotations compared with that of the total magnetic field (TMF).
The gray bands correspond to missing data when SOHO lost its pointing.}
      \label{pBActiv10_20}
\end{figure}

\subsection{Coronal Electron Density} 
\label{Sub:Ne}

The two-dimensional (2D) distributions of the coronal electron density $N_{e}$ were obtained by extending to the C3 \fov the method developed by \cite{Quemerais2002} for the 2D inversion of the C2 $pB$ images as previous implemented in the works of \cite{Lamy2002}, \cite{Lamy2014}, \cite{Lamy2017}, and \cite{Lamy2020}.
However, the C3 $pB$ images are much noisier than those of C2 as can be seen on prominently the polar profiles displayed in Figure~\ref{Fig:C3_pBs}.
Our inversion proceeds along radial directions, from the outer limit of the \fov towards the center of the Sun and requires a smooth gradient so that the local distribution of electrons $N_{e}(r)$ can be represented by power laws. 
The fluctuations at high spatial frequencies present in the outer part of the C3 $pB$ images must therefore be eliminated.
This was achieved by applying a Lee filter to the $pB$ images with the largest window (15$\times$15 pixels) compatible with the preservation of the large scale gradient of the coronal signal throughout the whole \fovnospace.

\begin{figure}[htpb!]
\centering
\includegraphics[width=\textwidth]{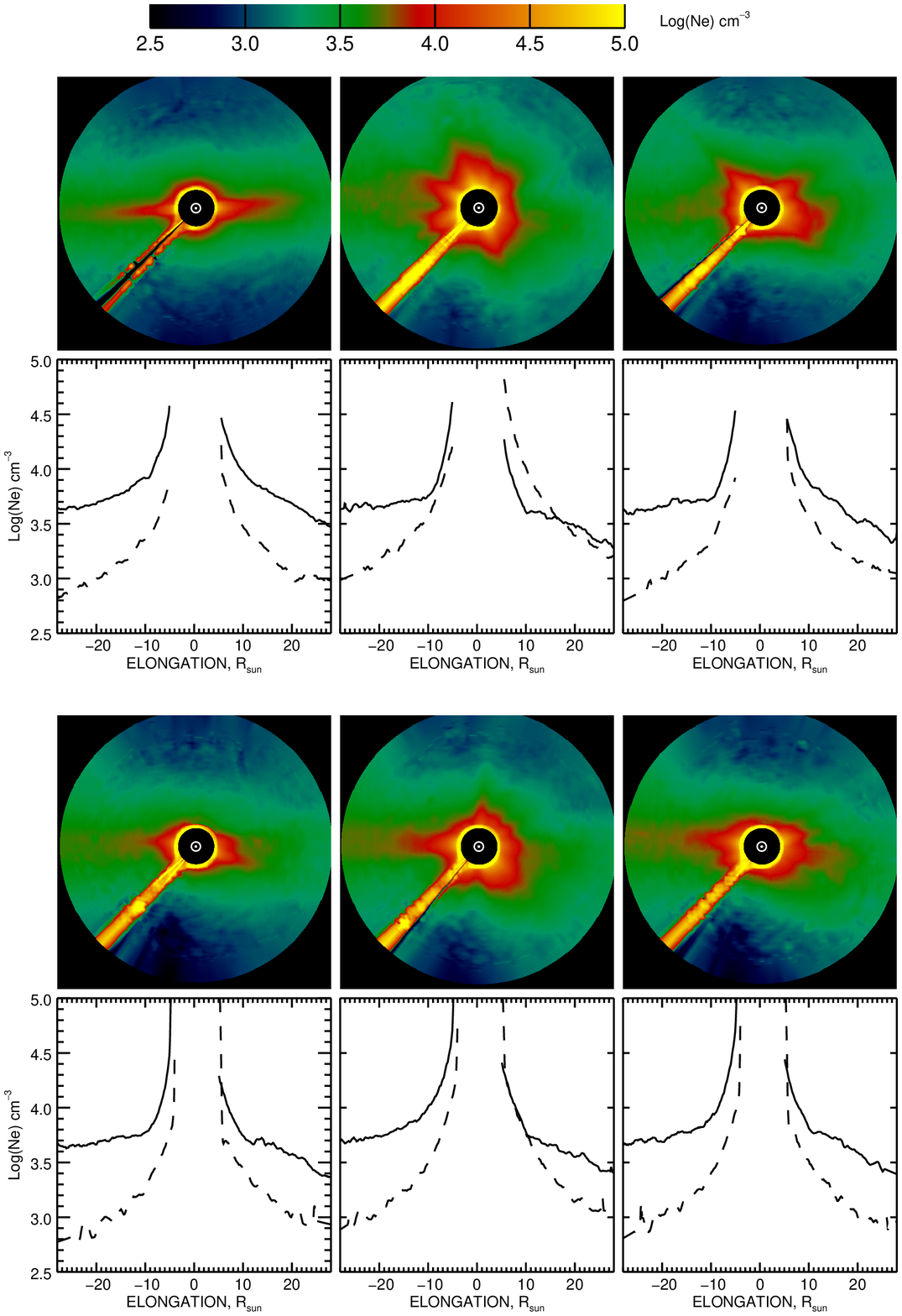}
\caption{Maps and radial profiles of the coronal electron density at three phases of activity of SC 23 (upper two rows) and of SC 24 (lower two rows).
The dates are identical to those of Figure~\ref{Fig:C3_pBs}: 16 May 1996, 15 Dec. 1999, and 15 May 2004 for SC 23 and 14 Jul. 2008, 16 Apr. 2014, and 31 Aug. 2016 for SC 24.  
The white circles represent the solar disk and solar north is up except for the 14 Jul. 2008, and 16 Apr. 2014 images where it is down.
The profiles are extracted along the equatorial (solid lines) and polar (dotted lines) directions.}
\label{NeSample}
\end{figure}

\begin{figure}[htpb!]
\centering
\includegraphics[width=\textwidth]{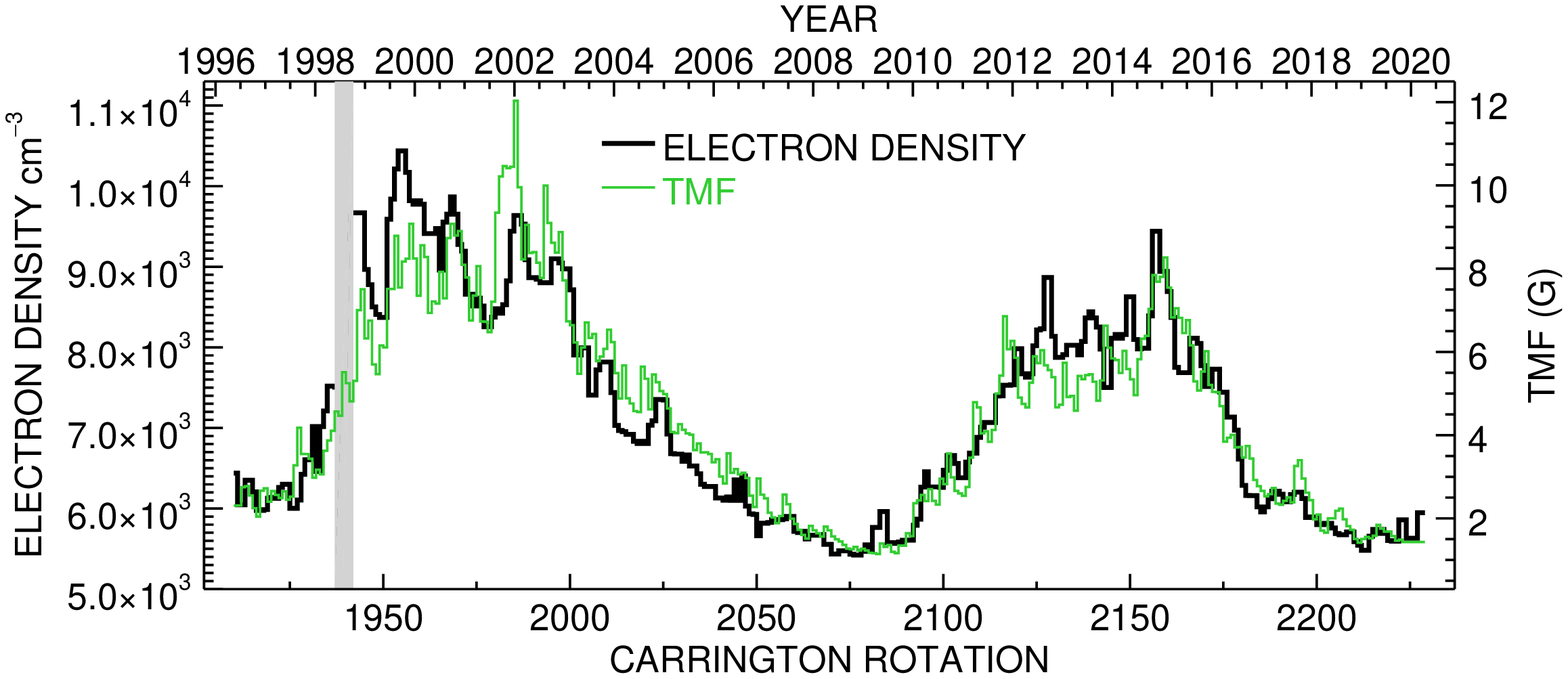}
\caption{Temporal variation of the electron density at 10\,R${}_\odot$ averaged over Carrington rotations compared with that of the total magnetic field (TMF). 
The gray bands correspond to missing data when SOHO lost its pointing.}
\label{NeActiv100}
\end{figure}

We display the resulting electron density in a set of figures similar to those presented above for the polarized radiance:
i) maps at three phases of solar activity during SC 23 and SC 24 with the corresponding profiles along the equatorial and polar directions (Figure~\ref{NeSample}), and
ii) temporal variation limited to the case of an elongation of 10\,R${}_\odot$ for brevity (Figure~\ref{NeActiv100}).

In principle, the images and profiles of the electron density should closely resemble those of the polarized radiance.
This is clearly the case in the polar regions but less so in the equatorial regions and we strongly suspect that the stray reinforcements bias the determination of the power laws when the inversion proceeds inward starting from the outer parts which are the most affected by the stray effects. 
The temporal variation of the electron density at 10\,R${}_\odot$ (Figure~\ref{Fig:PActiv10_20}) closely tracks those of the TMF very much like the polarized radiance with however slightly larger miss-matches during the two maxima of activity.

\subsection{K and F coronae} 
\label{Sub:K}
Maps of the K-corona were calculated according to the method described in Section~\ref{Sub:Separation}, \ie imposing $p_{K}(r)=0.64$.
We display a set of figures similar to those presented above for the electron density.

\begin{figure}[htpb!]
\centering
\includegraphics[width=\textwidth]{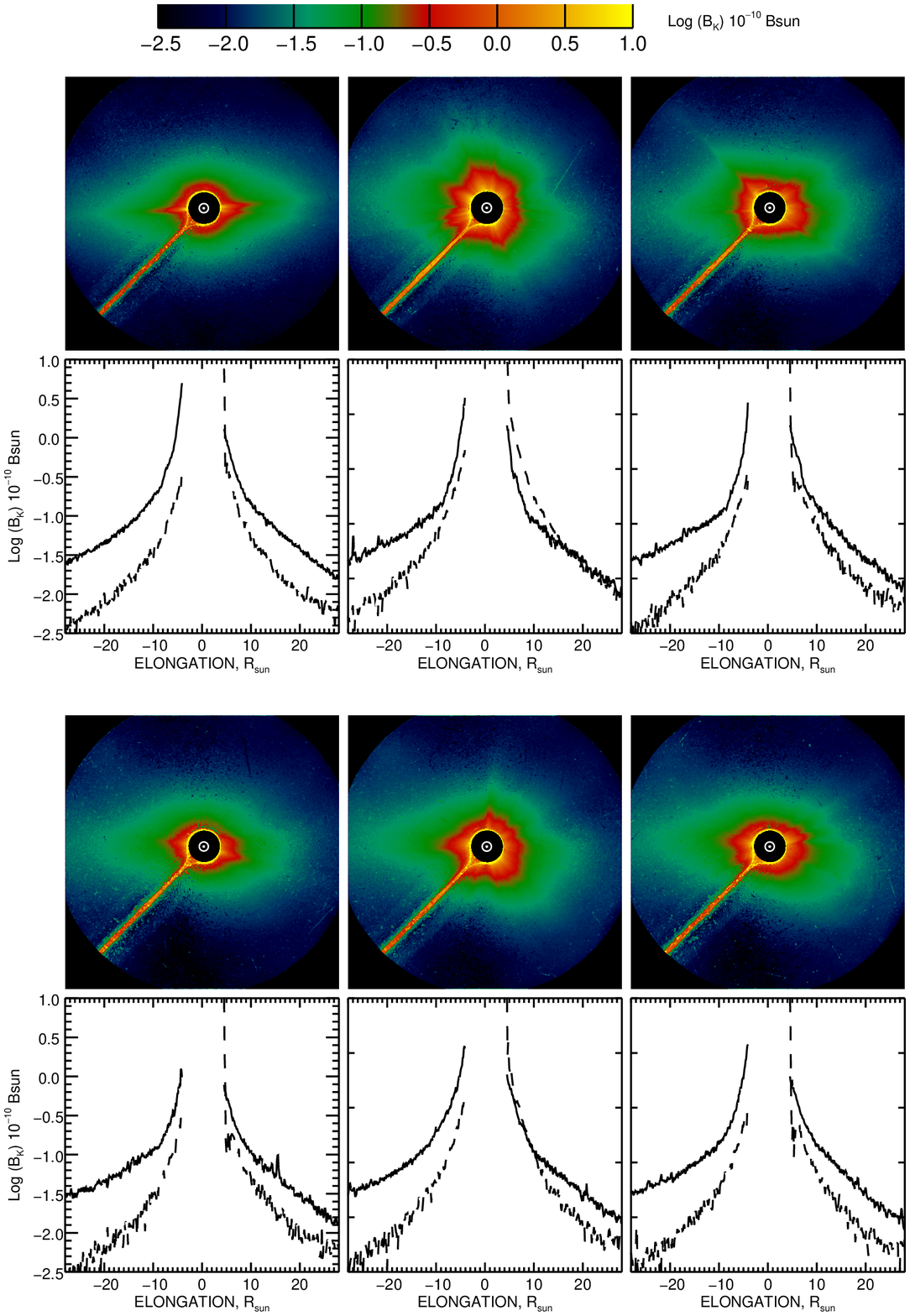}
\caption{Maps and radial profiles of the K-corona at three phases of activity of SC 23 (upper two rows) and SC 24 (lower two rows).
The dates are identical to those of Figure~\ref{Fig:C3_pBs}: 16 May 1996, 15 Dec. 1999, and 15 May 2004 for SC 23 and 14 Jul. 2008, 16 Apr. 2014, and 31 Aug. 2016 for SC 24.   
The white circles represent the solar disk and solar north is up except for the 14 Jul. 2008, 16 Apr. 2014 images where it is down.
The profiles are extracted along the equatorial (solid lines) and polar (dotted lines) directions.}
\label{C3_KSample}
\end{figure}

\begin{figure}[htpb!]
\centering
\includegraphics[width=\textwidth]{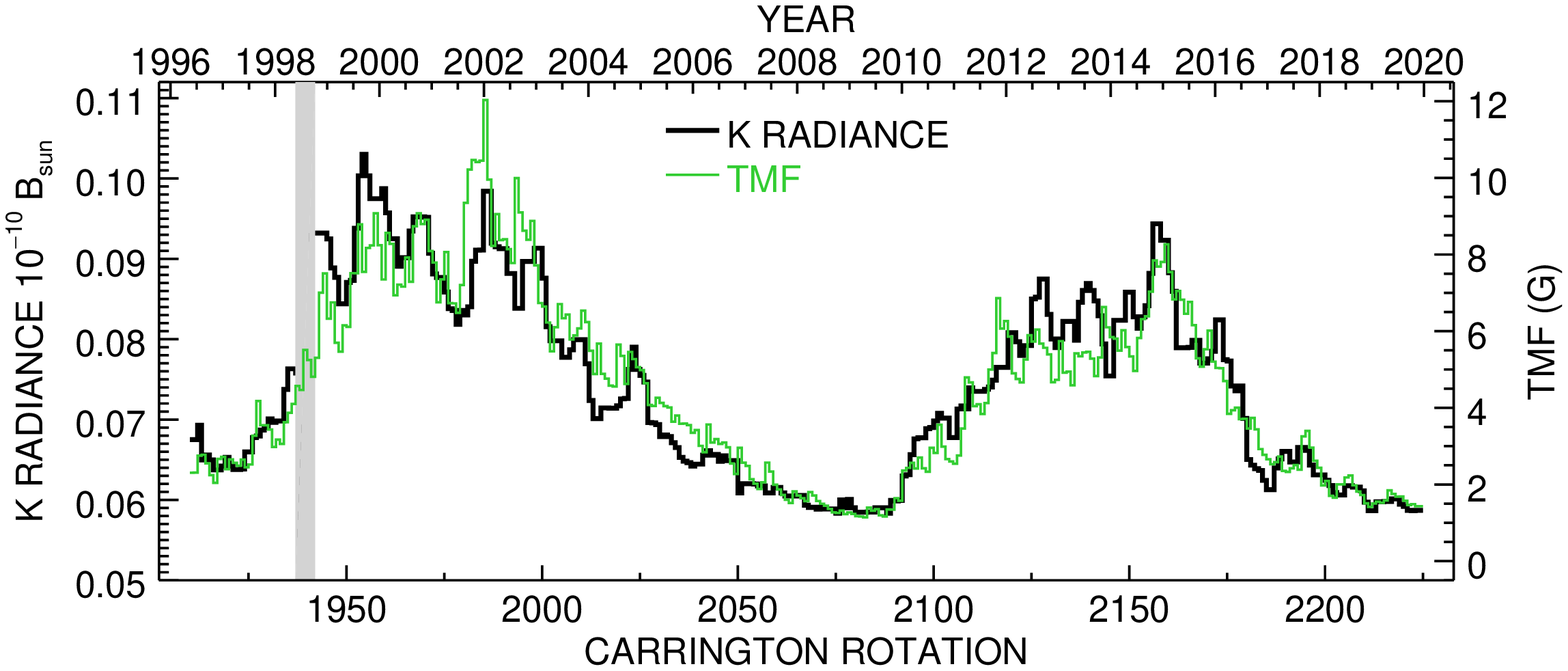}
\caption{Temporal variation of the radiance of the K-corona at 10\,R${}_\odot$ averaged over Carrington rotations compared with that of the total magnetic field (TMF). 
The gray bands correspond to missing data when SOHO lost its pointing.}			
\label{KActiv100}
\end{figure}

\begin{itemize}
	\item Maps of the K-corona at three phases of solar activity during SC 23 and SC 24 as well as the corresponding profiles along the equatorial and polar directions (Figure~\ref{C3_KSample});
	\item Monthly averaged temporal variation at 10\,R${}_\odot$ (Figure~\ref{KActiv100}).
\end{itemize}

Unsurprisingly, the maps and profiles of the K-corona and its temporal variations at 10\,R${}_\odot$ closely match those of the polarized radiance thus exhibiting the same features which are not repeated here.

F-corona maps were simply obtained by subtracting the K-corona maps from the K+F maps.
Their in-depth discussions is left to a forthcoming article that will combine results of the F-corona coming from C2 and C3 altogether.

\section{Uncertainty Estimates}
\label{Sec:Uncertainty}

As argued in Paper I, the error analysis for the LASCO-C2 polarization data was particularly complex in view of the many sources of error. 
The situation is even more difficult for C3 due to the lack of relevant laboratory calibrations (Section~\ref{Sec:Analysis}) and the early loss of the 0$\deg$ polarizer.
Nevertheless, we follow the method of Paper I and attempt to estimate the uncertainties affecting the different quantities.
According to Figures~\ref{Fig:EvolHistC3} and \ref{Fig:StatParamPolC3}, the measured angle of polarization deviated from the theoretical value of 90$\deg$ by only 0.5$\deg$ during the first three years and by $\leq2\deg$ thereafter remaining remarkably stable. 
This slight offset may result from improper correction of the global transmission of the polarizers or of the exposure times or both since they have the same effect of unbalancing the relative contributions of the polarization channels.
It could have been corrected by fine-tuning either of the above parameters, but we did not do so because of a negligible impact on the polarization.
Even the standard deviation at a level of approximately 5$\deg$ does not affect much the polarization.
The absolute value of the global transmissions introduces a larger uncertainty which cannot be assessed due to the lack of laboratory calibration, but a relative level of $\approx$8 \% measured on the C2 polarizers provides a reasonable estimate. 
This translates to typical uncertainties of $\pm$0.0016 for $p$=0.02 and $\pm$0.003 for $p$=0.04.
Unlike C2, local inhomogeneities in the principal transmittance k$_{1}$ of the polarizers could not be corrected but based on the C2 measurements, they should not exceed $\approx$8 \% thus leading to uncertainties similar to those resulting from the global transmissions. 
The accuracy of the correction function $S(x,y)$ is of major concern.   
The SOHO roll sequence of September 1997 was such that the pylon was oriented along the east and north-east directions at roll angles of 45$\deg$ and 90$\deg$, respectively thus disturbing the images in the sector where major stray effects were precisely found and affecting the determination of $S(x,y)$, a difficulty absent in the case of C2.
Moreover, this function determined early on, may well evolve with time. 
Clearly, all these effects cannot be quantified.
The problem is less severe with the ramps since they are very low-level corrections, prominently affecting the outermost region of the \fov where other effects may be present as well. 

The absolute calibration of $pB$ implements a two-step procedure.
In the first step, we make use of the ratio $I_{0}/I_{cp}$ which is determined with an accuracy of $\pm$0.01 during the first four years and $\pm$0.002 thereafter (Figure~\ref{Fig:C3CalCoef}).
The second step relies on the calibration of the unpolarized $B$ images based on photometric measurements of stars present in the C3 \fovnospace.
We used the calibration coefficient averaged over the first 7.5 years of LASCO operation determined by \cite{Thernisien2006}.
However, a detailed investigation of the evolution of the ``clear'' channel revealed a constant degradation at a rate of 0.44$\pm$ 0.1 \% per year thus leading to a possible variation of 11\% over 24 years.
We will show below that, on the basis of the C2-C3 intercomparison, such an extrapolated variation does not appear to be justified.

Finally, the separation of the three components K, F, and stray light S to retrieve the radiance of the K-corona requires several assumptions, namely that the F-corona and the stray light are unpolarized and that $p_{K}$ obeys a prescribed model. 
It turns out that the bulk of stray light in C3 results from the diffraction by the occulters -- thus ensuring that it is unpolarized -- with an influence limited to 5\,R${}_\odot$.
The other assumptions are obviously all sources of uncertainties affecting the determination of the $B_{K}$ maps, particularly $p_{F} = 0$ which becomes more and more questionable at large elongations.
Alternatively, this may offer a mean of probing how far this assumption is valid.

In view of the complexity of analyzing the uncertainties affecting polarized measurements of the corona, \cite{Frazin2012} proposed an intercomparison of the results coming from different coronagraphs both space- and ground-based, an approach which was feasible between 1.6 and 6\,R${}_\odot$.
It was indeed implemented in Paper I using eclipse measurements obtained during the LASCO lifetime.
Unfortunately, similar measurements are not available in the C3 \fov and we are left with the few data obtained decades before the LASCO era as presented in the Introduction and also a few models. 
An intercomparison of the C2 and C3 results is however possible thanks to the overlap between their fields of view and is presented in the next section

\section{Intercomparison of the C2 and C3 Results} 
\label{Sec:C2C3}

The overlap between the C2 and C3 fields of view is practically limited to an annulus extending from 5 to 6\,R${}_\odot$ in order to avoid possible stray light residuals in the innermost part of the C3 images and we selected an elongation of 5.5\,R${}_\odot$ as the best compromise to perform the comparison of circular profiles.
We also considered radial profiles along the east-west ans south-north directions extending from -10 to +10\,R${}_\odot$ as well as mean profiles calculated over the full 360$\deg$ range of position angle and extending from 2 to 10\,R${}_\odot$.
The comparisons are performed for the seven dates used in Figures\,\ref{Fig:polar_vec_C3_MinActiv}, \ref{Fig:polar_vec_C3_MaxActiv}, and \ref{Fig:polar_vec_C3_DecliningActiv}: 16 May 1996, 15 December 1999, 15 May 2004, 14 July 2008, 16 April 2014, 31 August 2016, and 15 August 2019 and for the four quantities: polarization, polarized brighness, and the radiances of the K and F coronae.

\begin{figure}[htpb!]
	\centering
	\includegraphics[width=\textwidth]{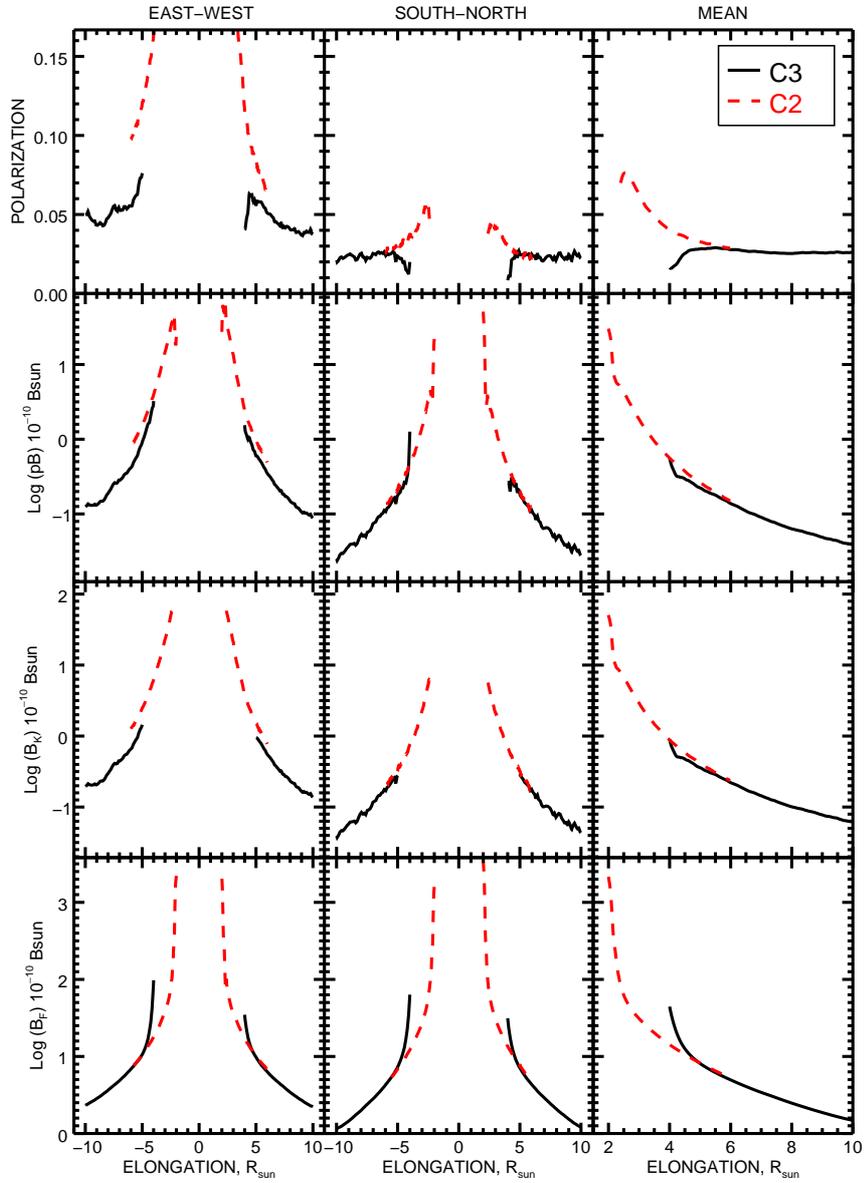}
	\caption{Comparison of the results of C2 and C3 on 16 May 1996 for, from top to bottom, the polarization, the polarized brightness, the radiance of the K-corona, and that of the F-corona.
	The radial profiles are displayed along the east-west (left column) and the south-north directions (central column), complemented by a mean profile (right column).} 
	\label{Fig:radial_profiles_all_c3_19960516}
\end{figure}

\begin{figure}[htpb!]
	\centering
	\includegraphics[width=\textwidth]{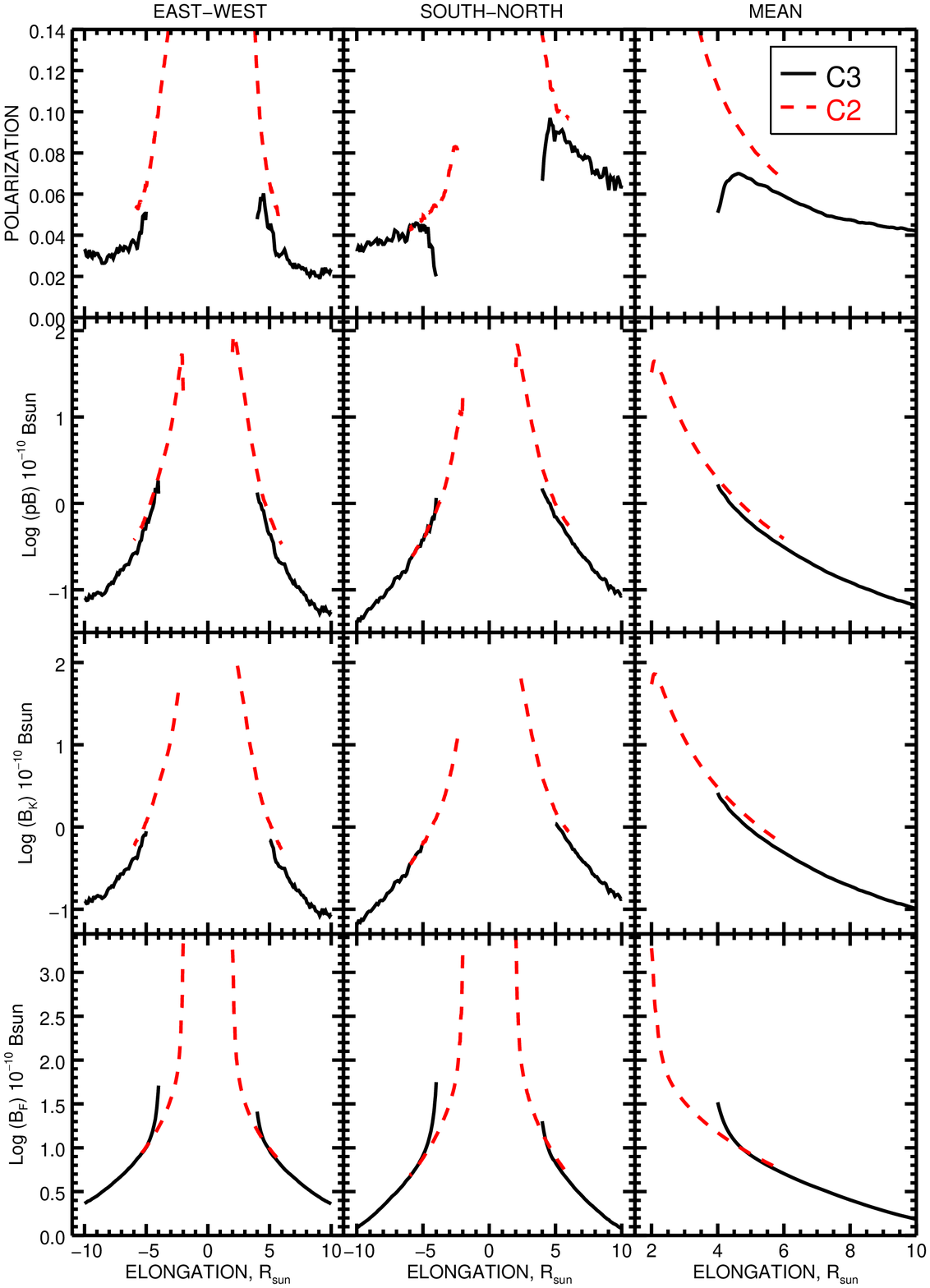}
	\caption{Same as Figures\,\ref{Fig:radial_profiles_all_c3_19960516} for 15 December 1999.} 
	\label{Fig:radial_profiles_all_c3_19991215}
\end{figure}

\begin{figure}[htpb!]
	\centering
	\includegraphics[width=\textwidth]{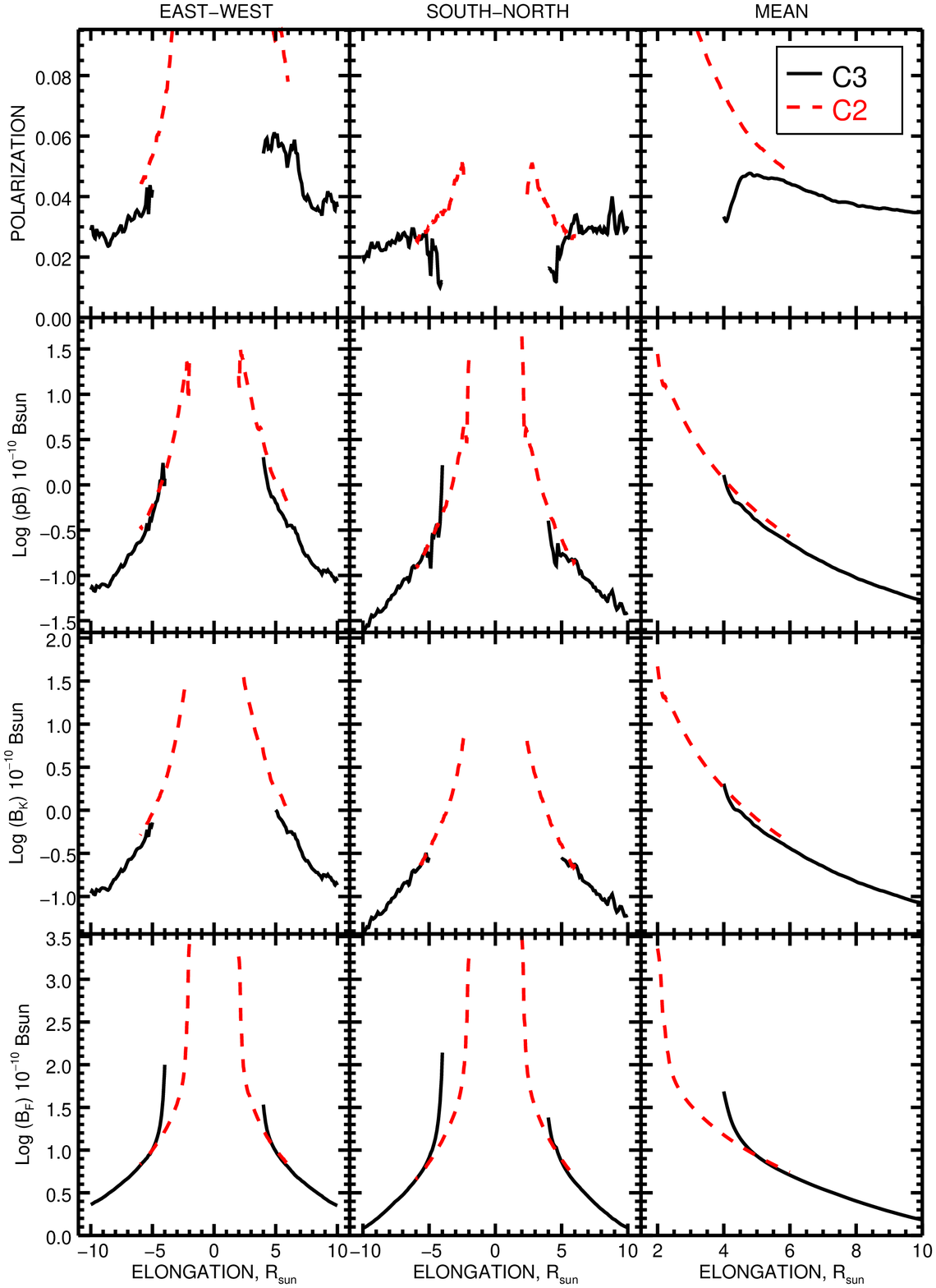}
	\caption{Same as Figures\,\ref{Fig:radial_profiles_all_c3_19960516} for 15 May 2004.} 
	\label{Fig:radial_profiles_all_c3_20040515}
\end{figure}

\begin{figure}[htpb!]
	\centering
	\includegraphics[width=\textwidth]{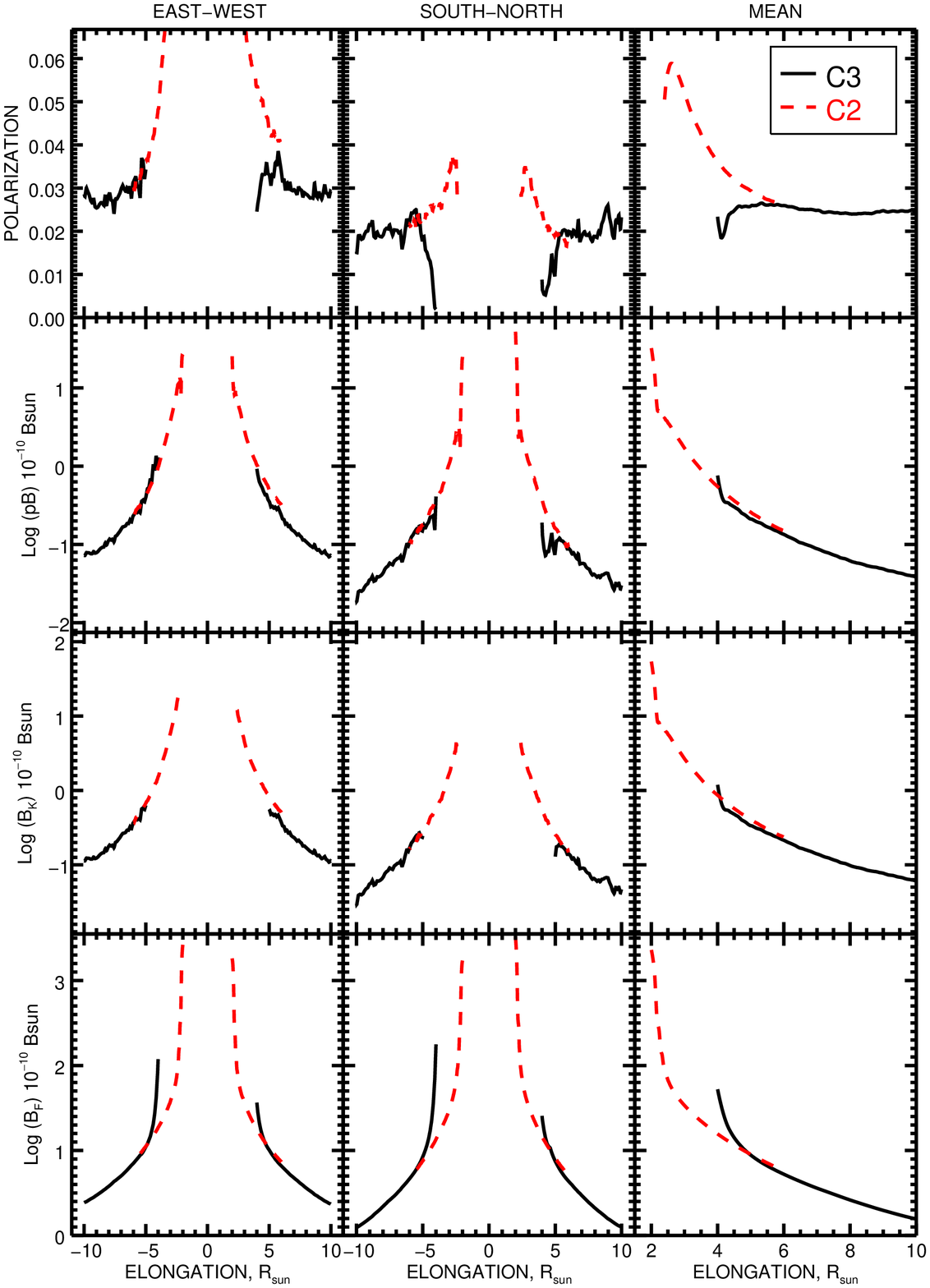}
	\caption{Same as Figures\,\ref{Fig:radial_profiles_all_c3_19960516} for 14 July 2008.} 
	\label{Fig:radial_profiles_all_c3_20080714}
\end{figure}

\begin{figure}[htpb!]
	\centering
	\includegraphics[width=\textwidth]{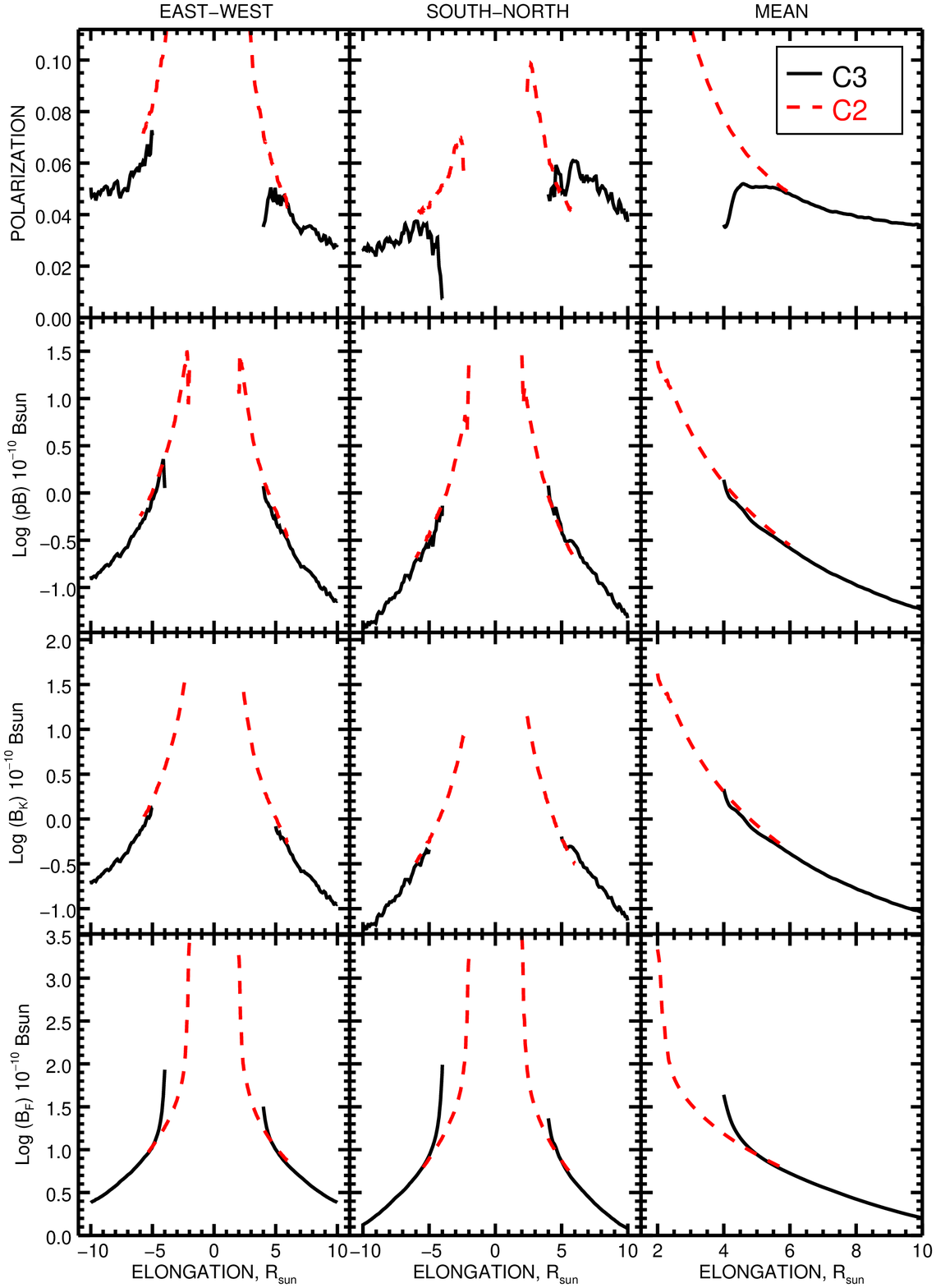}
	\caption{Same as Figures\,\ref{Fig:radial_profiles_all_c3_19960516} for 16 April 2014.} 
	\label{Fig:radial_profiles_all_c3_20140416}
\end{figure}

\begin{figure}[htpb!]
	\centering
	\includegraphics[width=\textwidth]{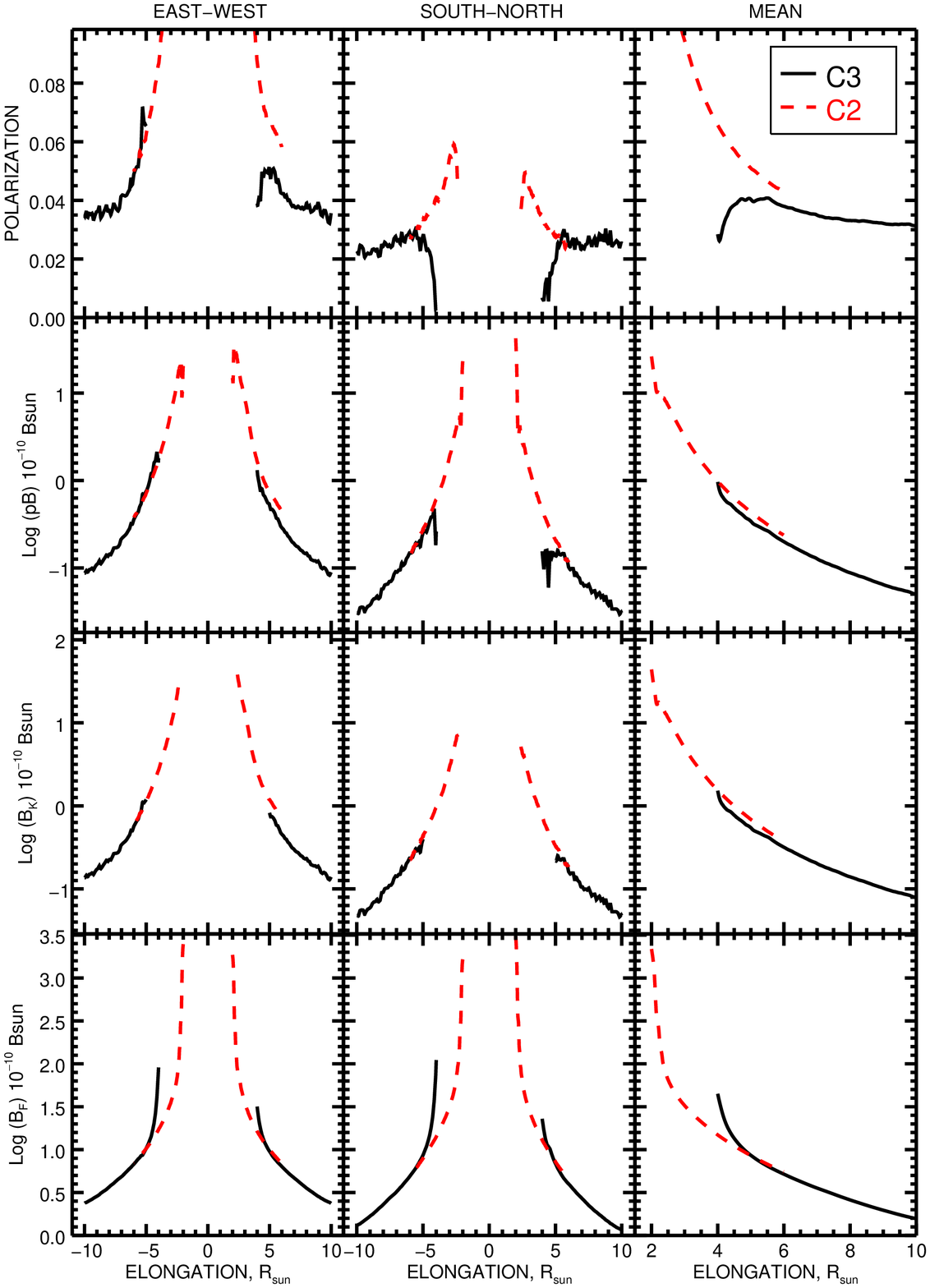}
	\caption{Same as Figures\,\ref{Fig:radial_profiles_all_c3_19960516} for 31 August 2016.} 
	\label{Fig:radial_profiles_all_c3_20160831}
\end{figure}

\begin{figure}[htpb!]
	\centering
	\includegraphics[width=\textwidth]{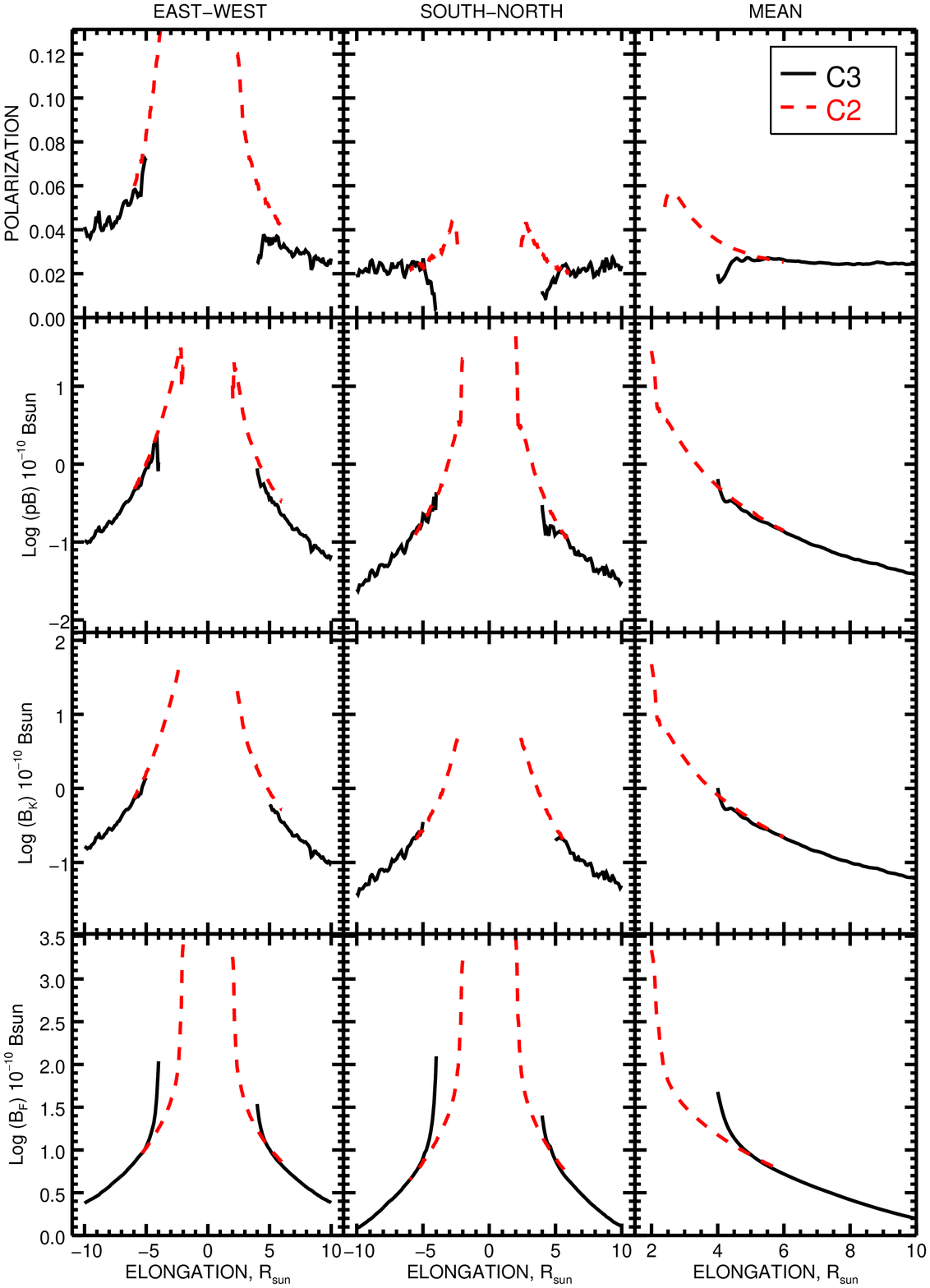}
	\caption{Same as Figures\,\ref{Fig:radial_profiles_all_c3_19960516} for 15 August 2019.} 
	\label{Fig:radial_profiles_all_c3_20190815}
\end{figure}

Figures~\ref{Fig:radial_profiles_all_c3_19960516} to \ref{Fig:radial_profiles_all_c3_20190815} display the radial profiles showing a nearly perfect agreement between C2 and C3. 
There exists a few exceptions, but then the differences remain very small.
This situation implies that the independent calibrations of C2 and C3 are correct and that the sensitivity of C3 in particular has only marginally evolved during the last 16 years.
A careful inspection of the profiles reveals that the inner stray light in C3 is unpolarized as expected from the diffraction by the occulters.
Indeed, stray light effects are totally absent in the $pB$ and $B_{K}$ profiles and only present in the $B_{F}$ profiles up to an elongation of $\approx$5\,R${}_\odot$.

\begin{figure}[htpb!]
	\centering
	\includegraphics[width=\textwidth]{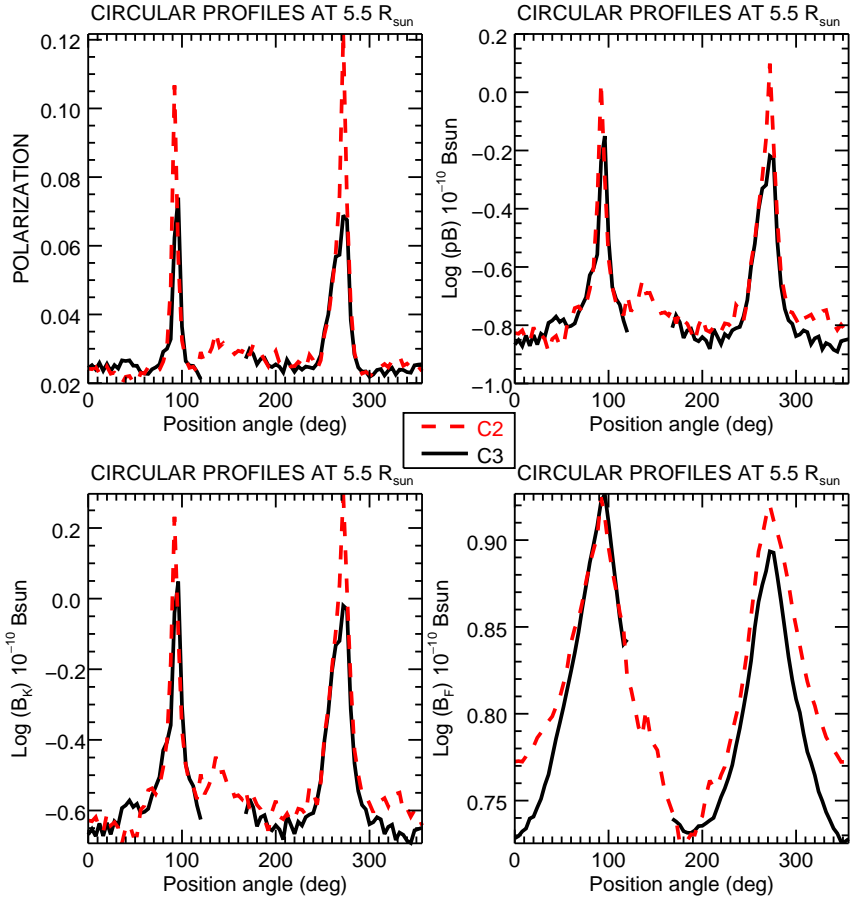}
	\caption{Comparison of the results of C2 and C3 on 16 May 1996 for the polarization (upper left panel), the polarized brightness (upper right panel), the radiance of the K-corona (lower left panel), and that of the F-corona (lower right panel).
	The circular profiles are taken at 5.5\,R${}_\odot$.} 
	\label{Fig:circular_profiles_all_c3_19960516}
\end{figure}

\begin{figure}[htpb!]
	\centering
	\includegraphics[width=\textwidth]{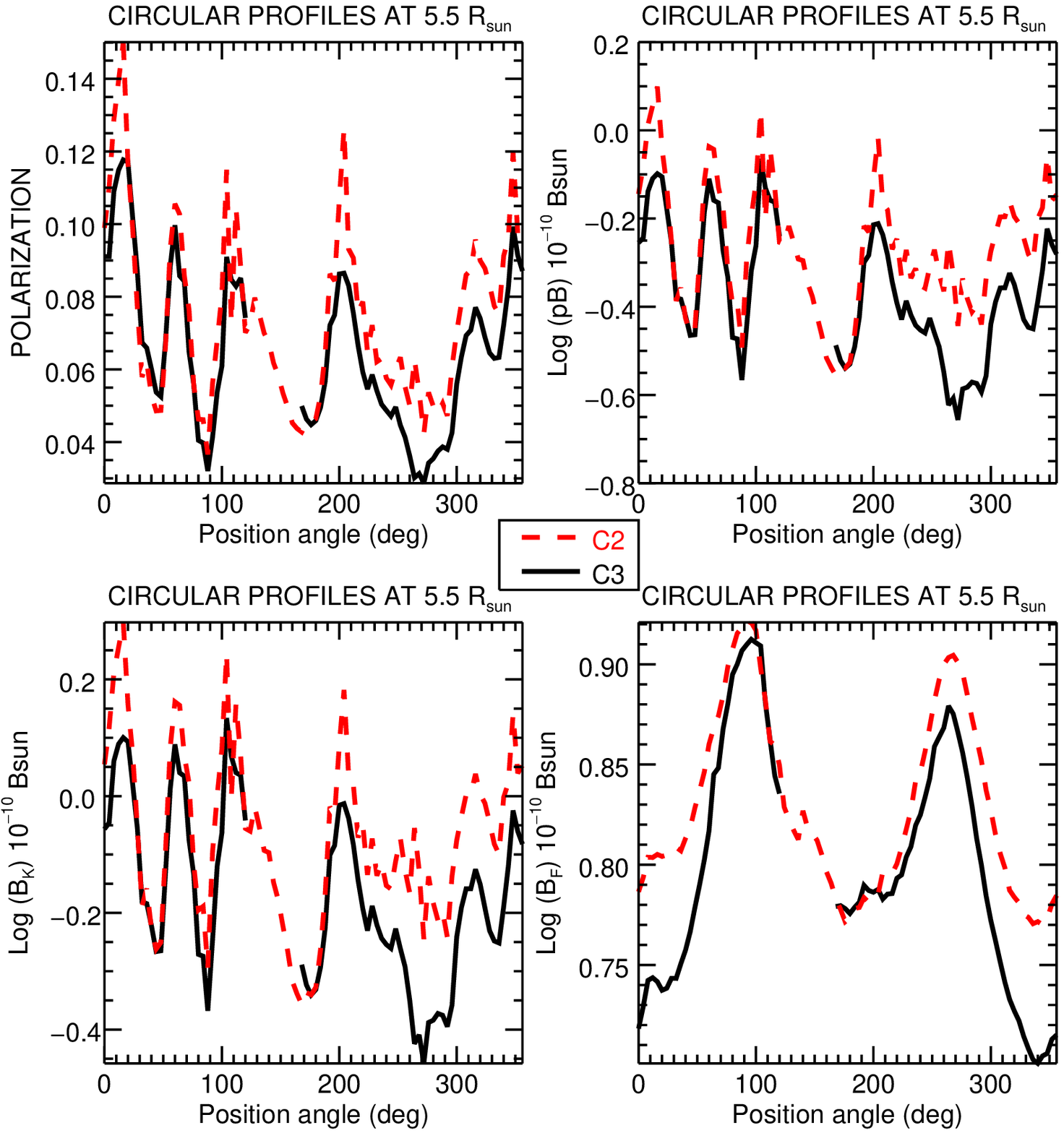}
	\caption{Same as Figures\,\ref{Fig:circular_profiles_all_c3_19960516} for 15 December 1999.} 
	\label{Fig:circular_profiles_all_c3_19991215}
\end{figure}

\begin{figure}[htpb!]
	\centering
	\includegraphics[width=\textwidth]{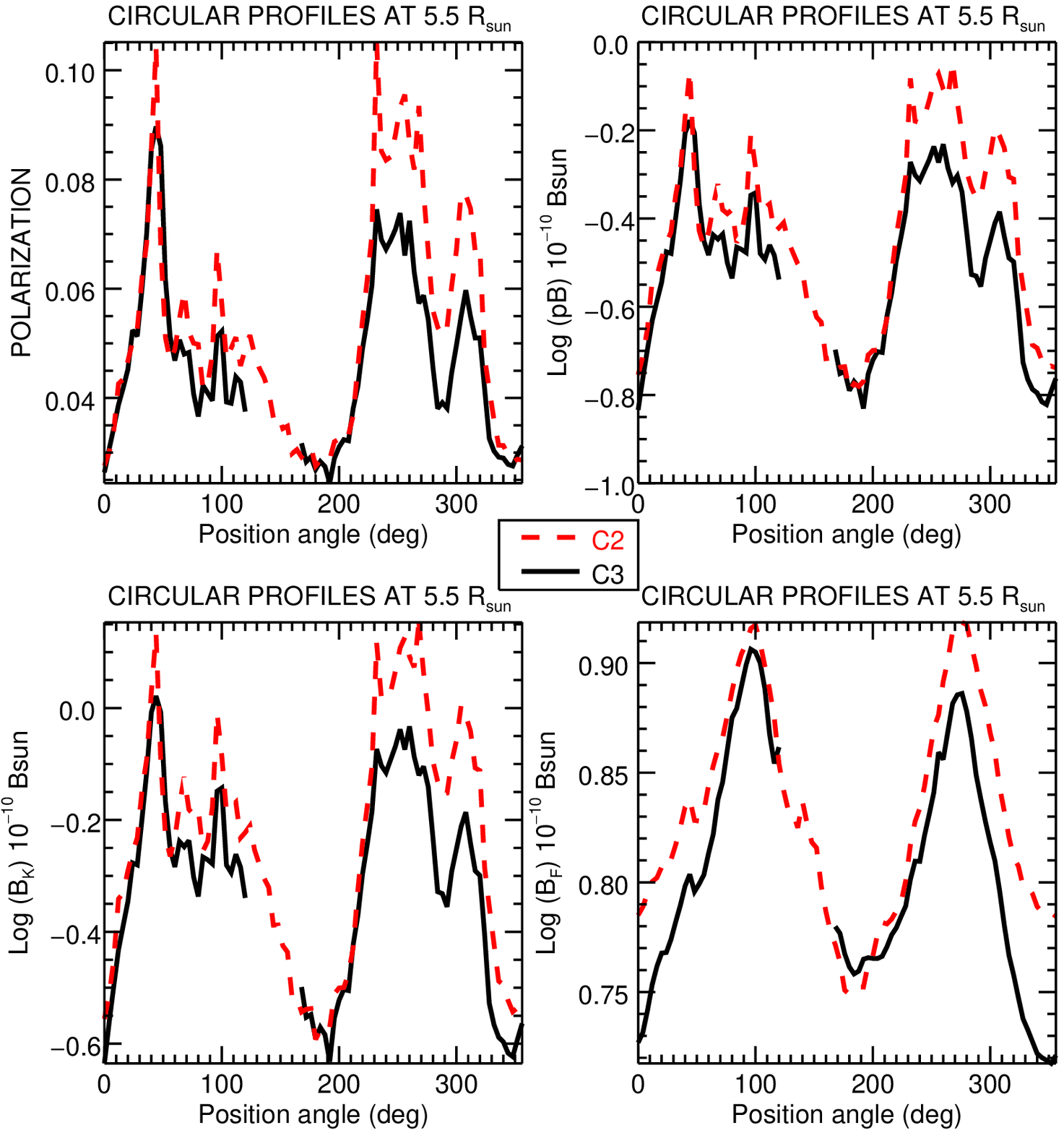}
	\caption{Same as Figures\,\ref{Fig:circular_profiles_all_c3_19960516} for 15 May 2004.} 
	\label{Fig:circular_profiles_all_c3_20040515}
\end{figure}

\begin{figure}[htpb!]
	\centering
	\includegraphics[width=\textwidth]{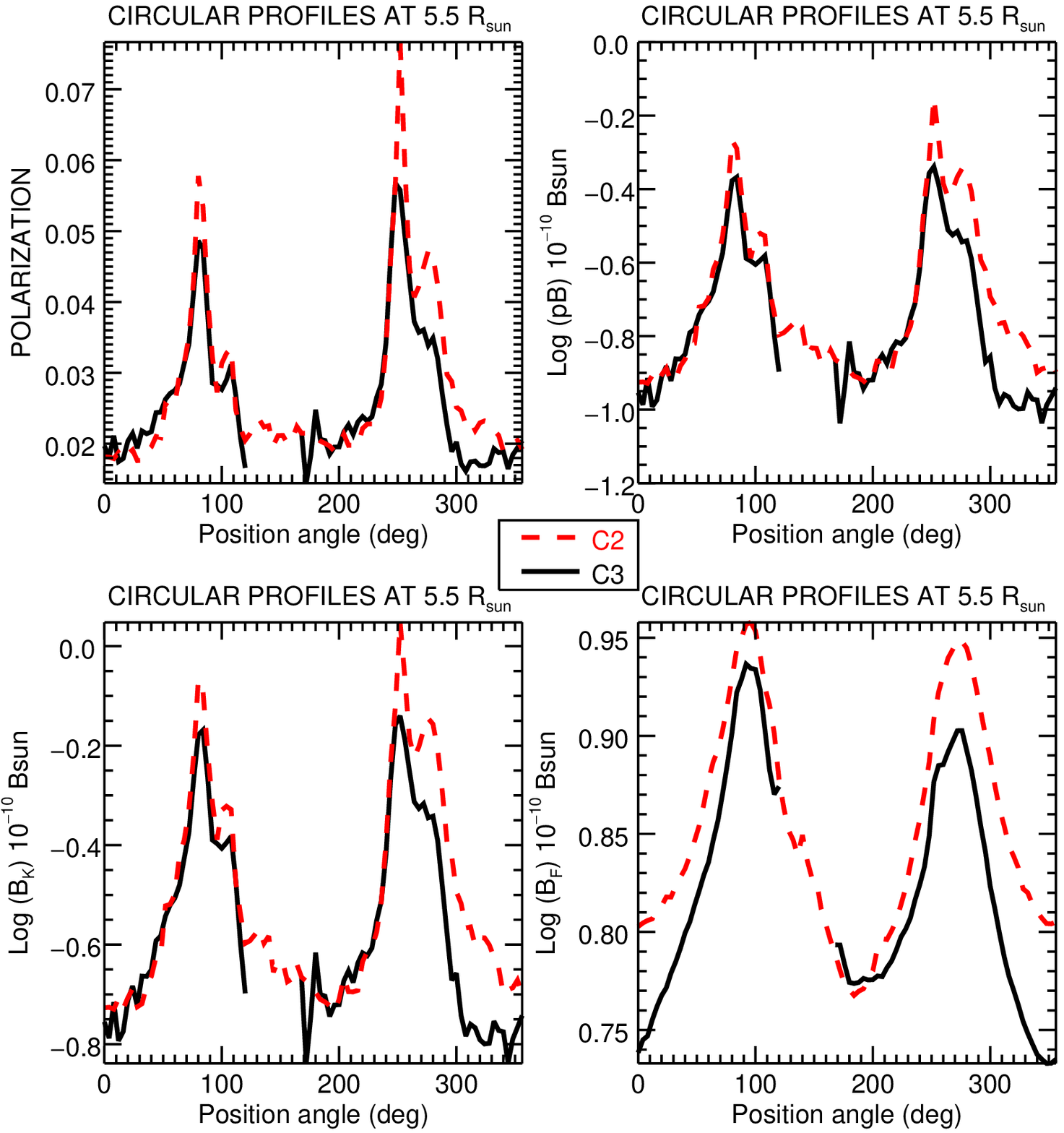}
	\caption{Same as Figures\,\ref{Fig:circular_profiles_all_c3_19960516} for 14 July 2008.} 
	\label{Fig:circular_profiles_all_c3_20080714}
\end{figure}

\begin{figure}[htpb!]
	\centering
	\includegraphics[width=\textwidth]{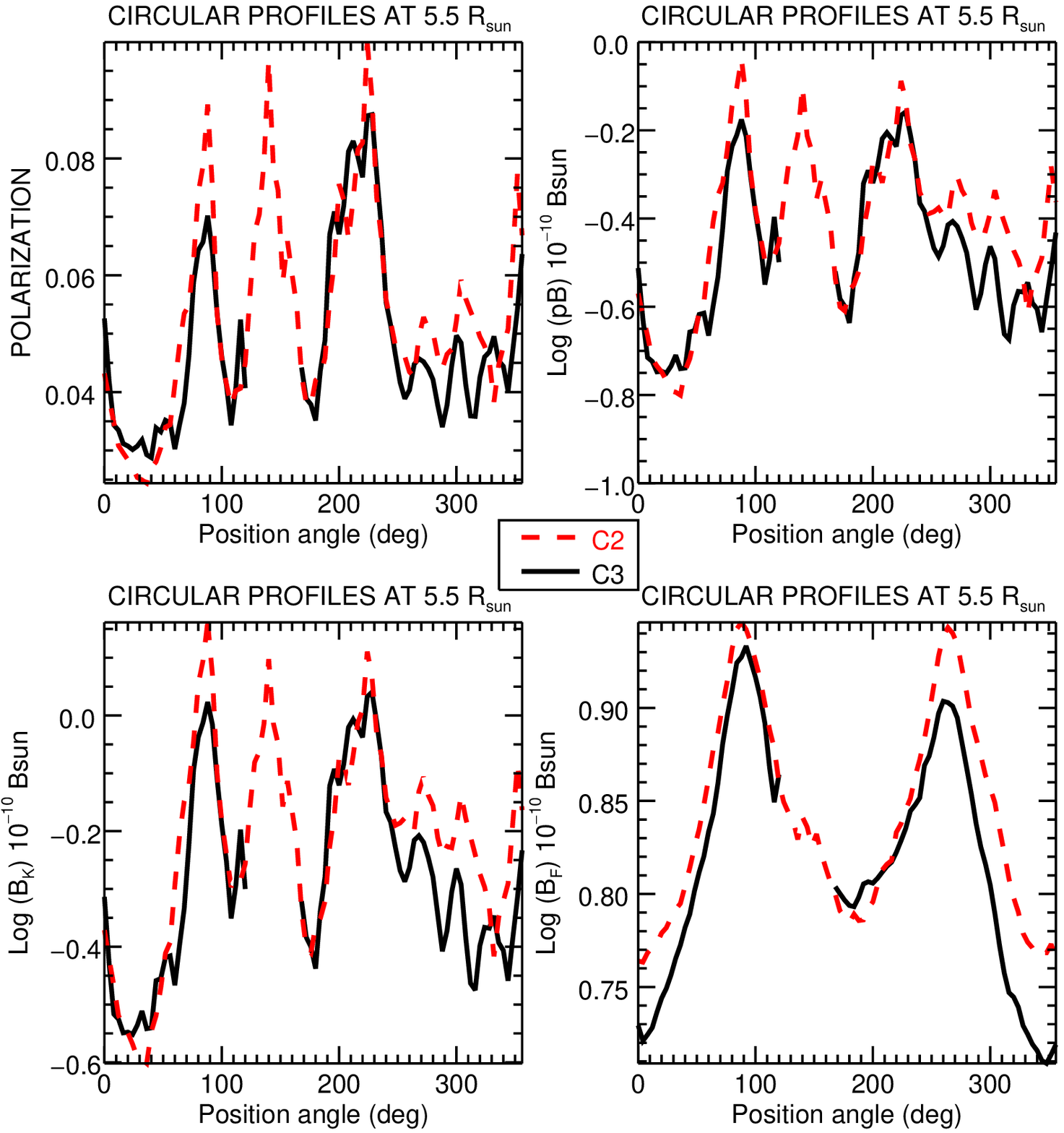}
	\caption{Same as Figures\,\ref{Fig:circular_profiles_all_c3_19960516} for 16 April 2014.} 
	\label{Fig:circular_profiles_all_c3_20140416}
\end{figure}

\begin{figure}[htpb!]
	\centering
	\includegraphics[width=\textwidth]{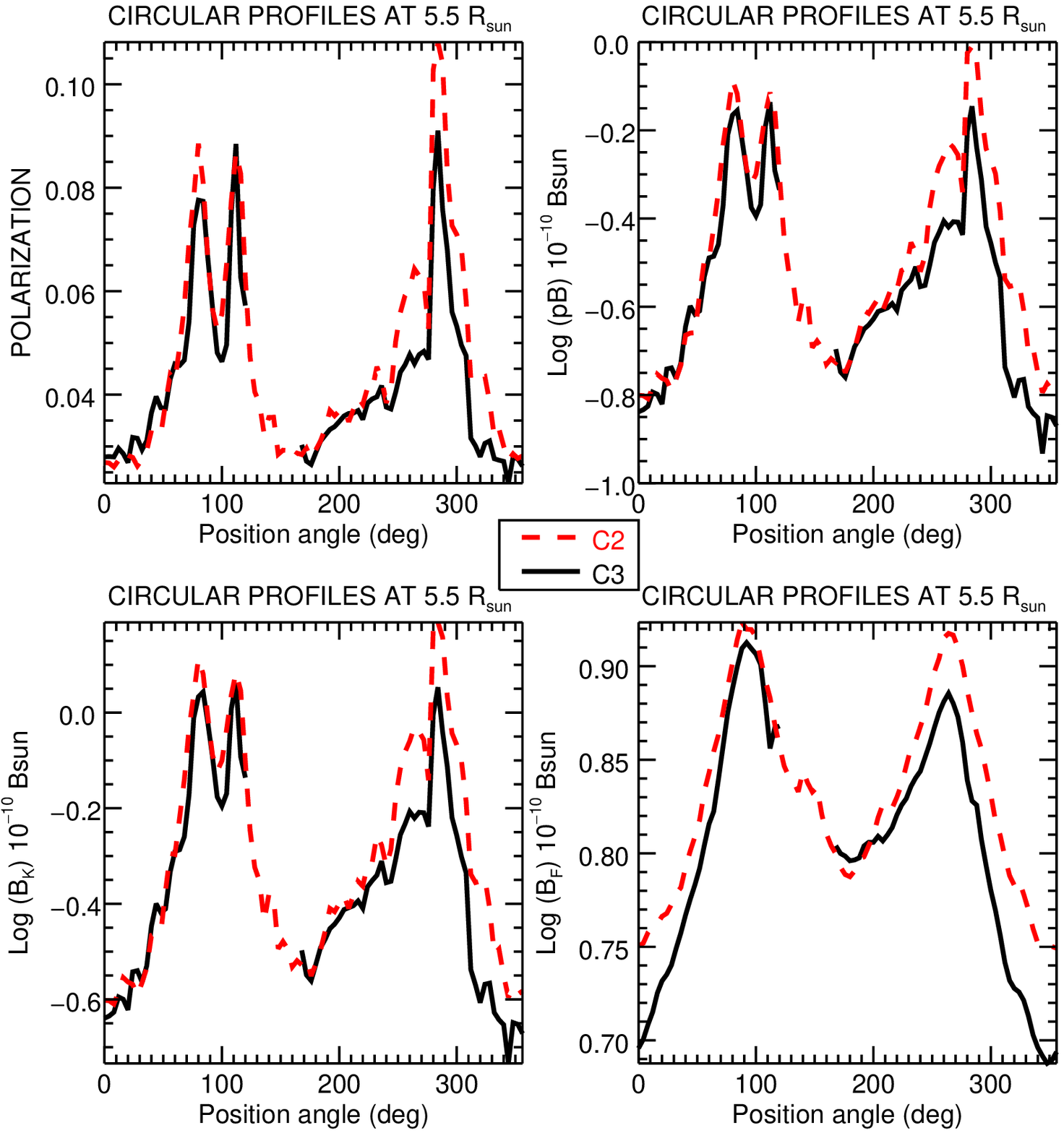}
	\caption{Same as Figures\,\ref{Fig:circular_profiles_all_c3_19960516} for 31 August 2016.} 
	\label{Fig:circular_profiles_all_c3_20160831}
\end{figure}

\begin{figure}[htpb!]
	\centering
	\includegraphics[width=\textwidth]{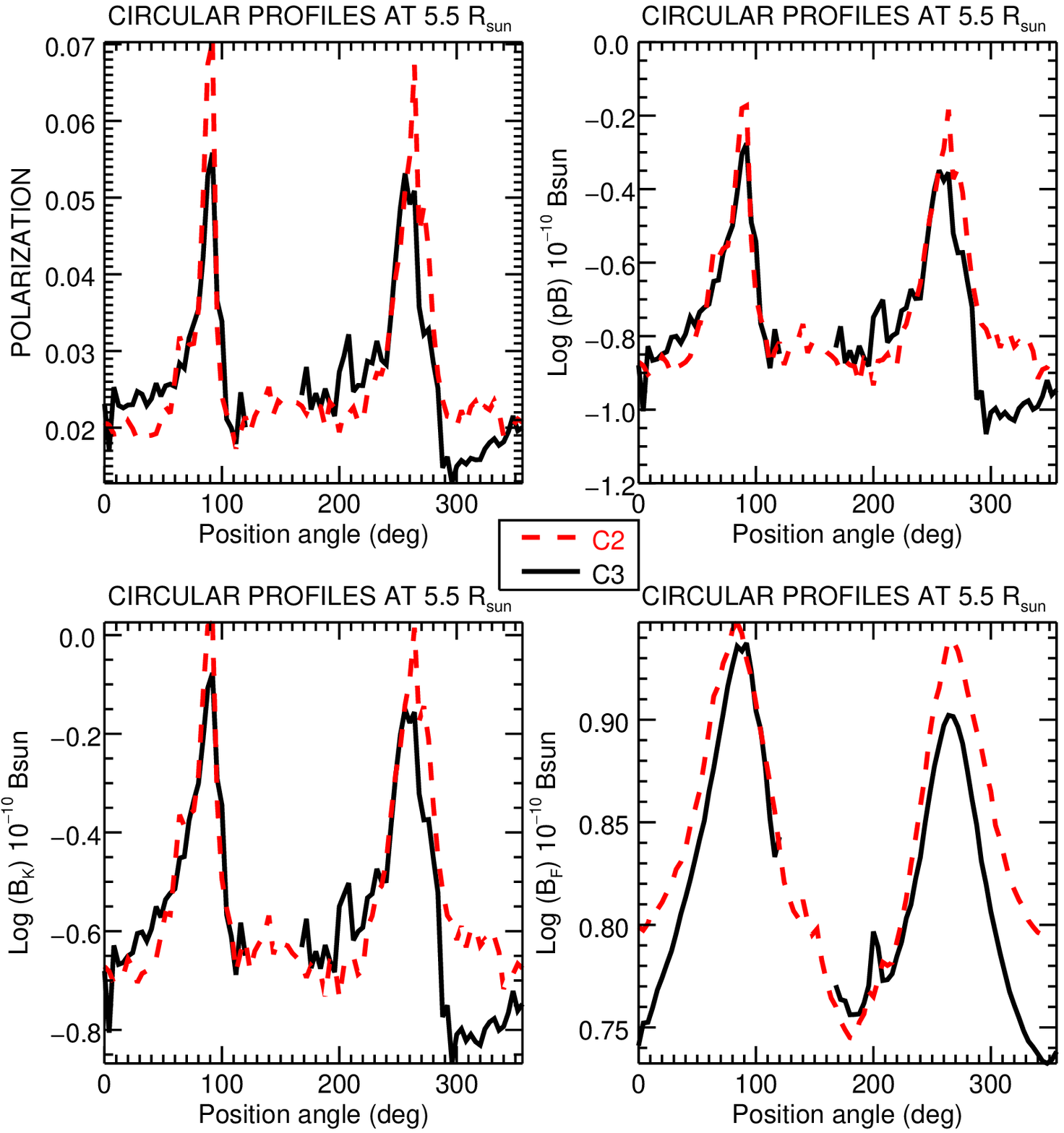}
	\caption{Same as Figures\,\ref{Fig:circular_profiles_all_c3_19960516} for 15 August 2019.} 
	\label{Fig:circular_profiles_all_c3_20190815}
\end{figure}

Figures~\ref{Fig:circular_profiles_all_c3_19960516} to \ref{Fig:circular_profiles_all_c3_20190815} display the circular profiles at 5.5\,R${}_\odot$.
In the case of C3, a large sector 50$\deg$ wide centered on the pylon is ignored to avoid the stray light pattern associated with it.
Leaving aside the case of the F-corona which may be affected by stray light effects, the coincidence of the coronal structures, prominently streamers, is perfect and the low values outside the streamer belt are generally in good agreement.
This is not the case of the peak values as those of C3 are systematically smaller than those of C2, a likely consequence of the very different spatial resolutions of the two instruments, namely by a factor of 5.
Except for the first case of 16 May 1996 (Figure~\ref{Fig:circular_profiles_all_c3_19960516}), there appears a problem in the range of position angle [220$\deg$ -- 360$\deg$] where systematic differences are present between the C2 and C3 profiles of the polarization, the polarized radiance, and the radiance of the K-corona. 
These differences are variable and typically amount to 0.01 to 0.02 for the polarization and a factor of $\approx$1.6 for both $pB$ and $B_{K}$.
It is very difficult, if not impossible, to track these discrepancies to a particular problem in the two instruments, but we did not notice any problem with the C2 data when we compared them with eclipse data in Paper I.


In spite of this problem, we attempted to build composites combining C2 and C3 images.
In fact, we selected two cases of Paper I for which we already combined eclipse and C2 images.
We completed them by introducing the contemporary C3 images whose \fov is however limited to 15\,R${}_\odot$ in order to ensure the visibility of the inner eclipse+C2 images.
A simple linear interpolation between the C2 and C3 images in a narrow ring centered at 5.5\,R${}_\odot$ was implemented to iron out discrepancies, but this ``stitching'' is barely visible on the composites.
In each of the two cases, we present two composites, one for the polarization and the other for the polarized radiance.
 
\begin{itemize}
	\item 
	This first case illustrates a corona of the minimum type and makes use of the observation of the eclipse during 26 February 1998 by a team of the High Altitude Observatory (HAO) with their Polarimetric Imager for Solar Eclipse 98 (POISE98).
	\item 
	This second case illustrates a corona of the maximum type and makes use of the observation of the eclipse during 11 August 1999 by the first author.
\end{itemize}

\begin{figure}[htpb!]
	\centering
	\includegraphics[width=\textwidth]{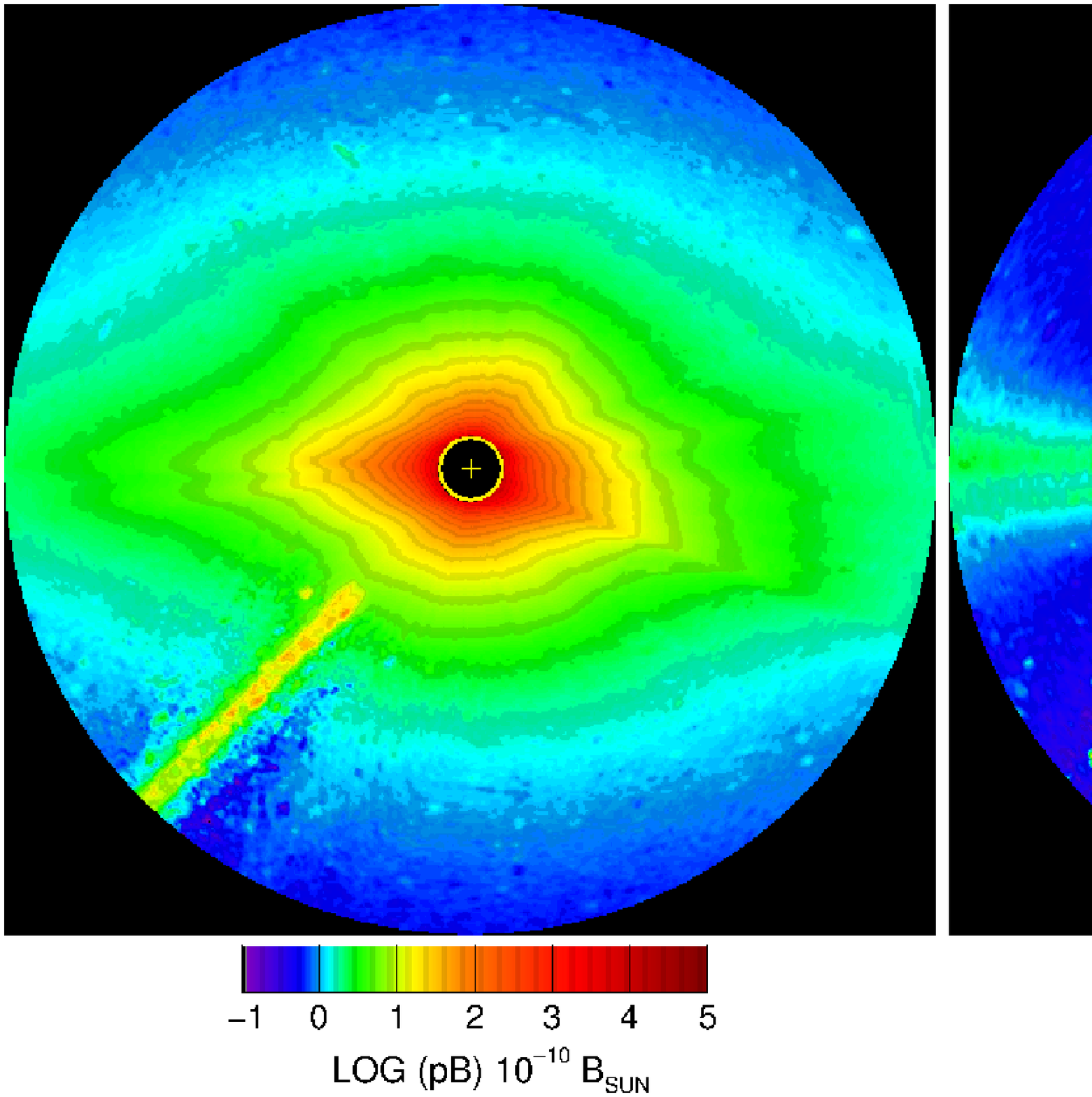}
	\caption{Composites of eclipse (26 February 1998), C2, and C3 images:  polarized brightness (left panel) and polarization (right panel).
	The yellow circles represent the solar disk with a cross at its center and the \fov extends over $\pm$15\,R${}_\odot$. 
	Solar north is up.} 
	\label{Fig:compo1998}
\end{figure}

\begin{figure}[htpb!]
	\centering
	\includegraphics[width=\textwidth]{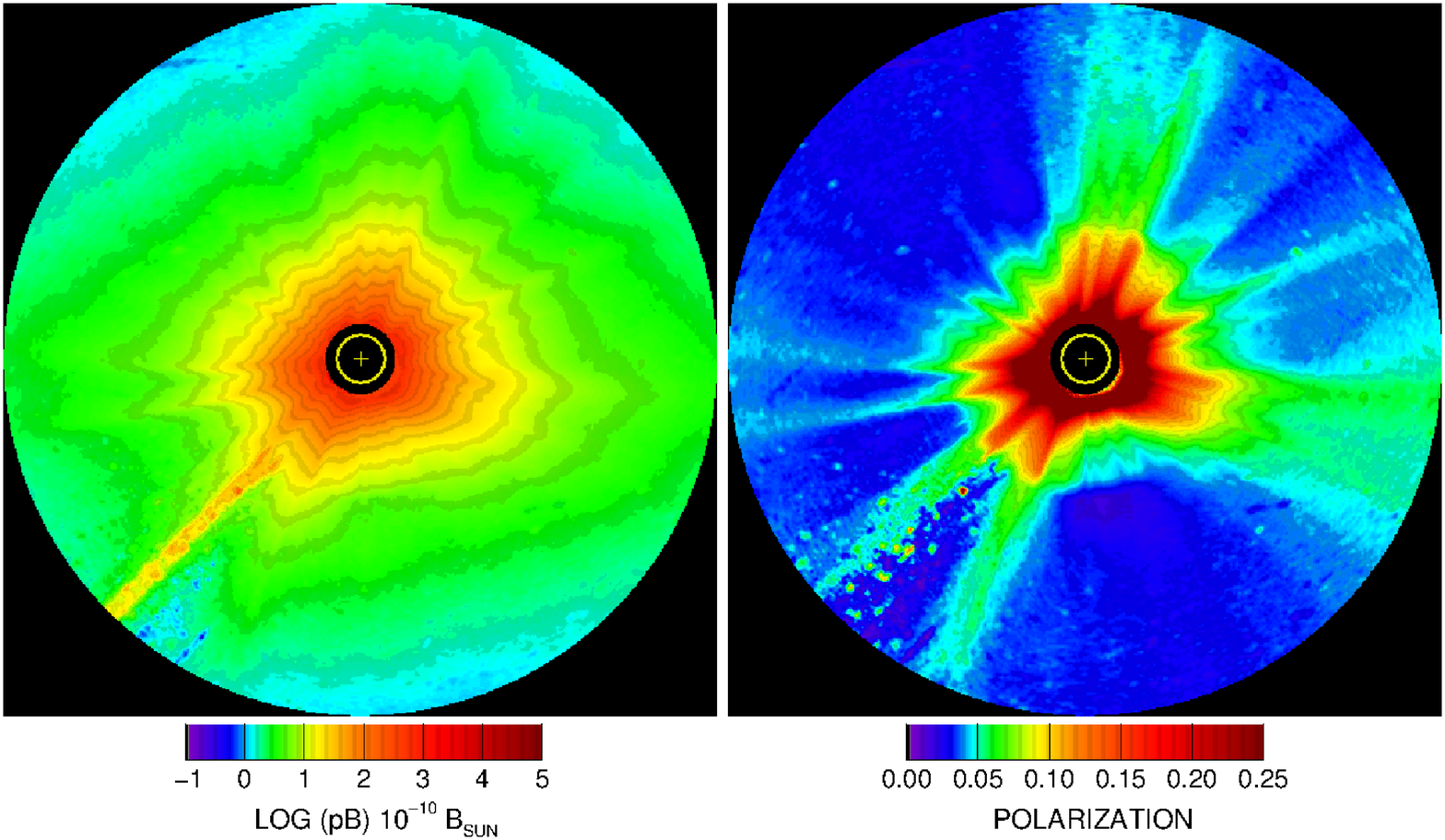}
	\caption{Composites of eclipse (11 August 1999), C2, and C3 images: polarized brightness (left panel) and polarization (right panel).
	The yellow circles represent the solar disk with a cross at its center and the \fov extends over $\pm$15\,R${}_\odot$. 
	Solar north is up.} 
	\label{Fig:compo1999}
\end{figure}

The four composites are displayed in Figures~\ref{Fig:compo1998} and \ref{Fig:compo1999} using the same scale for the $p$ and $pB$ images, respectively, so as to best reveal the contrast between the two states of the corona.
The polarization maps allow us tracking major coronal structures from the innermost region out to 15\,R${}_\odot$ and suggest their extension further out. 
That corresponding to the eclipse of 1998 conspicuously displays the large spatial extent and the depth of the coronal holes.

\begin{figure}[htpb!]
	\centering
	\includegraphics[width=1\textwidth]{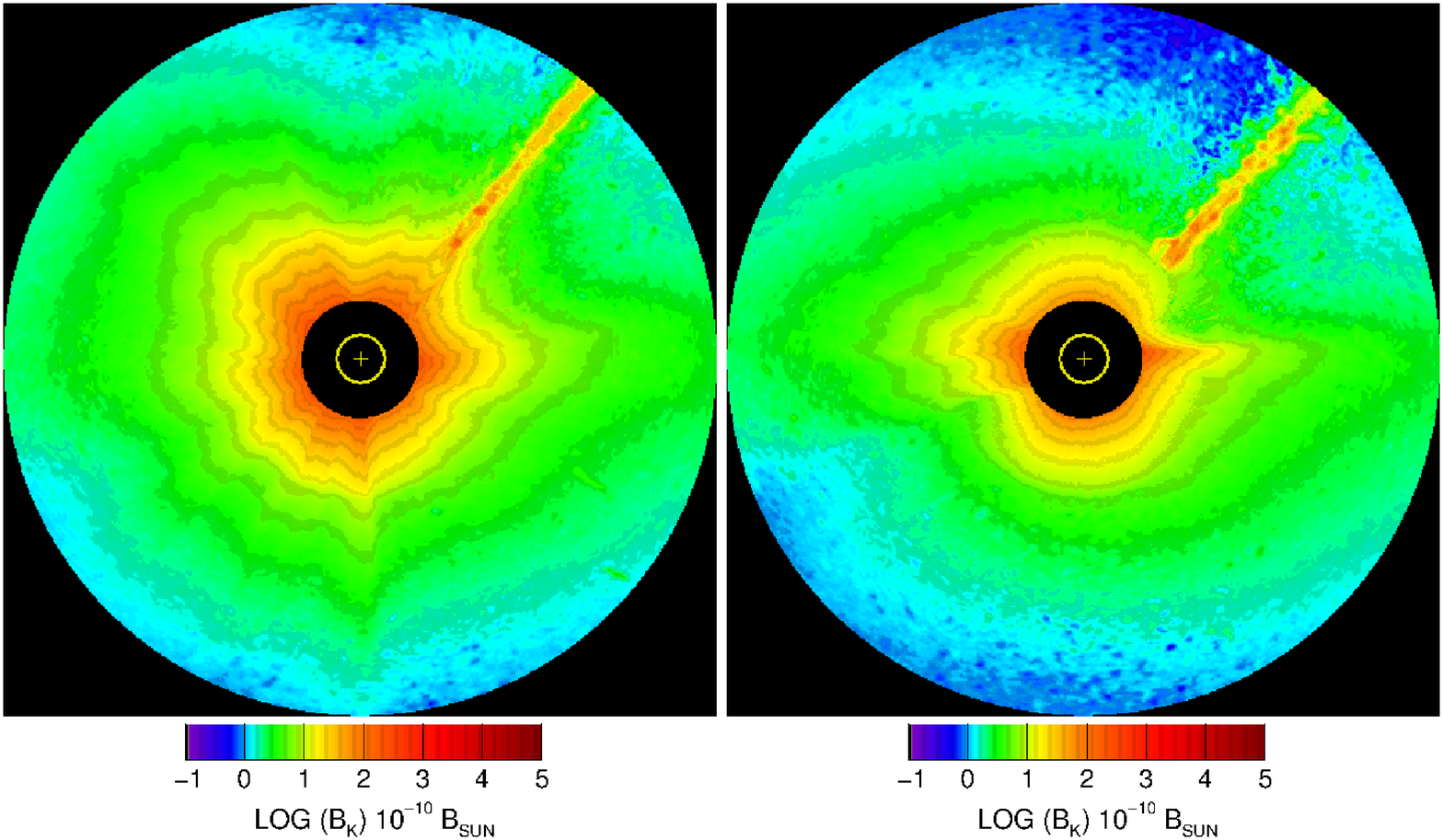}
	\caption{Composites of C2 and C3 images of the K-corona obtained on 16 April 2014 (left panel) and on 15 August 2019 (right panel).
	The yellow circles represent the solar disk with a cross at its center and the \fov extends over $\pm$15\,R${}_\odot$. 
	Solar north is down.} 
	\label{Fig:compo_K}
\end{figure}

We finally present two composites combining C2 and C3 images of the K-corona and we used two dates of Section~\ref{Sub:Ne}, namely 16 April 2014 for a corona of the maximum type and 15 August 2019 for a corona of the minimum type (Figure~\ref{Fig:compo_K}).
We further performed a quantitative comparison of characteristic radial profiles with those of standard models of a homogeneous corona of \cite{Allen1976}. 
To do so, we proceeded as follows.
\begin{itemize}
	\item 
	In the case of a corona of the maximum type, there is a single standard model.
	Consequently, we compare it with the mean radial profile from the C3 image of 16 April 2014, however excluding a wide sector centered on the pylon and encompassing its stray light pattern (two bands on each side of the pylon).
	\item
	In the case of a corona of the minimum type, there are two standard profiles, equatorial and polar, relevant to a non-spherical axisymmetric corona \citep{Saito1970}.
	Accordingly, we constructed two different profiles: 
	i) the equatorial one is the mean over a 40$\deg$ wide sector centered on the equatorial direction opposite to the stray reinforcement (Section~\ref{Sub:Polar}), and 
	ii) the polar one is the mean over a 40$\deg$ wide sector centered on the polar direction.
	The above two angular extents are typical of what has been used by past observers to tabulate their results (\eg \cite{Blackwell1955}; \cite{Michard1956}).
\end{itemize}

\begin{figure}[htpb!]
	\centering
	\includegraphics[width=\textwidth]{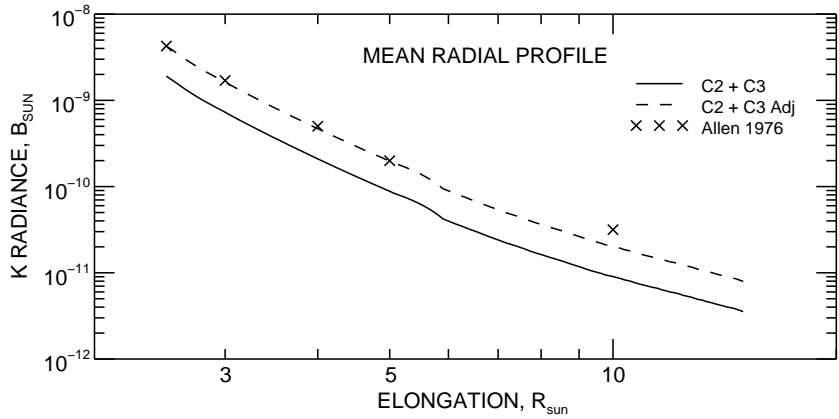}
	\caption{Mean profile from the C2+C3 composite image of the K-corona of the maximum type (16 April 2014) displayed in Figure~\ref{Fig:compo_K} compared with the standard model of \cite{Allen1976}.
	The dashed line corresponds to an adjusted version of the C2+C3 data up-scaled by a factor of 2.4.} 
	\label{Fig:profil_K_max}
\end{figure}

\begin{figure}[htpb!]
	\centering
	\includegraphics[width=\textwidth]{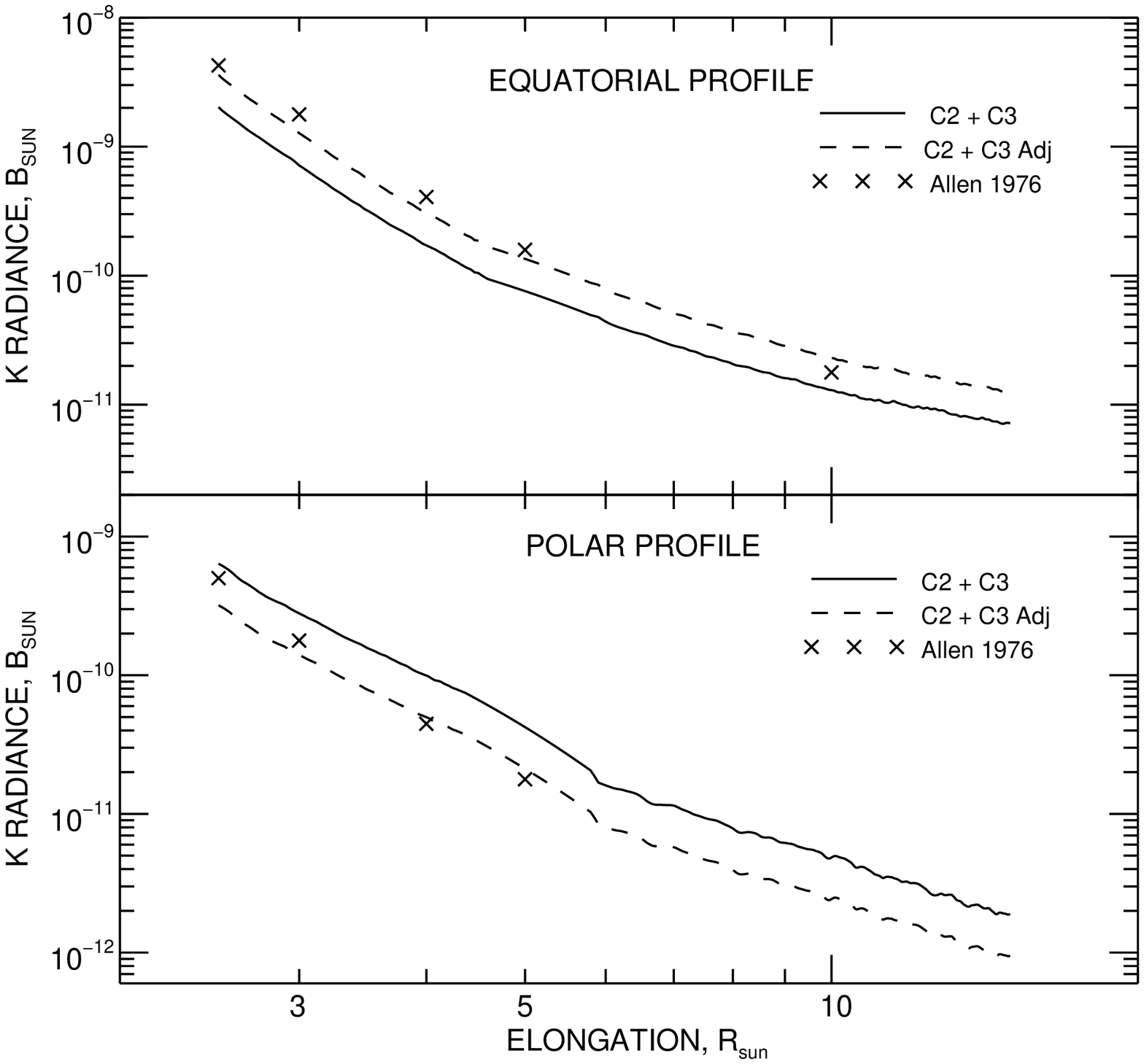}
	\caption{Equatorial (upper panel) and polar (lower panel) radial profiles from the C2+C3 composite images of the K-corona of the minimum type (15 August 2019) displayed in Figure~\ref{Fig:compo_K} compared with the corresponding standard models of \cite{Allen1976}.
	The dashed lines correspond to adjusted versions of the C2+C3 data, an up-scaling by a factor of 1.8 for the equatorial profile and a down-scaling by a factor of 2 for the polar profile.}
	\label{Fig:profil_K_min}
\end{figure}

As a general remark, Figures~\ref{Fig:profil_K_max} and \ref{Fig:profil_K_min} show the excellent match between the C2 and C3 profiles.
The ``stitching'' is perceptible but cannot affect the following conclusions.
In the case of the corona of the maximum type (16 April 2014), the gradient of the C2+C3 profile is remarkably similar to that of the model, but the radiance is globally weaker by a factor of $\approx$2.4 (Figure~\ref{Fig:profil_K_max}).
To some extent, this may be understood by considering that this model was built from $\approx$60 years old observations when the Sun and consequently the corona were more active than during the maximum  of the recent SC 24.
Scaling the C2+C3 profile by the above factor brings it in near perfect agreement with the standard one as illustrated by the dashed line.
The same conclusion holds true for the equatorial profiles of the minimum corona (15 August 2019) except that a smaller factor of $\approx$1.8 is needed to bring the two profiles in near agreement (Figure~\ref{Fig:profil_K_min}).
The situation is reversed for the polar profiles with the C2+C3 radiance being larger than the model by a factor of $\approx$2.

To put these comparisons in perspective, one should keep in mind the variability of the K-corona inside solar cycles and between cycles so that we should not expect that a particular observation will accurately correspond to some ``average'' model.
In that respect, the scaling factors quoted above are well in line with the typical modulations factors which have been reported in the past as summarized by \cite{Barlyaeva2015}.

\section{Comparison of C3 Polarization with Eclipse Measurements} 
\label{Sec:Pol-Ecl}

We first compare the C3 polarization results with the eclipse measurements presented in the Introduction (Table~\ref{Tab:Past_results}).
These measurements were obtained over a time interval of $\approx$21 years (almost two solar cycles), mostly close to minima of solar activity and their authors provided tabulated data for the equatorial and polar directions.
We note that the data of \cite{Blackwell1955} and of \cite{Blackwell1966} are identical except that the latter reference extends further out beyond 20\,R${}_\odot$.
We ignore values $<$0.01 usually given in parentheses by the authors thus indicating limited reliability. 
Regarding C3, we naturally considered the three dates of 16 May 1996, 14 July 2008, and 15 August 2019 illustrating three cases of corona of the minimum type (Figure~\ref{Fig:polar_vec_C3_MinActiv}).

\begin{figure}[htpb!]
	\centering
	\includegraphics[width=\textwidth]{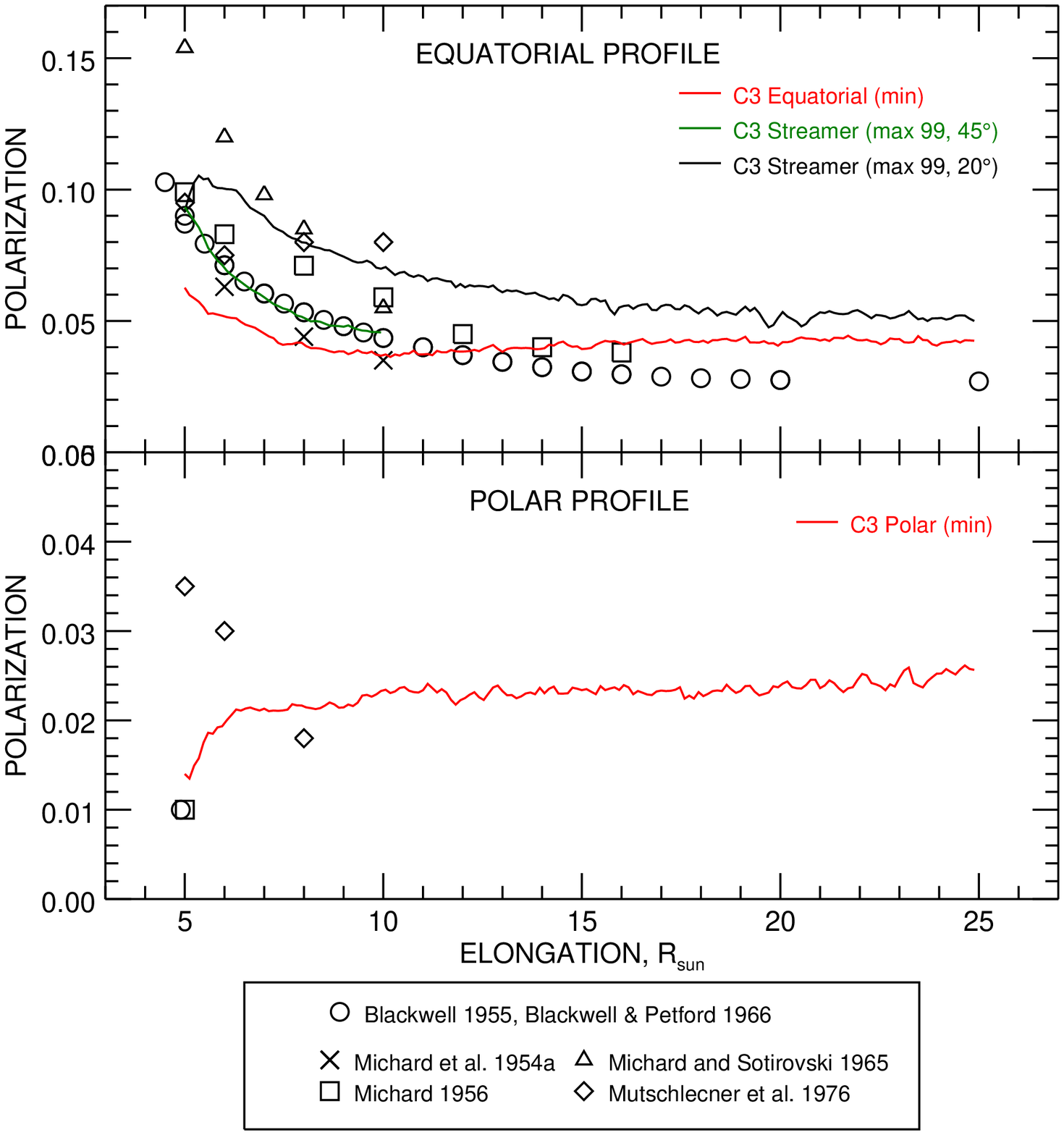}
	\caption{Variation of the polarization with elongation along the equatorial (upper panel) and polar (lower panel) directions.
	The LASCO-C3 are plotted as continuous lines and the eclipse results as symbols.} 
	\label{Fig:Ecl_C3}
\end{figure}

A polar profile was constructed by taking the means over sectors 40$\deg$ wide centered on the polar directions and averaging the resulting three individual profiles.
Figure~\ref{Fig:Ecl_C3} shows that the polarization is nearly constant at a value of 0.023 very slightly increasing with elongation, except for a rapid decrease within 6\,R${}_\odot$.
Although this looks compatible with the value of 0.01 at 5\,R${}_\odot$ reported by both \cite{Michard1954a} and \cite{Blackwell1955}, we suspect a pure coincidence and an artifact at the inner edge of the C3 \fov possibly associated with the diffraction fringe.
The data of \cite{Mutschlecner1976} indicate a very different behaviour, hardly compatible with the above results.

An equatorial profile was constructed by considering the streamers on the right side of the \fov as displayed in Figure~\ref{Fig:polar_vec_C3_MinActiv}, thus avoiding the stray reinforcement discussed in Section~\ref{Sub:Polar}, taking the means over their angular extent and averaging the three individual profiles.
This equatorial profile is compatible with the eclipse data except in the inner part of the C3 \fov where the polarization appears too low (but by 0.01 to 0.02) and beyond 15\,R${}_\odot$ where it levels off very much like the data of \cite{Blackwell1955}, but at a value slightly larger by 0.015.
However, these differences remain modest and well within the dispersion of the various eclipse measurements, further granting that the uncertainties are likely to be at least $\pm$0.01.

To broaden the scope of the C3--eclipse comparison, we included in the equatorial case of Figure~\ref{Fig:Ecl_C3} the polarization of two bright streamers present on 15 August 1999 at the maximum of SC 23.
They can conspicuously be seen on the corresponding upper panel of Figure~\ref{Fig:polar_vec_C3_MaxActiv} at PA=20$\deg$ and 45$\deg$. 
The fainter jet at PA=45$\deg$ exhibits a polarization profile (restricted to $\leq$10\,R${}_\odot$ to avoid the stray reinforcement) in remarkable agreement with the data of \cite{Blackwell1955}, indicating that this jet was representative of an equatorial mimimum when the eclipse measurements were performed.
The brighter jet at PA=20$\deg$ exhibits a higher polarization profile whose inner part ($\leq$10\,R${}_\odot$) is however compatible with some eclipse results. 
It is interesting to note that this profile is nearly parallel to the data of \cite{Blackwell1955}, except at elongations $<$6\,R${}_\odot$, and share the similar trend of  a constant polarization beyond 16\,R${}_\odot$.

\section{Comparison of C3 K-corona Radiance with Eclipse Measurements} 
\label{Sec:K-Ecl}

\begin{figure}[htpb!]
	\centering
	\includegraphics[width=\textwidth]{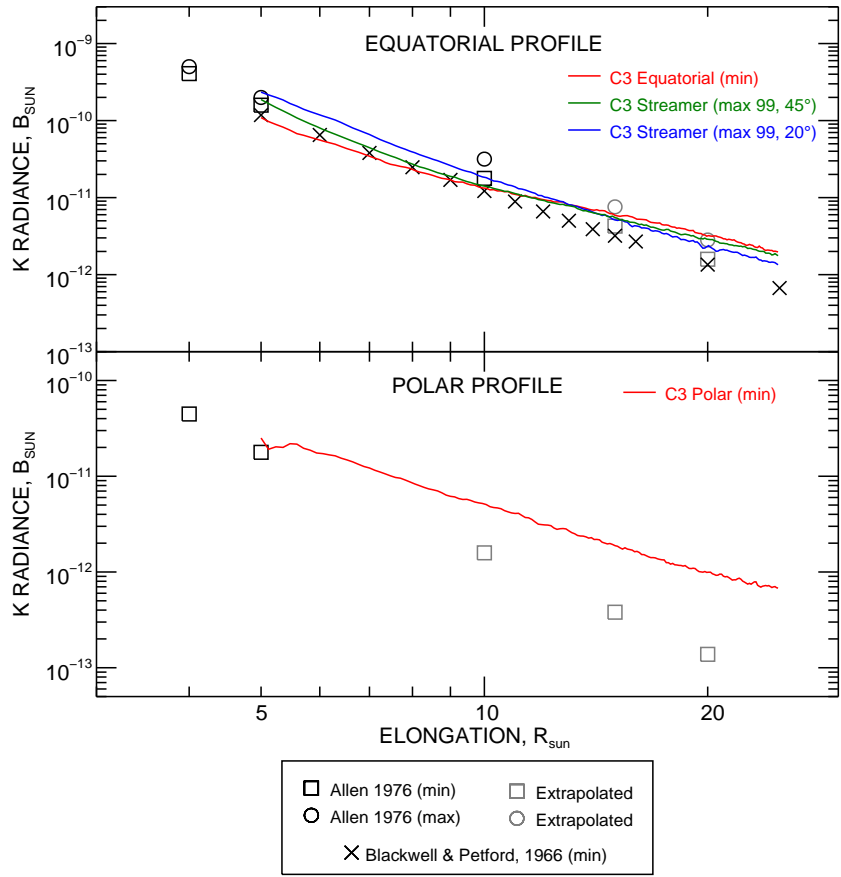}
	\caption{Equatorial (upper panel) and polar (lower panel) radial profiles of the radiance of the K-corona.
	The LASCO-C3 are plotted as continuous lines and the models and eclipse results as symbols.
	Gray symbols are extrapolated from the Allen models, see text for detail.}
	\label{Fig:profil_K}
\end{figure}

In Section~\ref{Sec:C2C3}, we compared the profiles of the C2+C3 composites with the \cite{Allen1976} model of the K-corona at two different dates.
We examine this question in more details, concentrating on the C3 results and considering mean radial profiles constructed likewise those of the polarization in the previous section.
These profiles correspond to a corona of the minimum type and they are compared with the results of \cite{Blackwell1966} and the minimum equatorial and polar models of \cite{Allen1976} as illustrated in Figure~\ref{Fig:profil_K}.
We further include his maximum model for comparison with the  two bright streamers present on the C3 image of 15 August 1999 at the maximum of SC 23.
When limited to 10\,R${}_\odot$ or even 5\,R${}_\odot$ in the case of the polar profile, the Allen models were extrapolated beyond by applying scaled power laws with an exponent of $-3.5$ as suggested by \cite{KoutchmyLamy1985}.

There is an excellent agreement between the C3 equatorial profile and the \cite{Blackwell1966} data up to 10\,R${}_\odot$ and they both follow a power law with an exponent of $-3.2$.
They diverge further out as the C3 profile experiences a decrease of its gradient with a power exponent of $\approx-2$ whereas it remains constant for the \cite{Blackwell1966} data.
The profile of the brightest streamer at PA=20$\deg$ is quite remarkable as it closely matches the Allen maximum model.
The other streamer at PA=45$\deg$ exhibits an intermediate behaviour being fainter within $\approx$10\,R${}_\odot$ and slightly brighter further out .

The situation with the polar profiles is less straightforward. 
A linear inward extrapolation (on a log-log scale) of the C3 profile leads to a fair agreement with the two Allen values at 4 and 5\,R${}_\odot$, but its power index of $-2.44$ appears at odd with the trend suggested by these two points and with the index of $-3.5$ adopted for their outer extrapolation.
At this stage, it is difficult to draw a conclusion.
On the one hand, the Allen model (strictly speaking two data points) and our extrapolation may be incorrect and the decrease in the polar direction may be less steep than assumed.
On the other hand, the presence of a faint stray light background may distort the C3 profile, or the K/F separation may be imperfect with the K-component being contaminated by the F-component, or ultimately, both adverse effects may be present.

\section{Comparison of C3 Electron Density with Eclipse and Radio Ranging Measurements} 
\label{Sub:Ne-Ecl}

Many eclipse data have been inverted to construct radial profiles of the electron density, but most of them are restricted to the inner corona, typically within a few solar radii.
We collected a set of results which extend over all or part of the C3 \fov which we briefly summarize.
\cite{Michard1954b} produced two profiles from his observation of the eclipse of 25 February 1952 (17 months before the SC 18/19 minimum), an equatorial one extending to 10\,R${}_\odot$ and a polar one to 4.75\,R${}_\odot$.
From their observation of the eclipse of 20 July 1963 (13 months before the SC 19/20 minimum), \cite{Blackwell1966} produced an equatorial profile up to 16\,R${}_\odot$ and then a model of the outer corona that extends to 40\,R${}_\odot$ (their Table III) to which they fitted a simple power law: $N_{e}(r) = 1.46 \times 10^6\,r^{-2.3}$ where $N_{e}$ is in units of cm$^{-3}$. 
\cite{Newkirk1967} combined a set of eclipse data and constructed an equatorial profile at solar minimum extending to 215\,R${}_\odot$.
\cite{Saito1970} also combined many eclipse results to generate a non-spherical axi-symmetric corona of the minimum type where the two-dimensional electron density is a function of elongation and heliolatitude.
This function appears valid up to 215\,R${}_\odot$ in the equatorial direction,.
This is not the case of the polar direction as the term ensuring the behavior at large elongations ($r^{-2.5}$) disappears causing the density to decrease abruptly.
This problem was fixed by \cite{Lemaire2011} who reintroduced such a term with a slightly different exponent ($r^{-2}$).
We finally include the classical model of \cite{Allen1976}, see also \cite{Cox2015}, which tabulates three profiles, one for a maximum corona up to 20\,R${}_\odot$, and two for a minimum corona: equatorial up to 10\,R${}_\odot$ and polar up to 5\,R${}_\odot$.

\begin{figure}[htpb!]
	\centering
	\includegraphics[width=\textwidth]{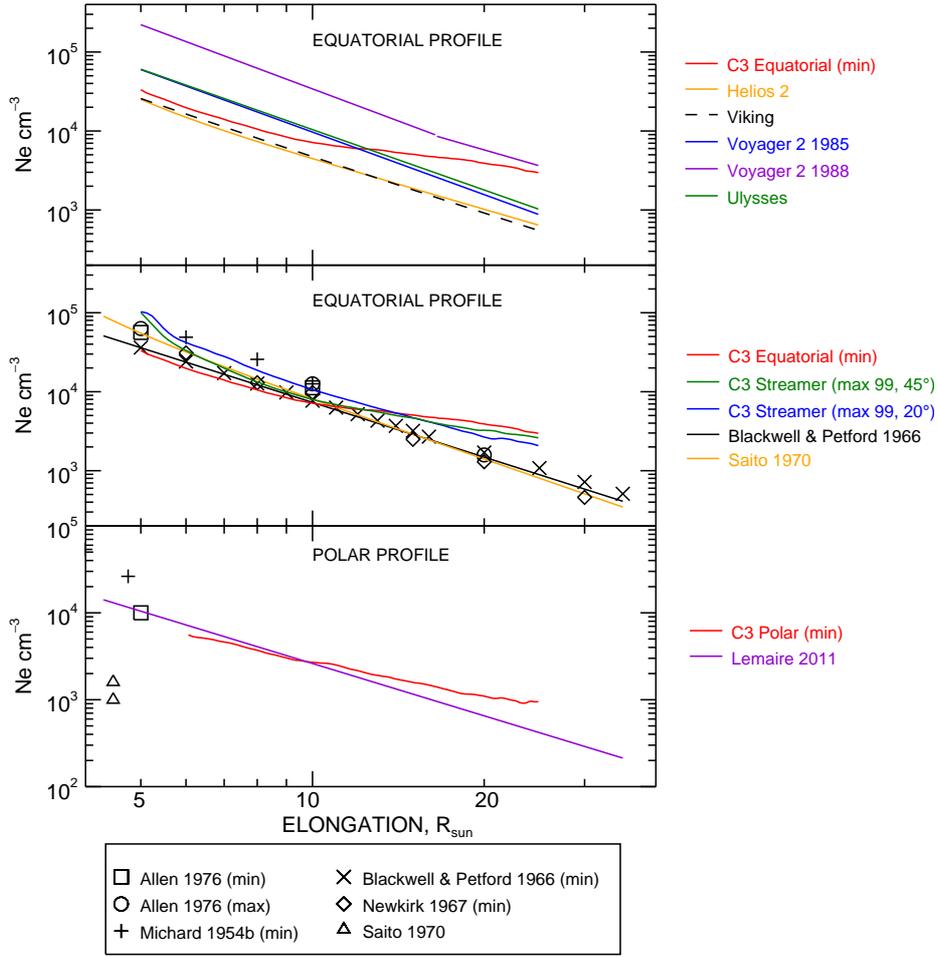}
	\caption{Comparisons of radial profiles of the electron density derived from LASCO-C3 observations with those from radio ranging (upper panel) and from eclipses measurements and 	 models along the equatorial (middle panel) and polar (lower panel) directions.}
	\label{Fig:profil_Ne}
\end{figure}

The comparison of these results with the C3 mean radial profiles constructed likewise those of the polarization and the K-corona in the previous sections are presented in Figure~\ref{Fig:profil_Ne}.
Within $\approx$10\,R${}_\odot$, the C3 equatorial and streamers profiles are well in line with the eclipse measurements and derived models, both in terms of density values and gradient of the radial variation.
Beyond this elongation, the diverging behavior of the C3 profiles is similar to those of both the polarized radiance and the K-corona radiance (Figure~\ref{Fig:profil_K}) and they certainly have the same origin(s) as discussed in the above section on the K-corona.
It is interesting to note that the two selected streamers were rather modest, with a density of 1 to $1.5 \times 10^4$ cm$^{-3}$ at 10\,R${}_\odot$, although they occurred during the maximum of SC 23,  whereas the one studied by \cite{Michard1954b} close to a solar minimum peaked at $3 \times 10^5$ cm$^{-3}$ at the same elongation.

The \fov of C3 extends over a region which is traditionally probed by spacecraft using the technique of radio ranging when they experience a solar conjunction.
Single-frequency or the more accurate dual-frequency measurements of time delay on the travel time of the carrier signal (Earth-to-spacecraft and return) sometime complemented by Doppler measurements offer a direct measure of the total electron content between spacecraft and ground stations. 
Electron density distributions are then derived assuming linear power-law representations with one or two terms. 
Table~\ref{Tab:Radio_results} summarizes past missions whose results are suitable for a comparison with those of C3.
The first three occultations occurred at or close to solar minima whereas the last two occurred close to a solar maximum.
The path along which the radio ranging is performed, and consequently its sky projection on the corona, depends upon the trajectory of the spacecraft. 
As the corona is generally not symmetric, two separate density distributions are generally reported corresponding to the two legs of the path known as ingress (\ie before occultation or solar proximate distance) and egress (\ie after these circumstances).
In addition, a latitude effect may have to be accounted for when the ray path makes a wide excursion in solar latitude (as it was the case of Viking during egress) when the authors report profiles along the equatorial direction as presented in the upper panel of Figure~\ref{Fig:profil_Ne}.
The C3 equatorial profile is shown to be consistent with the radio ranging profiles of Helios 2 and Viking and within a factor of 2 with those of Voyager 2 in 1985 and Ulysses up to an elongation of $\approx$10\,R${}_\odot$.
Beyond, we note a divergence of the C3 profile which has already been discussed above when comparing with eclipse data and in Section~\ref{Sec:K-Ecl}.
Incidentally, the agreement with the Ulysses result is somewhat surprising as it was obtained 13 months after a solar maximum.
On the contrary, the result of Voyager 2 in 1988 obtained 7 months before a solar maximum is as expected, significantly larger than those obtained at solar minima.

\begin{table*}
\begin{threeparttable}
\caption{Summary of past results of radio ranging determination of the coronal electron density .}
\vspace{0.2cm}
\label{Tab:Radio_results}
\begin{tabular}{cccc}
\hline 
Date 				& Solar activity		& Spacecraft 					& Ref	\\
\hline 
1976-05			& Min								& Helios 2			&	a		\\
1976-11			& Min + 6 Months		& Viking				& b   \\
1985-12    	& Min - 9 Months		& Voyager 2		  & c 	\\
1988-12    	& Max - 7 Months		& Voyager 2		  & d 	\\
1991-08    	& Max + 13 Months		& Ulysses		    & e 	\\	
\hline
\end{tabular}
\begin{tablenotes}
\item (a) \cite{Esposito1980}
\item (b) \cite{Muhleman1981}
\item (c) \cite{Anderson1987}
\item (d) \cite{Krisher1991}
\item (e) \cite{Bird1994}
\end{tablenotes}
\end{threeparttable}
\end{table*}

\section{Conclusion}
\label{Sec:Conclusion}

Likewise LASCO-C2, the capability of LASCO-C3 to perform accurate polarimetric measurements is hampered by the limitations inherent to the very basic polarization analyzer system (three Polaroid foils) and the fluctuations in exposure times.
On the one hand, it benefits from the absence of two folding mirrors which complicated the Mueller analysis in the case of C2.
On the other hand, it is affected by an extra source of stray light (in addition to the diffraction by the occulters), the early loss of the $0\deg$ polarizer, and incomplete calibration.
It is therefore quite encouraging that we could overcome these shortcoming, thus successfully carrying out the polarimetric analysis and obtaining maps of the polarization, the polarized radiance, the radiance of the K-corona, and of the electron density in a region poorly characterized so far and furthermore, with a quasi continuous monitoring over a time interval of 24 years covering two complete Solar Cycles 23 and 24.
This in itself represents a major advance in the characterization of the outer corona.

The four composites of images of the polarization and of the polarized radiance of the corona constructed by combining eclipse, C2, and C3 images supplemented by C2+C3 composites of images of the K-corona at two phases of solar activity, minimum and maximum, provide a strong validation of the polarization analysis, at least in the inner part of the C3 \fovnospace.
This is further reinforced by the direct comparison of the C3 results with the eclipse and radio ranging measurements.
However, the divergences found at elongations larger than $\approx$10\,R${}_\odot$ raise questions for which we presently have no clear-cut answers.
A critical puzzle concerns what we called the stray reinforcement prominently affecting the left part of the outer \fov and increasing with time; best perceived on the polarization maps, it contaminates the other maps as well. 
Whereas we could conspicuously reveal the degradation of the $0\deg$ polarizer, we found no evidence of such an effect on the other two that could explain this stray effect.
Another difficulty which must be fully appreciated lies in the faintness of the radiance and of the polarization of the corona in the outer part of the C3 \fovnospace.

In spite of these limitations, the C3 results do confirm the expected behavior of the polarization as illustrated in Figure~\ref{Fig:Black1966}, namely its progressive decrease with increasing elongation followed by a low plateau extending from $\approx$15 to 30\,R${}_\odot$.
This stems from the combination of the decreasing contribution of the highly polarized K-corona and the take-over of the brighter F-corona whose polarization is progressively  increasing with increasing elongation.
This balance obviously depends upon the solar cycle which determines the level and the geometry of the K-corona and therefore, its variation with solar latitude.
The plateau levels of polarization at $\approx$0.04 along the equatorial direction and $\approx$0.02 along the polar one during solar minima averaging to 0.03 is remarkably consistent with the model of \cite{Blackwell1966}.
These levels naturally increase with solar activity, in particular in the polar regions.

To which elongation does the polarization of the F-corona come into play and compromise the K/F separation?
On the one hand, we showed that the coronal polarization remains strongly controlled by the K-corona to an elongation of at least 20\,R${}_\odot$ on the basis of its high correlation with the temporal variation of the total photospheric magnetic field (TMF). 
On the other hand, we noted a progressive reduction of the gradient of the radial profiles of the K-corona and its streamers starting at approximately 10\,R${}_\odot$ whereas a constant gradient is expected on the basis of eclipse observations.
A similar trend naturally affects the electron density and is furthermore at odd with the radio ranging measurements.
This may in fact result from the increasing polarization of the F-corona and of the violation of the assumption of zero polarization beyond $\approx$10\,R${}_\odot$, thus answering the above question.
However, we cannot exclude the presence of a faint background of stray light whose effect would preferably be felt in the outer part of the \fov where the coronal radiance becomes very low.

The properties of the F-corona combining the outcomes of the photopolarimetric analysis of the C2 images (Paper I) and of the present analysis of the C3 images will be the subject of a forthcoming article.


\begin{acknowledgments}
We thank our former colleagues at the Laboratoire d'Astrophysique de Marseille, B. Boclet, M. Bout, and M. Burtin for their contributions to the early phases of this study.
We are grateful to Y.-M. Wang for providing the Total Magnetic Field (TMF) data.
The LASCO-C2 project at the Laboratoire d'Astrophysique de Marseille and the Laboratoire Atmosph\`eres, Milieux et Observations Spatiales is funded by the Centre National d'Etudes Spatiales (CNES).
LASCO was built by a consortium of the Naval Research Laboratory, USA, the Laboratoire d'Astrophysique de Marseille (formerly Laboratoire d'Astronomie Spatiale), France, the Max-Planck-Institut f\"ur Sonnensystemforschung (formerly Max Planck Institute f\"ur Aeronomie), Germany, and the School of Physics and Astronomy, University of Birmingham, UK.
SOHO is a project of international cooperation between ESA and NASA.
\end{acknowledgments}

\vspace{\baselineskip} 
\noindent
\textbf{Disclosure of Potential Conflicts of Interest} The authors declare that they have no conflicts of interest.


\bibliographystyle{spr-mp-sola}
\bibliography{LASCO-C3-Biblio}

\begin{thebibliography}{51}
\ifx\bisbn     \undefined \def\bisbn  #1{ISBN #1}\fi
\ifx\binits    \undefined \def\binits#1{#1}\fi
\ifx\bauthor   \undefined \def\bauthor#1{#1}\fi
\ifx\batitle   \undefined \def\batitle#1{#1}\fi
\ifx\bjtitle   \undefined \def\bjtitle#1{\textit{#1}}\fi
\ifx\bvolume   \undefined \def\bvolume#1{\textbf{#1}}\fi
\ifx\byear     \undefined \def\byear#1{#1}\fi
\ifx\bissue    \undefined \def\bissue#1{#1}\fi
\ifx\bfpage    \undefined \def\bfpage#1{#1}\fi
\ifx\blpage    \undefined \def\blpage #1{#1}\fi
\ifx\burl      \undefined \def\burl#1{#1}\fi
\ifx\href      \undefined \def\href#1#2{#2}\fi
\ifx\betal     \undefined \def\betal{et al.}\fi
\ifx\bctitle   \undefined \def\bctitle#1{#1}\fi
\ifx\beditor   \undefined \def\beditor#1{#1}\fi
\ifx\bbtitle   \undefined \def\bbtitle#1{\textit{#1}}\fi
\ifx\bedition  \undefined \def\bedition#1{#1}\fi
\ifx\bseriesno \undefined \def\bseriesno#1{\textbf{#1}}\fi
\ifx\blocation \undefined \def\blocation#1{#1}\fi
\ifx\bsertitle \undefined \def\bsertitle#1{\textit{#1}}\fi
\ifx\bsnm      \undefined \def\bsnm#1{#1}\fi
\ifx\bsuffix   \undefined \def\bsuffix#1{#1}\fi
\ifx\bparticle \undefined \def\bparticle#1{#1}\fi
\ifx\barticle  \undefined \def\barticle#1{}\fi
\ifx\binstitute  \undefined \def\binstitute#1{#1}\fi
\ifx\bpublisher  \undefined \def\bpublisher#1{#1}\fi
\ifx\doiurl    \undefined \def\doiurl#1{\href{#1}{DOI}}\fi
\makeatletter
\def\safeHref#1#2#3{\in@{http}{#2}\ifin@\href{#2}{#3}\else\href{#1#2}{#3}\fi}
\makeatother
\ifx\adsurl    \undefined
  \def\adsurl#1{\safeHref{https://ui.adsabs.harvard.edu/abs/}{#1}{ADS}}\fi
\ifx\arxivurl  \undefined
  \def\arxivurl#1{\safeHref{http://arxiv.org/abs/}{#1}{arXiv}}\fi
\ifx\botherref \undefined \def\botherref#1{}\fi
\ifx\url       \undefined \def\url#1{#1}\fi
\ifx\bchapter  \undefined \def\bchapter#1{}\fi
\ifx\bbook     \undefined \def\bbook#1{}\fi
\ifx\bcomment  \undefined \def\bcomment#1{#1}\fi
\ifx\oauthor   \undefined \def\oauthor#1{#1}\fi
\ifx\citeauthoryear \undefined\def \citeauthoryear#1{#1}\fi
\def\endbibitem {}
\ifx\bconflocation  \undefined \def\bconflocation#1{#1} \fi

\bibitem[\protect\citeauthoryear{Allen}{1976}]{Allen1976}
\begin{botherref}
\oauthor{\bsnm{Allen}, \binits{C.W.}}:
1976,
Astrophysical quantities.
\textit{Astrophysical Quantities, London: Athlone (3rd edition), 1976}
\textbf{1}.
\adsurl{1976asqu.book.....A}.
\end{botherref}
\endbibitem

\bibitem[\protect\citeauthoryear{{Anderson} et~al.}{1987}]{Anderson1987}
\begin{barticle}
\bauthor{\bsnm{{Anderson}}, \binits{J.D.}},
\bauthor{\bsnm{{Krisher}}, \binits{T.P.}},
\bauthor{\bsnm{{Borutzki}}, \binits{S.E.}},
\bauthor{\bsnm{{Connally}}, \binits{M.J.}},
\bauthor{\bsnm{{Eshe}}, \binits{P.M.}},
\bauthor{\bsnm{{Hotz}}, \binits{H.B.}},
\bauthor{\bsnm{{Kinslow}}, \binits{S.}},
\bauthor{\bsnm{{Kursinski}}, \binits{E.R.}},
\bauthor{\bsnm{{Light}}, \binits{L.B.}},
\bauthor{\bsnm{{Matousek}}, \binits{S.E.}},
\bauthor{\bsnm{{Moyd}}, \binits{K.I.}},
\bauthor{\bsnm{{Roth}}, \binits{D.C.}},
\bauthor{\bsnm{{Sweetnam}}, \binits{D.N.}},
\bauthor{\bsnm{{Taylor}}, \binits{A.H.}},
\bauthor{\bsnm{{Tyler}}, \binits{G.L.}},
\bauthor{\bsnm{{Gresh}}, \binits{D.L.}},
\bauthor{\bsnm{{Rosen}}, \binits{P.A.}}:
\byear{1987},
\batitle{{Radio Range Measurements of Coronal Electron Densities at 13 and 3.6
  Centimeter Wavelengths during the 1985 Solar Conjunction of Voyager 2}}.
\bjtitle{\apjl}
\bvolume{323},
\bfpage{L141}.
\doiurl{https://doi.org/10.1086/185074}.
\adsurl{1987ApJ...323L.141A}.
\end{barticle}
\endbibitem

\bibitem[\protect\citeauthoryear{Barlyaeva, Lamy, and
  Llebaria}{2015}]{Barlyaeva2015}
\begin{barticle}
\bauthor{\bsnm{Barlyaeva}, \binits{T.}},
\bauthor{\bsnm{Lamy}, \binits{P.}},
\bauthor{\bsnm{Llebaria}, \binits{A.}}:
\byear{2015},
\batitle{Mid-term quasi-periodicities and solar cycle variation of the
  white-light corona from 18.5 years (1996.0-2014.5) of lasco observations}.
\bjtitle{Solar Physics}
\bvolume{290},
\bfpage{2117}.
\doiurl{https://doi.org/10.1007/s11207-015-0736-6}.
\end{barticle}
\endbibitem

\bibitem[\protect\citeauthoryear{{Bird} et~al.}{1994}]{Bird1994}
\begin{barticle}
\bauthor{\bsnm{{Bird}}, \binits{M.K.}},
\bauthor{\bsnm{{Volland}}, \binits{H.}},
\bauthor{\bsnm{{Paetzold}}, \binits{M.}},
\bauthor{\bsnm{{Edenhofer}}, \binits{P.}},
\bauthor{\bsnm{{Asmar}}, \binits{S.W.}},
\bauthor{\bsnm{{Brenkle}}, \binits{J.P.}}:
\byear{1994},
\batitle{{The Coronal Electron Density Distribution Determined from
  Dual-Frequency Ranging Measurements during the 1991 Solar Conjunction of the
  ULYSSES Spacecraft}}.
\bjtitle{\apj}
\bvolume{426},
\bfpage{373}.
\doiurl{https://doi.org/10.1086/174073}.
\adsurl{1994ApJ...426..373B}.
\end{barticle}
\endbibitem

\bibitem[\protect\citeauthoryear{{Blackwell}}{1955}]{Blackwell1955}
\begin{barticle}
\bauthor{\bsnm{{Blackwell}}, \binits{D.E.}}:
\byear{1955},
\batitle{{A study of the outer corona from a high altitude aircraft at the
  eclipse of 1954 June 30. I. Observational data}}.
\bjtitle{\mnras}
\bvolume{115},
\bfpage{629}.
\doiurl{https://doi.org/10.1093/mnras/115.6.629}.
\adsurl{1955MNRAS.115..629B}.
\end{barticle}
\endbibitem

\bibitem[\protect\citeauthoryear{{Blackwell} and
  {Petford}}{1966}]{Blackwell1966}
\begin{barticle}
\bauthor{\bsnm{{Blackwell}}, \binits{D.E.}},
\bauthor{\bsnm{{Petford}}, \binits{A.D.}}:
\byear{1966},
\batitle{Observations of the 1963 july 20 solar eclipse. ii, the electron
  density in the solar corona in the region 5 \&lt;r/r\&lt; 16 obtained from
  measurements of fraunhofer line depth and the polarization of the f corona}.
\bjtitle{\mnras}
\bvolume{131},
\bfpage{399}.
\doiurl{https://doi.org/10.1093/mnras/131.3.399}.
\adsurl{1966MNRAS.131..399B}.
\end{barticle}
\endbibitem

\bibitem[\protect\citeauthoryear{{Brueckner} et~al.}{1995}]{Brueckner1995}
\begin{barticle}
\bauthor{\bsnm{{Brueckner}}, \binits{G.E.}},
\bauthor{\bsnm{{Howard}}, \binits{R.A.}},
\bauthor{\bsnm{{Koomen}}, \binits{M.J.}},
\bauthor{\bsnm{{Korendyke}}, \binits{C.M.}},
\bauthor{\bsnm{{Michels}}, \binits{D.J.}},
\bauthor{\bsnm{{Moses}}, \binits{J.D.}},
\bauthor{\bsnm{{Socker}}, \binits{D.G.}},
\bauthor{\bsnm{{Dere}}, \binits{K.P.}},
\bauthor{\bsnm{{Lamy}}, \binits{P.L.}},
\bauthor{\bsnm{{Llebaria}}, \binits{A.}},
\bauthor{\bsnm{{Bout}}, \binits{M.V.}},
\bauthor{\bsnm{{Schwenn}}, \binits{R.}},
\bauthor{\bsnm{{Simnett}}, \binits{G.M.}},
\bauthor{\bsnm{{Bedford}}, \binits{D.K.}},
\bauthor{\bsnm{{Eyles}}, \binits{C.J.}}:
\byear{1995},
\batitle{The large angle spectroscopic coronagraph (lasco)}.
\bjtitle{\solphys}
\bvolume{162},
\bfpage{357}.
\doiurl{https://doi.org/10.1007/BF00733434}.
\adsurl{1995SoPh..162..357B}.
\end{barticle}
\endbibitem

\bibitem[\protect\citeauthoryear{Cox}{2015}]{Cox2015}
\begin{bbook}
\bauthor{\bsnm{Cox}, \binits{A.N.}}:
\byear{2015},
\bbtitle{Allen’s astrophysical quantities fourth edition},
\bpublisher{Springer}.
\end{bbook}
\endbibitem

\bibitem[\protect\citeauthoryear{{Esposito}, {Edenhofer}, and
  {Lueneburg}}{1980}]{Esposito1980}
\begin{barticle}
\bauthor{\bsnm{{Esposito}}, \binits{P.B.}},
\bauthor{\bsnm{{Edenhofer}}, \binits{P.}},
\bauthor{\bsnm{{Lueneburg}}, \binits{E.}}:
\byear{1980},
\batitle{{Solar corona electron density distribution}}.
\bjtitle{\jgr}
\bvolume{85},
\bfpage{3414}.
\doiurl{https://doi.org/10.1029/JA085iA07p03414}.
\adsurl{1980JGR....85.3414E}.
\end{barticle}
\endbibitem

\bibitem[\protect\citeauthoryear{Frazin et~al.}{2012}]{Frazin2012}
\begin{barticle}
\bauthor{\bsnm{Frazin}, \binits{R.A.}},
\bauthor{\bsnm{V{\'a}squez}, \binits{A.M.}},
\bauthor{\bsnm{Thompson}, \binits{W.T.}},
\bauthor{\bsnm{Hewett}, \binits{R.J.}},
\bauthor{\bsnm{Lamy}, \binits{P.}},
\bauthor{\bsnm{Llebaria}, \binits{A.}},
\bauthor{\bsnm{Vourlidas}, \binits{A.}},
\bauthor{\bsnm{Burkepile}, \binits{J.}}:
\byear{2012},
\batitle{Intercomparison of the lasco-c2, secchi-cor1, secchi-cor2, and mk4
  coronagraphs}.
\bjtitle{Solar Physics}
\bvolume{280},
\bfpage{273}.
\doiurl{https://doi.org/10.1007/s11207-012-0028-3}.
\end{barticle}
\endbibitem

\bibitem[\protect\citeauthoryear{{Hahn} et~al.}{2002}]{Hahn2002}
\begin{barticle}
\bauthor{\bsnm{{Hahn}}, \binits{J.M.}},
\bauthor{\bsnm{{Zook}}, \binits{H.A.}},
\bauthor{\bsnm{{Cooper}}, \binits{B.}},
\bauthor{\bsnm{{Sunkara}}, \binits{B.}}:
\byear{2002},
\batitle{Clementine observations of the zodiacal light and the dust content of
  the inner solar system}.
\bjtitle{Icarus}
\bvolume{158},
\bfpage{360}.
\doiurl{https://doi.org/10.1006/icar.2002.6881}.
\adsurl{2002Icar..158..360H}.
\end{barticle}
\endbibitem

\bibitem[\protect\citeauthoryear{{Koomen} et~al.}{1975}]{Koomen1975}
\begin{barticle}
\bauthor{\bsnm{{Koomen}}, \binits{M.J.}},
\bauthor{\bsnm{{Detwiler}}, \binits{C.R.}},
\bauthor{\bsnm{{Brueckner}}, \binits{G.E.}},
\bauthor{\bsnm{{Cooper}}, \binits{H.W.}},
\bauthor{\bsnm{{Tousey}}, \binits{R.}}:
\byear{1975},
\batitle{White light coronagraph in oso-7}.
\bjtitle{Applied Optics}
\bvolume{14},
\bfpage{743}.
\doiurl{https://doi.org/10.1364/AO.14.000743}.
\adsurl{1975ApOpt..14..743K}.
\end{barticle}
\endbibitem

\bibitem[\protect\citeauthoryear{{Koutchmy} and
  {Lamy}}{1985}]{KoutchmyLamy1985}
\begin{bchapter}
\bauthor{\bsnm{{Koutchmy}}, \binits{S.}},
\bauthor{\bsnm{{Lamy}}, \binits{P.L.}}:
\byear{1985},
\bctitle{The f-corona and the circum-solar dust evidences and properties}.
In: \beditor{\bsnm{{Giese}}, \binits{R.H.}},
\beditor{\bsnm{{Lamy}}, \binits{P.}} (eds.)
\bbtitle{IAU Colloq. 85: Properties and Interactions of Interplanetary Dust},
\bsertitle{Astrophysics and Space Science Library}
\bseriesno{119},
\bfpage{63}.
\doiurl{https://doi.org/10.1007/978-94-009-5464-9_14}.
\adsurl{1985ASSL..119...63K}.
\end{bchapter}
\endbibitem

\bibitem[\protect\citeauthoryear{{Koutchmy} and
  {Livshits}}{1992}]{Koutchmy1992}
\begin{barticle}
\bauthor{\bsnm{{Koutchmy}}, \binits{S.}},
\bauthor{\bsnm{{Livshits}}, \binits{M.}}:
\byear{1992},
\batitle{Coronal streamers}.
\bjtitle{\ssr}
\bvolume{61},
\bfpage{393}.
\doiurl{https://doi.org/10.1007/BF00222313}.
\adsurl{1992SSRv...61..393K}.
\end{barticle}
\endbibitem

\bibitem[\protect\citeauthoryear{{Koutchmy}, {Picat}, and
  {Dantel}}{1977}]{Koutchmy1977}
\begin{barticle}
\bauthor{\bsnm{{Koutchmy}}, \binits{S.}},
\bauthor{\bsnm{{Picat}}, \binits{J.P.}},
\bauthor{\bsnm{{Dantel}}, \binits{M.}}:
\byear{1977},
\batitle{{Etude polarim{\'e}trique de la couronne solaire observe a
  l'{\`e}clipse totale du 30 juin 1973 {\`a} l'aide d'un filtre neutre
  radial.}}
\bjtitle{\aap}
\bvolume{59},
\bfpage{349}.
\adsurl{1977A&A....59..349K}.
\end{barticle}
\endbibitem

\bibitem[\protect\citeauthoryear{{Krisher} et~al.}{1991}]{Krisher1991}
\begin{barticle}
\bauthor{\bsnm{{Krisher}}, \binits{T.P.}},
\bauthor{\bsnm{{Anderson}}, \binits{J.D.}},
\bauthor{\bsnm{{Morabito}}, \binits{D.D.}},
\bauthor{\bsnm{{Asmar}}, \binits{S.W.}},
\bauthor{\bsnm{{Borutzki}}, \binits{S.E.}},
\bauthor{\bsnm{{Delitsky}}, \binits{M.L.}},
\bauthor{\bsnm{{Densmore}}, \binits{A.C.}},
\bauthor{\bsnm{{Eshe}}, \binits{P.M.}},
\bauthor{\bsnm{{Lewis}}, \binits{G.D.}},
\bauthor{\bsnm{{Maurer}}, \binits{M.J.}},
\bauthor{\bsnm{{Roth}}, \binits{D.C.}},
\bauthor{\bsnm{{Son}}, \binits{Y.H.}},
\bauthor{\bsnm{{Spilker}}, \binits{T.R.}},
\bauthor{\bsnm{{Sweetnam}}, \binits{D.N.}},
\bauthor{\bsnm{{Taylor}}, \binits{A.H.}},
\bauthor{\bsnm{{Tyler}}, \binits{G.L.}},
\bauthor{\bsnm{{Gresh}}, \binits{D.L.}},
\bauthor{\bsnm{{Rosen}}, \binits{P.A.}}:
\byear{1991},
\batitle{{Radio Range Measurements of Coronal Electron Densities at 13 and 3.6
  Centimeter Wavelengths during the 1988 Solar Conjunction of Voyager 2}}.
\bjtitle{\apjl}
\bvolume{375},
\bfpage{L57}.
\doiurl{https://doi.org/10.1086/186087}.
\adsurl{1991ApJ...375L..57K}.
\end{barticle}
\endbibitem

\bibitem[\protect\citeauthoryear{Lamy, Llebaria, and
  Quemerais}{2002}]{Lamy2002}
\begin{barticle}
\bauthor{\bsnm{Lamy}, \binits{P.}},
\bauthor{\bsnm{Llebaria}, \binits{A.}},
\bauthor{\bsnm{Quemerais}, \binits{E.}}:
\byear{2002},
\batitle{Solar cycle variation of the radiance and the global electron density
  of the solar corona}.
\bjtitle{Advances in Space Research}
\bvolume{29},
\bfpage{373}.
\doiurl{https://doi.org/10.1016/S0273-1177(01)00599-3}.
\end{barticle}
\endbibitem

\bibitem[\protect\citeauthoryear{Lamy and Perrin}{1986}]{Lamy1986}
\begin{barticle}
\bauthor{\bsnm{Lamy}, \binits{P.L.}},
\bauthor{\bsnm{Perrin}, \binits{J.-M.}}:
\byear{1986},
\batitle{Volume scattering function and space distribution of the
  interplanetary dust cloud}.
\bjtitle{Astronomy and Astrophysics}
\bvolume{163},
\bfpage{269}.
\adsurl{1986A\%26A...163..269L}.
\end{barticle}
\endbibitem

\bibitem[\protect\citeauthoryear{{Lamy} et~al.}{2014}]{Lamy2014}
\begin{barticle}
\bauthor{\bsnm{{Lamy}}, \binits{P.}},
\bauthor{\bsnm{{Barlyaeva}}, \binits{T.}},
\bauthor{\bsnm{{Llebaria}}, \binits{A.}},
\bauthor{\bsnm{{Floyd}}, \binits{O.}}:
\byear{2014},
\batitle{{Comparing the solar minima of cycles 22/23 and 23/24: The view from
  LASCO white light coronal images}}.
\bjtitle{Journal of Geophysical Research (Space Physics)}
\bvolume{119},
\bfpage{47}.
\doiurl{https://doi.org/10.1002/2013JA019468}.
\adsurl{2014JGRA..119...47L}.
\end{barticle}
\endbibitem

\bibitem[\protect\citeauthoryear{{Lamy} et~al.}{2017}]{Lamy2017}
\begin{barticle}
\bauthor{\bsnm{{Lamy}}, \binits{P.}},
\bauthor{\bsnm{{Boclet}}, \binits{B.}},
\bauthor{\bsnm{{Wojak}}, \binits{J.}},
\bauthor{\bsnm{{Vibert}}, \binits{D.}}:
\byear{2017},
\batitle{{Anomalous Surge of the White-Light Corona at the Onset of the
  Declining Phase of Solar Cycle 24}}.
\bjtitle{\solphys}
\bvolume{292},
\bfpage{60}.
\doiurl{https://doi.org/10.1007/s11207-017-1089-0}.
\adsurl{http://cdsads.u-strasbg.fr/abs/2017SoPh..292...60L}.
\end{barticle}
\endbibitem

\bibitem[\protect\citeauthoryear{{Lamy} et~al.}{2020}]{Lamy2020}
\begin{barticle}
\bauthor{\bsnm{{Lamy}}, \binits{P.}},
\bauthor{\bsnm{{Llebaria}}, \binits{A.}},
\bauthor{\bsnm{{Boclet}}, \binits{B.}},
\bauthor{\bsnm{{Gilardy}}, \binits{H.}},
\bauthor{\bsnm{{Burtin}}, \binits{M.}},
\bauthor{\bsnm{{Floyd}}, \binits{O.}}:
\byear{2020},
\batitle{{Coronal Photopolarimetry with the LASCO-C2 Coronagraph over 24 Years
  [1996 - 2019]}}.
\bjtitle{\solphys}
\bvolume{295},
\bfpage{89}.
\doiurl{https://doi.org/10.1007/s11207-020-01650-y}.
\adsurl{2020SoPh..295...89L}.
\end{barticle}
\endbibitem

\bibitem[\protect\citeauthoryear{{Leinert} and {Pitz}}{1989}]{Leinert1989}
\begin{barticle}
\bauthor{\bsnm{{Leinert}}, \binits{C.}},
\bauthor{\bsnm{{Pitz}}, \binits{E.}}:
\byear{1989},
\batitle{{Zodiacal light observed by HELIOS throughout solar cycle No 21:
  stable dust and varying plasma.}}
\bjtitle{\aap}
\bvolume{210},
\bfpage{399}.
\adsurl{1989A&A...210..399L}.
\end{barticle}
\endbibitem

\bibitem[\protect\citeauthoryear{{Leinert}, {Link}, and
  {Pitz}}{1974}]{Leinert1974}
\begin{barticle}
\bauthor{\bsnm{{Leinert}}, \binits{C.}},
\bauthor{\bsnm{{Link}}, \binits{H.}},
\bauthor{\bsnm{{Pitz}}, \binits{E.}}:
\byear{1974},
\batitle{Rocket photometry of the inner zodiacal light}.
\bjtitle{\aap}
\bvolume{30},
\bfpage{411}.
\adsurl{1974A&A....30..411L}.
\end{barticle}
\endbibitem

\bibitem[\protect\citeauthoryear{{Leinert} et~al.}{1976}]{Leinert1976}
\begin{barticle}
\bauthor{\bsnm{{Leinert}}, \binits{C.}},
\bauthor{\bsnm{{Link}}, \binits{H.}},
\bauthor{\bsnm{{Pitz}}, \binits{E.}},
\bauthor{\bsnm{{Giese}}, \binits{R.H.}}:
\byear{1976},
\batitle{Interpretation of a rocket photometry of the inner zodiacal light}.
\bjtitle{\aap}
\bvolume{47},
\bfpage{221}.
\adsurl{1976A\%26A....47..221L}.
\end{barticle}
\endbibitem

\bibitem[\protect\citeauthoryear{{Leinert} et~al.}{1980}]{Leinert1980}
\begin{barticle}
\bauthor{\bsnm{{Leinert}}, \binits{C.}},
\bauthor{\bsnm{{Hanner}}, \binits{M.}},
\bauthor{\bsnm{{Richter}}, \binits{I.}},
\bauthor{\bsnm{{Pitz}}, \binits{E.}}:
\byear{1980},
\batitle{{The plane of symmetry of interplanetary dust in the inner solar
  system.}}
\bjtitle{\aap}
\bvolume{82},
\bfpage{328}.
\adsurl{1980A&A....82..328L}.
\end{barticle}
\endbibitem

\bibitem[\protect\citeauthoryear{{Leinert} et~al.}{1981}]{Leinert1981}
\begin{barticle}
\bauthor{\bsnm{{Leinert}}, \binits{C.}},
\bauthor{\bsnm{{Richter}}, \binits{I.}},
\bauthor{\bsnm{{Pitz}}, \binits{E.}},
\bauthor{\bsnm{{Planck}}, \binits{B.}}:
\byear{1981},
\batitle{The zodiacal light from 1.0 to 0.3 a.u. as observed by the helios
  space probes}.
\bjtitle{\aap}
\bvolume{103},
\bfpage{177}.
\adsurl{1981A\%26A...103..177L}.
\end{barticle}
\endbibitem

\bibitem[\protect\citeauthoryear{{Leinert} et~al.}{1982}]{Leinert1982}
\begin{barticle}
\bauthor{\bsnm{{Leinert}}, \binits{C.}},
\bauthor{\bsnm{{Richter}}, \binits{I.}},
\bauthor{\bsnm{{Pitz}}, \binits{E.}},
\bauthor{\bsnm{{Hanner}}, \binits{M.}}:
\byear{1982},
\batitle{Helios zodiacal light measurements - a tabulated summary}.
\bjtitle{\aap}
\bvolume{110},
\bfpage{355}.
\adsurl{http://cdsads.u-strasbg.fr/abs/1982A\%26A...110..355L}.
\end{barticle}
\endbibitem

\bibitem[\protect\citeauthoryear{{Lemaire}}{2011}]{Lemaire2011}
\begin{botherref}
\oauthor{\bsnm{{Lemaire}}, \binits{J.F.}}:
2011,
{Determination of coronal temperatures from electron density profiles}.
\textit{arXiv e-prints},
arXiv:1112.3850.
\adsurl{2011arXiv1112.3850L}.
\end{botherref}
\endbibitem

\bibitem[\protect\citeauthoryear{Llebaria and Thernisien}{2001}]{Llebaria2001}
\begin{bchapter}
\bauthor{\bsnm{Llebaria}, \binits{A.}},
\bauthor{\bsnm{Thernisien}, \binits{A.}}:
\byear{2001},
\bctitle{Highly accurate photometric equalization of long sequences of coronal
  images}.
In: \bbtitle{International Symposium on Optical Science and Technology},
\bfpage{265}.
\bcomment{International Society for Optics and Photonics}.
\doiurl{https://doi.org/10.1117/12.447183}.
\end{bchapter}
\endbibitem

\bibitem[\protect\citeauthoryear{Llebaria, Lamy, and Bout}{2004}]{Llebaria2004}
\begin{bchapter}
\bauthor{\bsnm{Llebaria}, \binits{A.}},
\bauthor{\bsnm{Lamy}, \binits{P.}},
\bauthor{\bsnm{Bout}, \binits{M.}}:
\byear{2004},
\bctitle{Lessons learnt from the soho lasco-c2 calibration}.
In: \bbtitle{Proc. of SPIE Vol}
\bseriesno{5171},
\bfpage{27}.
\doiurl{https://doi.org/10.1117/12.506159}.
\end{bchapter}
\endbibitem

\bibitem[\protect\citeauthoryear{{MacQueen} and {Fisher}}{1983}]{MacQueen1983}
\begin{barticle}
\bauthor{\bsnm{{MacQueen}}, \binits{R.M.}},
\bauthor{\bsnm{{Fisher}}, \binits{R.R.}}:
\byear{1983},
\batitle{The kinematics of solar inner coronal transients}.
\bjtitle{\solphys}
\bvolume{89},
\bfpage{89}.
\doiurl{https://doi.org/10.1007/BF00211955}.
\adsurl{http://cdsads.u-strasbg.fr/abs/1983SoPh...89...89M}.
\end{barticle}
\endbibitem

\bibitem[\protect\citeauthoryear{{MacQueen} et~al.}{1980}]{MacQueen1980}
\begin{barticle}
\bauthor{\bsnm{{MacQueen}}, \binits{R.M.}},
\bauthor{\bsnm{{Csoeke-Poeckh}}, \binits{A.}},
\bauthor{\bsnm{{Hildner}}, \binits{E.}},
\bauthor{\bsnm{{House}}, \binits{L.}},
\bauthor{\bsnm{{Reynolds}}, \binits{R.}},
\bauthor{\bsnm{{Stanger}}, \binits{A.}},
\bauthor{\bsnm{{Tepoel}}, \binits{H.}},
\bauthor{\bsnm{{Wagner}}, \binits{W.}}:
\byear{1980},
\batitle{{The High Altitude Observatory coronagraph/polarimeter on the Solar
  Maximum Mission.}}
\bjtitle{\solphys}
\bvolume{65},
\bfpage{91}.
\doiurl{https://doi.org/10.1007/BF00151386}.
\adsurl{1980SoPh...65...91M}.
\end{barticle}
\endbibitem

\bibitem[\protect\citeauthoryear{{Michard}}{1954}]{Michard1954b}
\begin{barticle}
\bauthor{\bsnm{{Michard}}, \binits{R.}}:
\byear{1954},
\batitle{{Densit{\'e}s {\'e}lectroniques dans la Couronne externe du 25
  f{\'e}vrier 1952}}.
\bjtitle{Annales d'Astrophysique}
\bvolume{17},
\bfpage{429}.
\adsurl{1954AnAp...17..429M}.
\end{barticle}
\endbibitem

\bibitem[\protect\citeauthoryear{{Michard}}{1956}]{Michard1956}
\begin{barticle}
\bauthor{\bsnm{{Michard}}, \binits{R.}}:
\byear{1956},
\batitle{{Observations a{\'e}riennes de la couronne externe {\`a} l'{\'e}clipse
  du 20 juin 1955}}.
\bjtitle{Annales d'Astrophysique}
\bvolume{19},
\bfpage{229}.
\adsurl{1956AnAp...19..229M}.
\end{barticle}
\endbibitem

\bibitem[\protect\citeauthoryear{{Michard} and
  {Sotirovski}}{1965}]{Michard1965}
\begin{barticle}
\bauthor{\bsnm{{Michard}}, \binits{R.}},
\bauthor{\bsnm{{Sotirovski}}, \binits{P.}}:
\byear{1965},
\batitle{Contributions {\`a} l'{\'e}tude de la couronne lors de l'{\'e}clipse
  du 15 f{\'e}vrier 1961. ii. photom{\'e}trie et polarim{\`e}tre de la
  couronne}.
\bjtitle{Annales d'Astrophysique}
\bvolume{28},
\bfpage{96}.
\adsurl{1965AnAp...28...96M}.
\end{barticle}
\endbibitem

\bibitem[\protect\citeauthoryear{{Michard} et~al.}{1954}]{Michard1954a}
\begin{barticle}
\bauthor{\bsnm{{Michard}}, \binits{R.}},
\bauthor{\bsnm{{Dollfus}}, \binits{A.}},
\bauthor{\bsnm{{Pecker}}, \binits{J.C.}},
\bauthor{\bsnm{{Laffineur}}, \binits{M.}},
\bauthor{\bsnm{{D'Azambuja}}, \binits{M.}}:
\byear{1954},
\batitle{{I Observations photom{\'e}triques et polarim{\'e}triques de la
  Couronne externe}}.
\bjtitle{Annales d'Astrophysique}
\bvolume{17},
\bfpage{320}.
\adsurl{1954AnAp...17..320M}.
\end{barticle}
\endbibitem

\bibitem[\protect\citeauthoryear{{Moran} et~al.}{2006}]{Moran2006}
\begin{barticle}
\bauthor{\bsnm{{Moran}}, \binits{T.G.}},
\bauthor{\bsnm{{Davila}}, \binits{J.M.}},
\bauthor{\bsnm{{Morrill}}, \binits{J.S.}},
\bauthor{\bsnm{{Wang}}, \binits{D.}},
\bauthor{\bsnm{{Howard}}, \binits{R.}}:
\byear{2006},
\batitle{Solar and heliospheric observatory/large angle spectrometric
  coronagraph polarimetric calibration}.
\bjtitle{\solphys}
\bvolume{237},
\bfpage{211}.
\doiurl{https://doi.org/10.1007/s11207-006-0147-9}.
\adsurl{2006SoPh..237..211M}.
\end{barticle}
\endbibitem

\bibitem[\protect\citeauthoryear{{Morrill} et~al.}{2006}]{Morrill2006}
\begin{barticle}
\bauthor{\bsnm{{Morrill}}, \binits{J.S.}},
\bauthor{\bsnm{{Korendyke}}, \binits{C.M.}},
\bauthor{\bsnm{{Brueckner}}, \binits{G.E.}},
\bauthor{\bsnm{{Giovane}}, \binits{F.}},
\bauthor{\bsnm{{Howard}}, \binits{R.A.}},
\bauthor{\bsnm{{Koomen}}, \binits{M.}},
\bauthor{\bsnm{{Moses}}, \binits{D.}},
\bauthor{\bsnm{{Plunkett}}, \binits{S.P.}},
\bauthor{\bsnm{{Vourlidas}}, \binits{A.}},
\bauthor{\bsnm{{Esfandiari}}, \binits{E.}},
\bauthor{\bsnm{{Rich}}, \binits{N.}},
\bauthor{\bsnm{{Wang}}, \binits{D.}},
\bauthor{\bsnm{{Thernisien}}, \binits{A.F.}},
\bauthor{\bsnm{{Lamy}}, \binits{P.}},
\bauthor{\bsnm{{Llebaria}}, \binits{A.}},
\bauthor{\bsnm{{Biesecker}}, \binits{D.}},
\bauthor{\bsnm{{Michels}}, \binits{D.}},
\bauthor{\bsnm{{Gong}}, \binits{Q.}},
\bauthor{\bsnm{{Andrews}}, \binits{M.}}:
\byear{2006},
\batitle{Calibration of the soho/lasco c3 white light coronagraph}.
\bjtitle{\solphys}
\bvolume{233},
\bfpage{331}.
\doiurl{https://doi.org/10.1007/s11207-006-2058-1}.
\adsurl{2006SoPh..233..331M}.
\end{barticle}
\endbibitem

\bibitem[\protect\citeauthoryear{Mueller}{1943}]{Mueller1943}
\begin{botherref}
\oauthor{\bsnm{Mueller}, \binits{H.}}:
1943,
\textit{Memorandum on the polarization optics of the photo elastic shutter osrd
  project oemsr-576 report no. 2},
November.
\end{botherref}
\endbibitem

\bibitem[\protect\citeauthoryear{{Muhleman} and
  {Anderson}}{1981}]{Muhleman1981}
\begin{barticle}
\bauthor{\bsnm{{Muhleman}}, \binits{D.O.}},
\bauthor{\bsnm{{Anderson}}, \binits{J.D.}}:
\byear{1981},
\batitle{{Solar wind electron densities from Viking dual-frequency radio
  measurements}}.
\bjtitle{\apj}
\bvolume{247},
\bfpage{1093}.
\doiurl{https://doi.org/10.1086/159119}.
\adsurl{1981ApJ...247.1093M}.
\end{barticle}
\endbibitem

\bibitem[\protect\citeauthoryear{{Mutschlecner}, {Keller}, and
  {Tabor}}{1976}]{Mutschlecner1976}
\begin{bchapter}
\bauthor{\bsnm{{Mutschlecner}}, \binits{J.P.}},
\bauthor{\bsnm{{Keller}}, \binits{C.F.}},
\bauthor{\bsnm{{Tabor}}, \binits{J.E.}}:
\byear{1976},
\bctitle{Electron density models of the solar corona.}
In: \bbtitle{Bulletin of the American Astronomical Society}
\bseriesno{8},
\bfpage{397}.
\adsurl{1976BAAS....8..397M}.
\end{bchapter}
\endbibitem

\bibitem[\protect\citeauthoryear{{Newkirk}}{1967}]{Newkirk1967}
\begin{barticle}
\bauthor{\bsnm{{Newkirk}}, \binits{J.} \bsuffix{Gordon}}:
\byear{1967},
\batitle{{Structure of the Solar Corona}}.
\bjtitle{\araa}
\bvolume{5},
\bfpage{213}.
\doiurl{https://doi.org/10.1146/annurev.aa.05.090167.001241}.
\adsurl{1967ARA&A...5..213N}.
\end{barticle}
\endbibitem

\bibitem[\protect\citeauthoryear{Pagot et~al.}{2014}]{Pagot2014}
\begin{barticle}
\bauthor{\bsnm{Pagot}, \binits{E.}},
\bauthor{\bsnm{Lamy}, \binits{P.}},
\bauthor{\bsnm{Llebaria}, \binits{A.}},
\bauthor{\bsnm{Boclet}, \binits{B.}}:
\byear{2014},
\batitle{Automated processing of lasco coronal images: Spurious
  point-source-filtering and missing-blocks correction}.
\bjtitle{Solar Physics}
\bvolume{289},
\bfpage{1433}.
\doiurl{https://doi.org/10.1007/s11207-013-0402-9}.
\end{barticle}
\endbibitem

\bibitem[\protect\citeauthoryear{{Pepin}}{1970}]{Pepin1970}
\begin{barticle}
\bauthor{\bsnm{{Pepin}}, \binits{T.J.}}:
\byear{1970},
\batitle{{Observations of the Brightness and Polarization of the Outer Corona
  during the 1966 November 12 Total Eclipse of the Sun}}.
\bjtitle{\apj}
\bvolume{159},
\bfpage{1067}.
\doiurl{https://doi.org/10.1086/150384}.
\adsurl{1970ApJ...159.1067P}.
\end{barticle}
\endbibitem

\bibitem[\protect\citeauthoryear{Qu{\'e}merais and Lamy}{2002}]{Quemerais2002}
\begin{barticle}
\bauthor{\bsnm{Qu{\'e}merais}, \binits{E.}},
\bauthor{\bsnm{Lamy}, \binits{P.}}:
\byear{2002},
\batitle{Two-dimensional electron density in the solar corona from inversion of
  white light images-application to soho/lasco-c2 observations}.
\bjtitle{Astronomy \& Astrophysics}
\bvolume{393},
\bfpage{295}.
\doiurl{https://doi.org/DOI: 10.1051/0004-6361:20021019}.
\end{barticle}
\endbibitem

\bibitem[\protect\citeauthoryear{{Richter}, {Leinert}, and
  {Planck}}{1982}]{Richter1982}
\begin{barticle}
\bauthor{\bsnm{{Richter}}, \binits{I.}},
\bauthor{\bsnm{{Leinert}}, \binits{C.}},
\bauthor{\bsnm{{Planck}}, \binits{B.}}:
\byear{1982},
\batitle{{Search for short term variations of zodiacal light and optical
  detection of interplanetary plasma clouds.}}
\bjtitle{\aap}
\bvolume{110},
\bfpage{115}.
\adsurl{1982A&A...110..115R}.
\end{barticle}
\endbibitem

\bibitem[\protect\citeauthoryear{{Saito}, {Hata}, and {Tojo}}{1972}]{Saito1972}
\begin{barticle}
\bauthor{\bsnm{{Saito}}, \binits{K.}},
\bauthor{\bsnm{{Hata}}, \binits{S.}},
\bauthor{\bsnm{{Tojo}}, \binits{A.}}:
\byear{1972},
\batitle{{Photometric and polarimetric analysis of the coronal streamers
  observed at the March 7, 1970 Mexican eclipse}}.
\bjtitle{Annals of the Tokyo Astronomical Observatory}
\bvolume{13},
\bfpage{91}.
\adsurl{1972AnTok..13...93S}.
\end{barticle}
\endbibitem

\bibitem[\protect\citeauthoryear{{Saito} et~al.}{1970}]{Saito1970}
\begin{barticle}
\bauthor{\bsnm{{Saito}}, \binits{K.}},
\bauthor{\bsnm{{Makita}}, \binits{M.}},
\bauthor{\bsnm{{Nishi}}, \binits{K.}},
\bauthor{\bsnm{{Hata}}, \binits{S.}}:
\byear{1970},
\batitle{{A non-spherical axisymmetric model of the solar K corona of the
  minimum type}}.
\bjtitle{Annals of the Tokyo Astronomical Observatory}
\bvolume{12},
\bfpage{51}.
\doiurl{https://doi.org/10.1007/BF00150879}.
\adsurl{1970AnTok..12...53S}.
\end{barticle}
\endbibitem

\bibitem[\protect\citeauthoryear{{Stenborg}, {Howard}, and
  {Stauffer}}{2018}]{Stenborg2018}
\begin{barticle}
\bauthor{\bsnm{{Stenborg}}, \binits{G.}},
\bauthor{\bsnm{{Howard}}, \binits{R.A.}},
\bauthor{\bsnm{{Stauffer}}, \binits{J.R.}}:
\byear{2018},
\batitle{Characterization of the white-light brightness of the f-corona between
  5 and 24 elongation}.
\bjtitle{\apj}
\bvolume{862},
\bfpage{168}.
\doiurl{https://doi.org/10.3847/1538-4357/aacea3}.
\adsurl{2018ApJ...862..168S}.
\end{barticle}
\endbibitem

\bibitem[\protect\citeauthoryear{Thernisien et~al.}{2006}]{Thernisien2006}
\begin{barticle}
\bauthor{\bsnm{Thernisien}, \binits{A.}},
\bauthor{\bsnm{Morrill}, \binits{J.}},
\bauthor{\bsnm{Howard}, \binits{R.}},
\bauthor{\bsnm{Wang}, \binits{D.}}:
\byear{2006},
\batitle{Photometric calibration of the lasco-c3 coronagraph using stars}.
\bjtitle{Solar Physics}
\bvolume{233},
\bfpage{155}.
\doiurl{https://doi.org/10.1007/s11207-006-2047-4}.
\end{barticle}
\endbibitem

\bibitem[\protect\citeauthoryear{{Wang} and {Sheeley}}{2003}]{Wang2003}
\begin{barticle}
\bauthor{\bsnm{{Wang}}, \binits{Y.-M.}},
\bauthor{\bsnm{{Sheeley}}, \binits{J.} \bsuffix{N.~R.}}:
\byear{2003},
\batitle{{On the Fluctuating Component of the Sun's Large-Scale Magnetic
  Field}}.
\bjtitle{\apj}
\bvolume{590},
\bfpage{1111}.
\doiurl{https://doi.org/10.1086/375026}.
\adsurl{2003ApJ...590.1111W}.
\end{barticle}
\endbibitem

\end{thebibliography}
\nocite{*}

\appendix

\section*{Appendix I: Determination of the C3 stray light ramp}
\label{AppendixRamp}

The C3 images suffer from a non-radially symmetric pattern of stray light which we identified very early in our analysis of the F-corona. 
Monitoring the temporal variation of the radiance in two windows above the north and south poles was expected to reveal the semi-annual variation as SOHO (and the Earth) oscillates about the plane of symmetry of the Zodiacal Cloud. 
The two radiance profiles, north and south, were expected to cross twice per year when SOHO crosses this plane and the corresponding epochs would constraint its orientation, namely the longitude of its ascending node.
However, and contrary to the equivalent C2 profiles which behaved as expected, the C3 profiles, both unpolarized and polarized were conspicuously offset and never crossed each other.

\cite{Morrill2006} presented a solution for determining the component of the C3 stray light responsible for this anomaly in the case of routine unpolarized images obtained with the ``clear'' broadband images of 1024$\times$1024 pixels.
They took advantage of the roll maneuver performed by SOHO on 19 March 1996 and used the images taken at roll positions of $0\deg$ and $180\deg$.
The difference between these two images removed the coronal scene and revealed a diagonally oriented pattern, so-called ``ramp'', shown in their Figure~20.
Retrospectively, there is a flaw in this procedure as it assumes that the coronal background, prominently the F-corona, is symmetric. 
This situation occurs only twice per year, in late June and December (see below), when SOHO crosses the plane of symmetry of the Zodiacal Cloud.
The second half of March is unfortunately the time when the asymmetry of the F-corona is maximized, thus biasing the determination of the ramp.

\begin{figure}[htbp!]
	\centering
	\includegraphics[width=\textwidth]{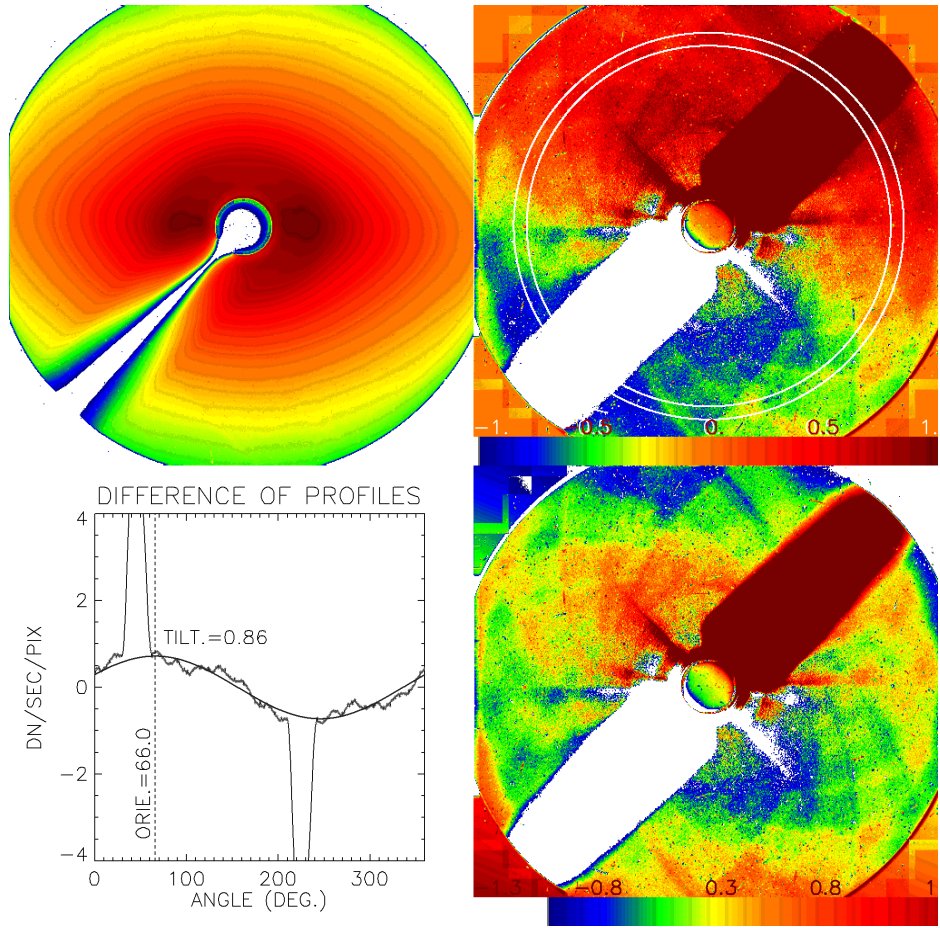}
	\caption{The upper left panel displays a processed image obtained with the LASCO-C3 coronagraph on 21 June 1997 when SOHO crossed the symmetry plane of the Zodiacal cloud. 
The upper  right panel displays the difference between this image and the same image rotated by 180$\deg$ around the center of the Sun. 
The two white circles define the ring used to analyze the ramp. 
The lower left panel displays the mean circular profile of the rings extracted from the difference image and the fit by a sinusoidal function. 
The lower right panel displays the difference image corrected by the ramp.}
	\label{Fig:Image_Diff}
\end{figure}

\begin{figure}[htbp!]
	\centering
	\includegraphics[width=\textwidth]{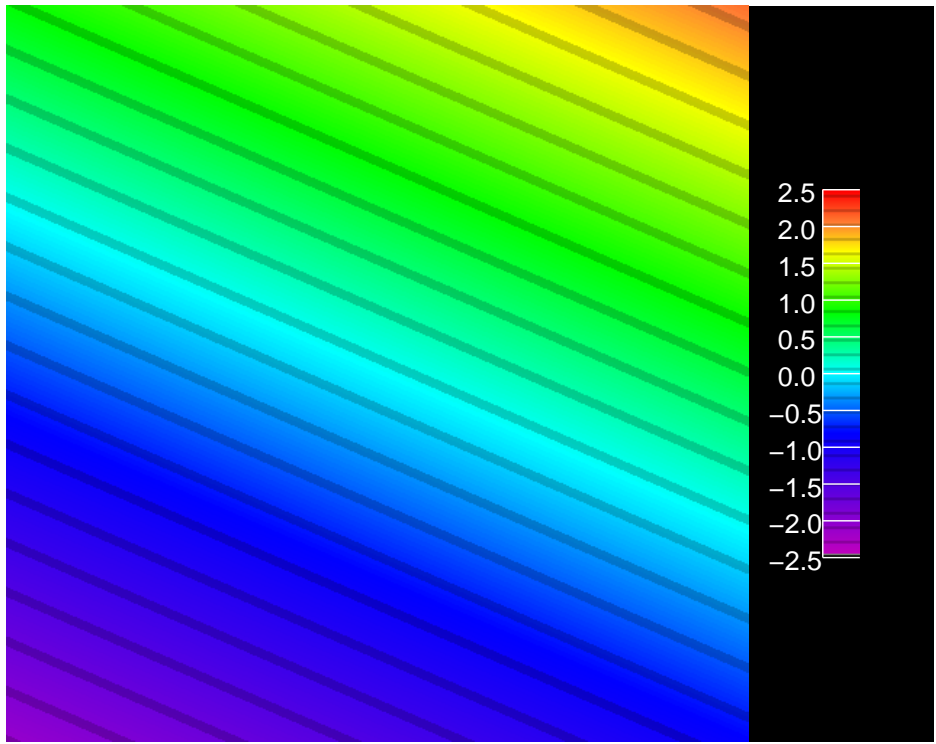}
	\caption{Illustration of the ramp constructed from the sinusoidal profile fitted to the difference image that defines its orientation and gradient.}
	\label{Fig:Ramp}
\end{figure}

We used a different approach that exploits these two plane crossings insuring that the F-corona is symmetric. 
The corresponding dates are known from the C2 profiles as described above, approximately 20 June and 20 December, very close to past determinations of the line of nodes of the symmetry plane (\eg \cite{Leinert1980}): 24 June and 23 December $\pm$2 days.
In practice, the extrema are sufficiently shallow that the symmetry condition holds over a time interval of approximately $\pm$10 days centered on the above dates.
This allowed averaging typically 20 polarized images (one polarization sequence per day) and many more unpolarized images, thus improving the determination of the ramps. 
These images were preprocessed as described in Section~\ref{Sub:DataProc} and averaged images were calculated at each node.
The center of symmetry of the F-corona was accurately determined (in practice, it is the same for all images). 
A $180\deg$ rotated image was generated and subtracted from the original image.
The difference image revealed the diagonally oriented ramp, faint arcs resulting from multiple internal reflections in the instrument, and remnants of the K-corona (Figure~\ref{Fig:Image_Diff}).
The pylon holding the occulter and its symmetric create a gap in the images appearing as a diagonal band which is blocked by an appropriate mask wide enough to encompass the specific stray light pattern associated with the pylon.
We found that the ramp could be confidently modeled by a plane whose parameters were determined by considering a thin ring that best capture its geometry while excluding the artifacts mentioned above as well as the outermost region affected by various stray effects.
A sinusoidal function was fitted to the average circular profile of this ring (Figure~\ref{Fig:Image_Diff}) and used to characterize the line of steepest descent of the plane: its phase defines its orientation 
and its amplitude yields the gradient or tilt of the plane, more precisely twice that of the ramp because of the difference (Figure~\ref{Fig:Ramp}). 
This procedure left the absolute level of the ramp undefined and the offsets were determined by imposing that the corrected F-corona be symmetric along the north-south direction.
 
The overall procedure was applied at the nodes spanning the 24 years of operation (limited to the first four years for the $0\deg$ polarizer because of its degradation thereafter) to possibly uncover temporal variations of the ramps.  
This in-depth analysis allowed us to reach the following conclusions.

\begin{figure}[htbp!]
	\centering
	\includegraphics[width=\textwidth]{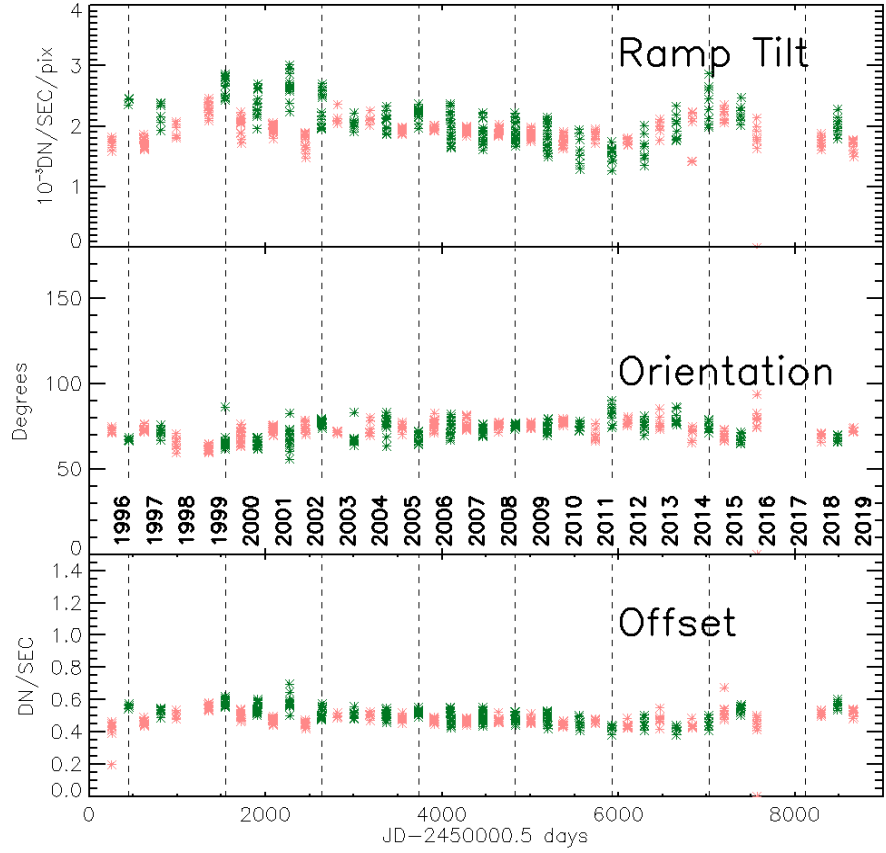}
	\caption{Temporal variations of the ramp parameters of the unpolarized images: tilt, orientation, and offset. 
	The colors distinguish the June ascending (pink) and December descending (green) nodes of the orbit of SOHO as it crosses the symmetry plane of the Zodiacal Cloud.
	At each node, several determinations of the parameters were obtained using different images.}
	\label{Fig:ramp_parameters}
\end{figure}

\begin{itemize}
	\item The planar ramp of the unpolarized images is extremely well determined and remarkably constant.
	The temporal variations of the ramp parameters remain within their 2$\sigma$ uncertainties and the $1\sigma$ uncertainties of the individual determinations do not exceed 10\% with only a few exceptions (Figure~\ref{Fig:ramp_parameters}).
	\item A similar behaviour is observed for the parameters of the ramps of the polarized images with however larger errors, typically 30\%. 
	\item The geometry of the ramps of the  polarized images is compatible with that of the unpolarized images.
\end{itemize}

These conclusions led us to adopt the ramp geometry of the unpolarized images for all polarized images with appropriate scaling of the parameters defined by the well-defined ratios of the individual polarized images to the unpolarized images.

\end{article} 

\end{document}